\documentclass[twocolumn,
nofootinbib,
 amsmath,amssymb,
 aps,
 prd,
 floatfix,
]{revtex4}
\usepackage{tabularray}
\usepackage{graphicx}
\usepackage{dcolumn}
\usepackage{bm}
\usepackage[normalem]{ulem}
\newcommand{\overbar}[1]{\mkern 1.5mu\overline{\mkern-1.5mu#1\mkern-1.5mu}\mkern 1.5mu}

\usepackage[center]{caption}
\usepackage{xcolor}

\usepackage{hyperref}
\usepackage{cleveref}
\usepackage{gensymb}
\begin{document}


\title{Violation of the equivalence principle induced by oscillating rest mass and transition frequency, and its detection in atom interferometers}

\author{Jordan Gu\'e,\footnote{jordan.gue@obspm.fr} Aur\'elien Hees, Peter Wolf}
\affiliation{%
 SYRTE, Observatoire de Paris, Universit\'e PSL, CNRS, Sorbonne Universit\'e, LNE, 61 avenue de l’Observatoire 75014 Paris, France\\
}%

\begin{abstract}
We present a theoretical investigation of the expected experimental signals produced by freely falling atoms with time oscillating mass and transition frequency. These oscillations could be produced in a variety of models, in particular, models of scalar dark matter (DM) non universally coupled to the standard matter (SM) such as axion-like particles (ALP) and dilatons. Performing complete and rigorous calculations, we show that, on one hand, two different atomic species would accelerate at a different rate, and on the other hand, they would produce a non-zero differential phase shift in atom interferometers (AI). The former would produce observable signals in equivalence principle tests like the recent MICROSCOPE mission, and we provide a corresponding sensitivity estimate, showing that MICROSCOPE can reach beyond the best existing searches in the ALP case. We also compare the expected sensitivity of two future AI experiments, namely the AION-10 gradiometer and an isotope differential AI considered for MAGIS-100, that we will refer to as SPID. We show that the SPID setup would be more sensitive to these dark matter fields compared to the gradiometer one, assuming equivalent experimental parameters.
\end{abstract}

\maketitle


\section{\label{sec:Intro}Introduction}

General Relativity (GR) is, alongside the Standard Model (SM) of particle physics, one of the building blocks of fundamental physics \cite{Weinberg}. One of the main assumptions of this theory is the Einstein Equivalence Principle (EEP), which includes the weak equivalence principle (WEP), local position invariance (LPI) and local Lorentz invariance (LLI). The WEP, also referred as the universality of free fall (UFF), states that gravitational and inertial masses are equivalent, implying that all bodies fall at the same rate in the same gravitational field. While being experimentally tested at extraordinary levels of precision by on ground measurements (e.g. \cite{Wagner12}) or space tests \cite{Microscope22}, many theories beyond GR and the SM predict its violation \cite{damour94,Fayet18,Fayet2019}. More generally, the EEP postulates that all types of energy (i.e all SM fields) are universally coupled to gravity, such that the underlying gravity theory must be metric \cite{Will14}. Since EEP is not based on a fundamental symmetry of the Universe, it is also expected to be broken at some scale, for example by the space-time variation of fundamental constants \cite{Uzan11}.

Spacetime varying fundamental constants arise in many theories \cite{Uzan11, damour:2010zr,damour:2010ve}, in particular models of massive scalar fields non-universally coupled to SM. Many of these new fields are good dark matter (DM) candidates, as they are oscillating at their Compton frequency $\omega$, thus behaving as cold DM at late cosmological times, when $\omega > H$, the Hubble constant. In particular, these oscillations can lead to spacetime variation of the rest mass and transition frequency of atoms, with magnitudes depending on various parameters including mass and charge numbers, making such oscillations not universal. Spacetime varying constants have been widely studied in the case of a scalar dilatonic field non universally coupled to SM fields, e.g in \cite{damour:2010zr}. Recently, \cite{Kim22} showed that the coupling between an axion-like particle (ALP), another promising ULDM candidate \cite{Marsh16}, and the gluon field induces a variation of the pion mass, leading to a variation of the rest mass and transition frequency of atoms as well. Therefore, we will be interested in the search for these two ULDM candidates through various couplings, on one hand using classical tests of UFF, and on the other hand through atom interferometry. Atom interferometers (AI) provide quantum tests of the EEP where one can measure the acceleration and/or transition frequency of atoms freely falling in the Earth gravitational field, by using the wave behavior of matter. In such experiments, light pulses are used to split, reflect and recombine the atomic wavepackets, replacing the beam-splitters and mirrors of usual optical interferometers. The phase shift between the two wavepackets after recombination relies on the momentum exchange between atom and light, the time between the various light pulses, the transition frequencies, and the acceleration of the wavepackets.

The first ambition of this paper is to derive carefully the classical differential acceleration between two tests masses (see e.g. \cite{Graham16}) and the phase shift in several AI setups, taking into account all oscillating effects. For the former, we show the signal is predominantly dependent on the atomic rest mass oscillation amplitude, while for the latter, it depends on the AI configuration setup. The first one under consideration is the two-photon transition configuration, whose signal is a combination of both rest mass and transition frequency oscillations, and whose dominant effect depends on the setup. One of these contributions, coming from the oscillation of the rest mass has already been derived in \cite{Geraci16} but we show that a non negligible contribution dependent on the velocity of the atoms in the galactic DM rest frame must be added, which would improve the constraints presented in \cite{Geraci16} by 4 orders of magnitude. The second case regards single photon transition interferometers, in particular gradiometers, as proposed in \cite{Graham13} and then considered for the search of ultralight dark matter (ULDM) fields in the AION (Atom Interferometer Observatory and Network) experiment \cite{Badurina20,Badurina22}. In that setup, two interferometric sequences are performed at different elevation, using the same atomic species in both of them. This makes possible the use of the same laser beam for both interactions, which strongly suppresses the effect of laser phase noise. In addition, a very large momentum transfer (LMT) from laser beams to atomic wavepackets is possible, making such experiments very sensitive to transition frequency oscillations, as explicitly shown in \cite{Arvanitaki18}. However, the signal is limited by the reduced time the atomic wavepackets spend in their excited state.
The second single photon transitions setup under consideration is based on the same LMT sequence as in gradiometers, but would consist on the differential phase shift measurement between two atomic isotopes which perform the AI sequence at the same elevation. We name this variation SPID for Single Photon transition Isotope Differential AI. A similar setup has already been proposed for the search of ULDM in MAGIS-100 \cite{Abe21}, but using two-photon Bragg transitions instead of single photon transition. However, as we shall see, both setups are equivalent in terms of sensitivity to ULDM, since they have identical signal and noise levels. 
Finally, for the two ULDM candidates studied, ALP and dilatons, we show that already existing experiments testing UFF or involving AI measurements can put constraints on their couplings with SM fields. Using the experimental parameters of AION-10, \cite{Badurina20,Badurina22}, we demonstrate that the SPID variation would be more sensitive to the couplings of these ULDM fields with SM fields compared to the gradiometer version of AION-10. Whilst a dual isotope mode with large LMT has been previously proposed by the MAGIS-100 collaboration \cite{Abe21}, no detailed calculations and modelling have been provided, and the sensitivity to couplings between the dilaton field and SM fields was not calculated, which we do in this paper. 

\section{Observational signatures induced by oscillations in atom's rest mass and transition frequency}
\label{sec:signatures}

In this section, we consider a very generic phenomenological model where 
both the rest mass of an atom $A$ and its internal energy oscillate. We derive the observable signatures of such oscillations in some systems of interest and we will then apply those results in two different theoretical models in Sec.~\ref{sec:charges_DM}. 

We start by introducing a phenomenological parametrization of the oscillation of both the atom rest mass $m_A$ and its transition frequency $\omega_A$ as
\begin{subequations}
\begin{align}
    m_A(t) &= m^0_A\left(1+Q^A_M \cos(\omega t + \phi_0) \right)\, \label{general_mass_osc}, \\
    \omega_A(t) &= \omega^0_A\left(1+Q^A_\omega \cos(\omega t  + \phi_0) \right) \, \label{general_freq_osc} ,
\end{align}
\label{general_mass_freq_osc}
\end{subequations}
where $\omega$ is the oscillation frequency, $\phi_0$ a phase, $m^0_A, \omega^0_A$ are respectively the unperturbed rest mass and transition frequency of $A$ and $Q^A_M, Q^A_\omega$ are respectively the mass and frequency charge of the atom $A$. These characterize the relative amplitude of oscillation, supposed much smaller than unity.

The motion of particles can be described by the point mass action
\begin{align}
    S_\mathrm{mat} = -\sum_A\int_A d\tau (m_A(t)c^2+E_\mathrm{int}(t)) \, ,
    \label{macro_action}
\end{align}
where the first term represents the rest mass and the second term is the internal energy contribution ($E_\mathrm{int}=0$ when the atom is in its ground state and $E_\mathrm{int}=\hbar \omega_A$ when the atom is in the excited state). $d\tau$ is the proper time interval defined as $c^2d\tau^2=-g_{\alpha\beta} dx^\alpha dx^\beta$, where $g_{\mu\nu}$ is the space-time metric. For an atom of rest mass $m_A$ and in an internal state characterized by an energy $E_\mathrm{int}$, the Lagrangian derived from Eq.\eqref{macro_action} becomes
\begin{equation}
    L = -\left(m_A(t)c^2+E_\mathrm{int}(t)\right)\Big(1-\frac{v_A^2(t)}{2c^2}\Big)
    \label{macro_lagrangian}\, ,
\end{equation}
to first order in $(v_A/c)^2$ (where $v_A$ is the coordinate velocity of the atom) and considering a flat space-time, i.e $g_{\mu\nu}=\eta_{\mu\nu}=\mathrm{diag}(-1,1,1,1)$, the Minkowskian metric. 

In the following, we show how oscillating rest mass and transition frequency characterized by Eq.~\eqref{general_mass_freq_osc} produce observational signatures in different experiments.

\subsection{\label{classical_general}Classical trajectories of test masses}

In the classical derivation, as long as the atom is not in its energetic ground state, both oscillations Eq.~\eqref{general_mass_freq_osc} contribute to a violation of the equivalence principle through Eq.~\eqref{macro_lagrangian}. The Euler-Lagrange equation applied to the Lagrangian from Eq.~\eqref{macro_lagrangian} leads to the conservation of the momentum of body $A$, i.e. $\left(m_A(t) + E_\mathrm{int}(t)/c^2 \right)\times v_A(t) = \mathrm{cst}$. The oscillation of the mass and internal energy of the atom will therefore induce a perturbation to the acceleration which, to first order in both the $Q$ factors and in $E_\mathrm{int}/m_A^{0}c^2$, is
\begin{align}
    \left[\vec a_A \right]_{\overbar{\mathrm{UFF}}} \approx \omega \vec v_A \left(Q^A_M +\frac{\hbar \omega^0_A}{m^0_Ac^2} Q^A_\omega\right)\sin(\omega t+\phi_0) \, ,
    \label{a_vUFF}
\end{align}
where $\vec a_A=d\vec v_A/dt$ is the acceleration. In general, $m^0_Ac^2/\hbar \omega^0_A > 10^{10}$, while $Q^A_\omega/Q^A_M$ is of the order of $10^3$ at maximum (see Table.~\ref{axionic_dilatonic_charge_table}) implying that the first term of Eq.~\eqref{a_vUFF} is dominant. 
It follows a differential acceleration between two macroscopic bodies of different composition $A$ and $B$ with the same initial velocities given by
\begin{align}
    \Delta \vec a &= \vec a_A - \vec a_B \approx \omega  v_0 \left(Q^A_M-Q^B_M\right)\sin(\omega t+\phi_0)\hat e_v \, ,
    \label{eq:delta_a_UFF} 
\end{align}
where $\vec v_A = \vec v_B \equiv \vec v_0 = v_0 \hat e_v$ is the unperturbed velocity, i.e. the  velocity at zeroth order in $Q_M$.

If the two test masses have different charges $Q_M$, Eq.~\eqref{eq:delta_a_UFF} implies a non zero differential acceleration between them, corresponding to a UFF violation. The signature induced by the differential acceleration from Eq.~\eqref{eq:delta_a_UFF} can be searched for using UFF measurements using macroscopic test masses such as torsion balances on Earth \cite{Wagner12}, the MICROSCOPE experiment \cite{Microscope17} or Lunar Laser Ranging \cite{Williams09}\footnote{Note that Lunar Laser Ranging measures a combination of both the weak and strong equivalence principle.}.

\subsection{\label{phase_AI_general}Phase observable in atom interferometry}

As derived in the previous section, the time-dependent mass and internal frequency Eq.~\eqref{general_mass_freq_osc} produce a differential acceleration between two atoms. Standard UFF tests search for such a differential acceleration of macroscopic test masses. Atom interferometry (AI) constitutes the quantum equivalent of such classical experiments. Therefore, as we shall see in this section, AI experiments are also sensitive to oscillations from Eq.~\eqref{general_mass_freq_osc}.

In this section, we will derive the observable signatures produced by oscillating atomic rest mass and transition energy on various AI configurations. In the most generic case, AI consists of generating interference between atomic wavepackets that are split and then recombined using atom-laser interactions. We classify the various AI schemes into two classes: (i) AI schemes using two-photon transitions and (ii) AI schemes using single photon transitions. In the former, the transition from a stable momentum state to another stable momentum state is done using at least two photons, while in the latter, only one photon is needed for such a transition.

We will present the impact of the rest mass and atomic transition frequency oscillations on the measurable phase of an AI. The detailed calculations are performed following the methodology described in \cite{Storey} and are presented in details in Appendix~\ref{ap:MZ_phase} while in the main text, we focus on the discussion of the results. 

\subsubsection{Two-photon transitions}
\begin{figure}[h!]
\centering
\includegraphics[scale=0.3]{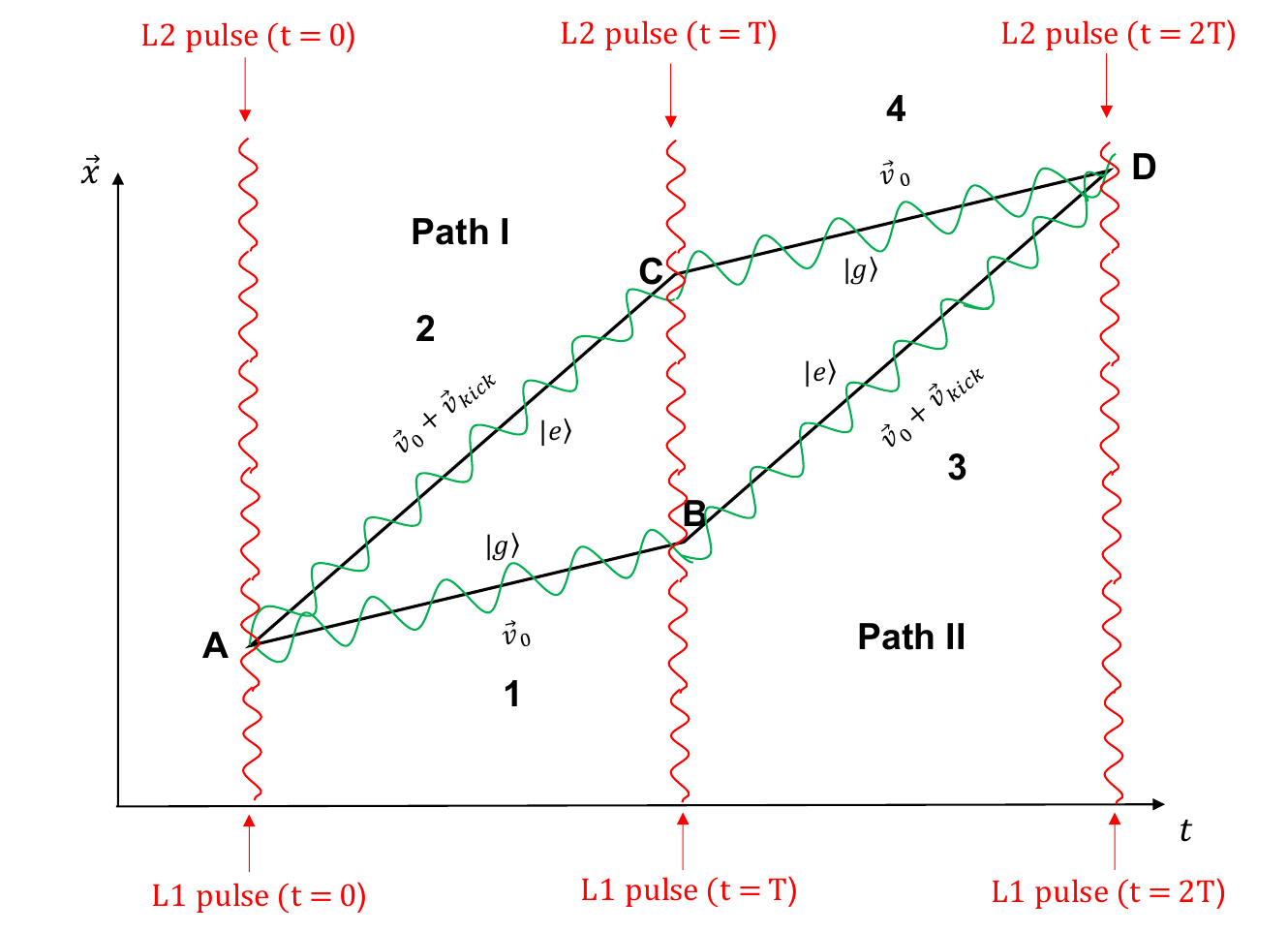}
\caption{Two-photon Raman transition interferometry sequence considered in this paper. The atomic wavepackets start in the ground state  and the laser pulses change the internal energy state. The Bragg equivalent has the same space-time diagram, but the wavepackets do not change their internal energy state. In black are shown the atomic paths without any perturbation, i.e straight lines. In green is shown the perturbed motion of the atoms induced by Eq.~\eqref{general_mass_freq_osc}, with exaggerated amplitude of oscillation.}
\label{Mach-Zehnder_osc}
\end{figure}
We start by studying two different two-photon transition interferometers. 
The first interferometric scheme considered is known as two-photon transition Raman interferometry. Its sequence is depicted in Fig.~\ref{Mach-Zehnder_osc}. In this setup, two-level free falling atoms $A$ enter the interferometer with an initial momentum $\hbar k$ and in their internal energetic ground state, noted $|g\rangle$ (i.e. their initial state is defined as $|g,\hbar k\rangle$). After entering the interferometer, they interact with a pair of laser waves $L_1$ and $L_2$ with respective frequencies $\omega_\mathrm{L_1}, \omega_\mathrm{L_2}$, whose energy difference is resonant with the transition of the $|g\rangle \rightarrow |e\rangle$ where $|e\rangle $ is the excited state of the two-levels atom, i.e $\omega^0_A = \omega_\mathrm{L_1}-\omega_\mathrm{L_2}$. This means that the atom first absorbs a photon from $L_1$ and stimulately emits a photon in $L_2$. This process splits the atoms into two spatially distant wavepackets, which means the state of each atom becomes the superposition of the two internal states: (i) the ground state that remains unchanged $|g,\hbar k\rangle$ and (ii) the excited state that has changed its momentum because of its interaction with the two lasers $|e,\hbar(k +k_\mathrm{eff})\rangle$, where $\hbar \vec k_\mathrm{eff} = \hbar (\vec k_\mathrm{L_1}-\vec k_\mathrm{L_2})$ denotes the effective momentum transfer, considering both lasers. This whole experimental manoeuvre is called a $\pi/2$ laser pulse since the transition amplitude from ground state to excited state corresponds to a probability $1/2$, i.e the state of the atoms is now a half-half superposition of the two different states.

The atom freely propagates inside the interferometer and at time $t=T$, both wavepackets undergo a $\pi$ pulse, which will invert the state of all atoms. In other words, depending on the internal state of the atom prior to the $\pi$ pulse, two states transitions happen : $|e,\hbar(k +k_\mathrm{eff})\rangle \rightarrow |g,\hbar k\rangle$ and $|g,\hbar k\rangle \rightarrow  |e,\hbar(k +k_\mathrm{eff})\rangle$. 

A final laser-atom interaction happens at time $t=2T$, where a second $\pi/2$ pulse divides the two incoming wavepackets into four different ones: two of them are in the state $|g,\hbar k\rangle$ and the remaining two are in the state $|e,\hbar(k +k_\mathrm{eff})\rangle$. The study of interference pattern between the wavepackets in the same state allows one to measure a phase shift difference.

The second two-photon transfer AI scheme considered is the two-photon transition Bragg-type interferometer. This scheme is similar to the Raman interferometry presented above except that the atoms remain in the same energy state during all the interferometric path, i.e, the laser pulses only change the momentum state of the atom.

For both interferometers described above, the effective wavevector $k_\mathrm{eff}$ depends on the setup of the experiment. If counterpropagating lasers are used, the atom absorbs a photon in one direction as a result of the interaction with the first laser, and emits another photon in the opposite direction during the interaction with the second laser, implying $k_\mathrm{eff}=k_\mathrm{L_1}+k_\mathrm{L_2}$. In general, this effective wave vector is multiple orders of magnitude larger than the transition frequency of the atom, i.e $k_\mathrm{eff} \gg \omega^0_A/c$. On the opposite, if co-propagating laser waves are used in the experimental setup, the absorption and emission directions are the same, implying $k_\mathrm{eff}=k_{L_1}-k_{L_2}$. In that case, $k_\mathrm{eff} = \omega^0_A/c$.

As already discussed in \cite{Geraci16}, the velocity kick experienced by the atom during its interaction with the laser beams is perturbed by the effective mass of the atom at the time of the kick. In addition to that effect, the beams being locked to a given frequency reference, their own frequency oscillates as $\omega_L(t) = \omega^0_L(1+Q^L_\omega \cos(\omega t + \phi_0))$, neglecting the additional phase coming from the propagation of photons to the atom (in the following sections, the travelling distance of the photon to reach the freely falling atoms will be of the order of 10 m maximum, which would induce a significant phase for oscillation frequencies $\omega \gtrapprox 10^6$ rad/s, way above the DM frequencies of interest for this paper, see Sec.~\ref{sec:sens_experiments}).
Therefore, if the atom interacts with the laser beam at time $t$, still considering $Q^A_M, Q^L_\omega \ll 1$,
\begin{align}
v^A_\mathrm{kick}(t) = \frac{\hbar k_\mathrm{eff}(t)}{m_A(t)} &\approx \frac{\hbar k_\mathrm{eff}}{m^0_A}\Big(1+(Q^L_\omega-Q^A_M)\cos(\omega t +\phi_0)\Big)\,\nonumber\\
&\equiv v^A_\mathrm{kick, \ 0}+\delta v^A_\mathrm{kick}(t)\, ,
\label{kick_modif}
\end{align}
where $v^A_\mathrm{kick, \ 0} = \hbar k_\mathrm{eff}/m^0_A$ is the unperturbed kick velocity and $\delta v^A_\mathrm{kick}(t)$ is the perturbed contribution to the total velocity kick imparted to the atom.

The calculation of the observable phase shift at the output of the interferometer follows closely the ones presented in \cite{Storey} and is detailed in Appendix.~\ref{ap:MZ_phase}. This method relies on Feynman path integrals and can be used for Lagrangians which are at most quadratic in the position and velocity \cite{Storey} and considering atomic plane waves at initial time $t=t_0$. As shown in Eq.~\eqref{macro_lagrangian}, in our case, the Lagrangian is quadratic in the velocity and there is no dependence on the position, hence this framework can be safely used. Note that in the case of oscillations induced by a coupling with a DM massive field, all calculations should formally be done in the galactocentric frame which is the field rest frame, where the Lagrangian describing the motion of the atom has no dependency on the position. In all other reference frames, the Lagrangian will exhibit a $\cos \left(\omega t-\vec k\cdot \vec x + \phi_0\right)$ dependency and the methodology from \cite{Storey} can formally not be used. However, as the propagation of the field is negligible at the frequencies of interest and for the size of experiments considered, one can still recover the right phase observable by working in the laboratory frame, see the discussion in Appendix.~\ref{ap:acc_lab}.

As derived in e.g. \cite{Wolf, Wolf04}, there are three independent contributions to the total phase shift in an AI: (i) the separation phase noted $\Phi_\mathrm{u}$ which corresponds to a spatial incoincidence between the two output wavepackets, (ii) the laser phase $\Phi_\mathrm{\ell}$ which gathers the additional phase factors of the laser, due to displacement of the light-matter interaction vertices and (iii) the propagation phase shift denoted $\Phi_\mathrm{s}$ which is essentially the phase accumulated by the atom wavepackets over the full interferometric path. All of these contributions must be calculated accurately to predict the phase shift at the output of the interferometer. In Raman-type interferometers, the oscillation of the internal energy impacts mainly the propagation phase of atoms when they are in their excited internal states, i.e on paths 2 and 3 in Fig.~\ref{Mach-Zehnder_osc}. As it can be seen in Eq.~\eqref{a_vUFF}, it also modifies the equations of motion of the atom, implying a small contribution on the laser and separation phases, but these will be suppressed by a factor $\hbar \omega^0_A/m^0_A c^2$. In Bragg-type interferometers, the oscillation of the internal energy does not contribute as the energy state of the atoms is unchanged. The rest mass contribution arises on all the three different contributions to the phase shift, and it can be shown that the total phase shift of an atom A at the end of the interferometric sequence shown in Fig.~\ref{Mach-Zehnder_osc} is (see Appendix.~\ref{ap:MZ_phase} for the detailed calculation)
\begin{widetext}
\begin{subequations}\label{phase_shift_exact_AI}
\begin{align}
    \Delta \Phi^\mathrm{Bragg}_\mathrm{A} &= \frac{4}{\omega}\left(k_\mathrm{eff}v_0(Q^A_M-Q^M_M)\hat e_v \cdot \hat e_\mathrm{kick} + \frac{\hbar k^2_\mathrm{eff}}{2m^0_A}Q^A_M\right)\sin^2\left(\frac{\omega T}{2}\right)\sin(\omega T+\phi_0)+\,\nonumber\\
    &4Q^L_\omega k_\mathrm{eff}\left(L+\frac{\hbar k_\mathrm{eff}T}{m^0_A}\right)\sin^2\left(\frac{\omega T}{2}\right)\cos(\omega T+\phi_0)\, , \label{phase_shift_exact_AI_Bragg}\\
    \Delta \Phi^\mathrm{Raman}_\mathrm{A} &= \frac{4}{\omega}\left(k_\mathrm{eff}v_0(Q^A_M-Q^M_M)\hat e_v \cdot \hat e_\mathrm{kick}-\omega^0_A (Q^A_\omega-Q^L_\omega) +\frac{\hbar k^2_\mathrm{eff}}{2m^0_A}Q^A_M\right)\sin^2\left(\frac{\omega T}{2}\right)\sin(\omega T+\phi_0)+\,\nonumber\\
    &4Q^L_\omega k_\mathrm{eff}\left(L\left(1-\frac{\omega^0_A}{k_\mathrm{eff}c}\right) +\frac{\hbar k_\mathrm{eff} T}{m^0_A}\right)\sin^2\left(\frac{\omega T}{2}\right)\cos(\omega T+\phi_0)\, ,
    \label{phase_shift_exact_AI_Raman}
\end{align}
\end{subequations}
\end{widetext}
at lowest order in $v_0/c$, where $L$ is the height of the retro-reflective mirror of the AI, $Q^M_M$ the mass charge of the Earth, and $\hat e_\mathrm{kick}$ is the direction of the kick imparted to atoms. In practice, the mirror is used to reflect the beams in order to create the counter-propagating waves. In the case of co-propagating beams, one needs to set $L=0$, since there is no retro-reflective mirror. 

Note that in double diffraction interferometers, i.e when two pairs of laser beams transfer opposite momentum to the atom \cite{Giese13}, such that the spatial separation between the two coherent wavepackets is twice as large, the total phase shift is the same with the change $k_\mathrm{eff} \rightarrow 2 k_\mathrm{eff}$, as expected. 

If the mass charges $Q_M$ are composition-dependent, the first term in Eq.~(\ref{phase_shift_exact_AI}), proportional to $v_{0}$, in the following denoted as the mass term, is a signature of the violation of the Einstein EP. If the charge is universal i.e $Q^M_M=Q^A_M$, this term vanishes as expected. However, both terms quadratic in the effective wavevector and respectively proportional to the mass charge of the atom A and the frequency charge of the laser still remain. In fact, they can be understood as a non-local measurement, whose macroscopic counterpart would be to compare the free fall of two test masses with different velocities. In the following, we will neglect these term, since we will consider a practical situation where $v_0 \sim 10^5 \: \mathrm{m/s} \gg 10^{-2} \: \mathrm{m/s} \sim \hbar k_\mathrm{eff}/m^0_A$. 
We will also drop the terms $\propto L$, the height of the mirror, in the following, as they are also much smaller than the leading order term, in particular when measuring a differential phase shift, see next paragraphs. Note that these terms appear only in counter-propagating schemes. Therefore, in the Bragg case, the mass term is the dominant contribution. The final term appearing in the Raman phase shift is related to the oscillation of the atomic and laser frequencies and will be denoted as the frequency term. Depending on the setup, $Q^L_\omega$ and $Q^A_\omega$ can be the same, cancelling this term completely. 

Two-photon transitions AI are commonly operating using hyperfine transitions of alkaline-Earth atoms (e.g the hyperfine transition of Rb, Cs, K atoms \cite{STE-QUEST, Gauguet08, Gillot16}). For both co-propagating and counter-propagating configurations, the laser beams are locked onto the optical transition of an atomic ensemble of the same species as the atoms in free fall inside the interferometer, and the frequency difference $\omega_\mathrm{L_1}-\omega_\mathrm{L_2} = \omega^0_A$ is provided by a radio-frequency source (in the GHz range). Then $Q^L_\omega$ is the charge of that source, which, depending on the experimental configuration, may or may not be the same as $Q^A_\omega$ the charge of the hyperfine transition of the atoms in the AI. As discussed previously, in the case of co-propagating lasers, the effective wavevector corresponds to the frequency transition of the atom, i.e $k_\mathrm{eff}=\omega^0_A/c$. Therefore, the mass term of Eq.~\eqref{phase_shift_exact_AI_Raman} is suppressed by a factor $v_0/c$ compared to the frequency term. For counter-propagating laser beams, the effective wavevector $k_\mathrm{eff} \gg \omega^0_A/c$, implying that the mass term of Eq.~\eqref{phase_shift_exact_AI_Raman} is much bigger than in the co-propagating case. In that case, both mass and frequency terms are relevant and need to be taken into account. The mass term has already been derived in \cite{Geraci16}, in the case of oscillating mass coming from a coupling between matter and a classical oscillating dark matter field. However, the calculation in \cite{Geraci16} was performed in the lab frame, but only considering the velocity of the atoms in this frame, i.e $v_0 \sim 10$ m/s, while we argue that another component, due to the galactic velocity $v_\mathrm{DM} \sim 10^5 \: \mathrm{m/s} \gg v_0$ should be taken into account, see Appendix.~\ref{ap:acc_lab}.
Adding this contribution would improve the expected sensitivity to the charge $Q^A_M$ by several orders of magnitude (see next sections for a practical example). In addition, our calculations take into account the Earth contribution, which was not the case in \cite{Geraci16}. This is an important contribution because in the case of universal rest mass oscillations, the mass term of Eq.~\eqref{phase_shift_exact_AI} cancels. 

In the following, we will assume that the AI experiment is performed in a laboratory frame which has a velocity $\vec v_\mathrm{lab}$ with respect to the reference frame where the Lagrangian takes the form given by Eqs.~(\ref{general_mass_freq_osc}) and (\ref{macro_lagrangian}), such that $\vec v_0 = \vec v_\mathrm{lab}+\vec {\tilde v}_0$ where the second term is the initial velocity of the atoms with respect to the laboratory reference frame (in the case of a coupling with a massive DM scalar field, it is given by the DM velocity in the lab frame, i.e. $\vec v_\mathrm{DM}$). $\tilde v_0$ is the velocity impacting systematic effects such as gravity gradients or second-order Doppler shift of the atoms when interacting with light beams (see e.g \cite{Dimopoulos07} for a comprehensive list of such effects). Transforming Eq.~(\ref{phase_shift_exact_AI}) to the laboratory reference frame requires in principle to transform all the quantities appearing in the equations (such as $\vec k_\mathrm{eff}$, $\omega_A^0$, $T$, \dots) to the lab frame. Such a Lorentz transformation would induce corrections of order $\mathcal O(v_\mathrm{lab}/c)$ which can safely be neglected as discussed in Appendix~\ref{ap:acc_lab}. One can notice that the leading term in Eq.~(\ref{phase_shift_exact_AI}) is proportional to $v_\mathrm{lab}$, a conclusion that differs from existing results in the literature \cite{Geraci16}. In appendix~\ref{ap:acc_lab}, we show that such a conclusion can also be obtained to first order in $v_\mathrm{lab}/c$ by working directly in the laboratory reference frame, strengthening our results.

Dual atom interferometers using two atomic species $A$ and $B$ with different mass and frequency charges will measure the difference of the interferometric phases, whose amplitude is given by
\begin{widetext}
\begin{subequations}\label{eq:delta_phase_shift}
\begin{align}
    \left|\Delta \Phi^\mathrm{Bragg}_\mathrm{AB}\right| &= \frac{4}{\omega}\left|\left[k^A_\mathrm{eff}(Q^A_M-Q^M_M)-k^B_\mathrm{eff}(Q^B_M-Q^M_M)\right]\left(v_\mathrm{lab}\hat e_{vL} \cdot \hat e_\mathrm{kick}+\tilde v_0 \hat e_{v0} \cdot \hat e_\mathrm{kick} \right)\right|\sin^2\left(\frac{\omega T}{2}\right)\, \label{delta_phase_shift_Bragg} , \\
     \left|\Delta \Phi^\mathrm{Raman}_\mathrm{AB}\right| &= \frac{4}{\omega}\left|\left[k^A_\mathrm{eff}(Q^A_M-Q^M_M)-k^B_\mathrm{eff}(Q^B_M-Q^M_M)\right]\left(v_\mathrm{lab}\hat e_{vL} \cdot \hat e_\mathrm{kick}+\tilde v_0 \hat e_{v0} \cdot \hat e_\mathrm{kick}\right)-\right.\,\nonumber\\
     &\left.\left(\omega^0_A(Q^A_\omega-Q^{L,A}_\omega) -\omega^0_B(Q^B_\omega-Q^{L,B}_\omega)\right)\right|\sin^2\left(\frac{\omega T}{2}\right)\, ,
    \label{delta_phase_shift_Raman}
\end{align}
\end{subequations}
\end{widetext}
where we neglected the subdominant terms $\propto k^2_\mathrm{eff},L$ and where $Q^{L,A}_\omega, Q^{L,B}_\omega$ are respectively the frequency charge of the beams used for the transition of the atomic species A and B in the Raman case. In addition, we explicitly set $v_0 \hat e_v= v_\mathrm{lab} \hat e_{vL} + \tilde v_0 \hat e_{v0}$. The mass terms of Eq.~\eqref{eq:delta_phase_shift} are the quantum equivalent of the classical calculation Eq.~\eqref{eq:delta_a_UFF}, while the second term of Eq.~\eqref{delta_phase_shift_Raman}, proportional to the frequency charges $Q_\omega$ has no classical counterpart. Assuming mass and frequency charges of same order of magnitude, the frequency terms of Eq.~\eqref{delta_phase_shift_Raman} will dominate in co-propagating laser Raman interferometers, while for counter-propagating laser Raman interferometers, both terms contribute to the phase shift. 

\subsubsection{Single photon transition}

We will now focus on interferometric setup that involves single photon transitions.  The first setup considered is known as a gradiometer and has already been studied in \cite{Graham13,Badurina22}. In addition, we will also study an isotope differential interferometric scheme that leads to an improved sensitivity to mass and frequency oscillations.

\paragraph{\label{gradio_general}Gradiometers}

\begin{figure}
    \centering
    \includegraphics[scale=0.3]{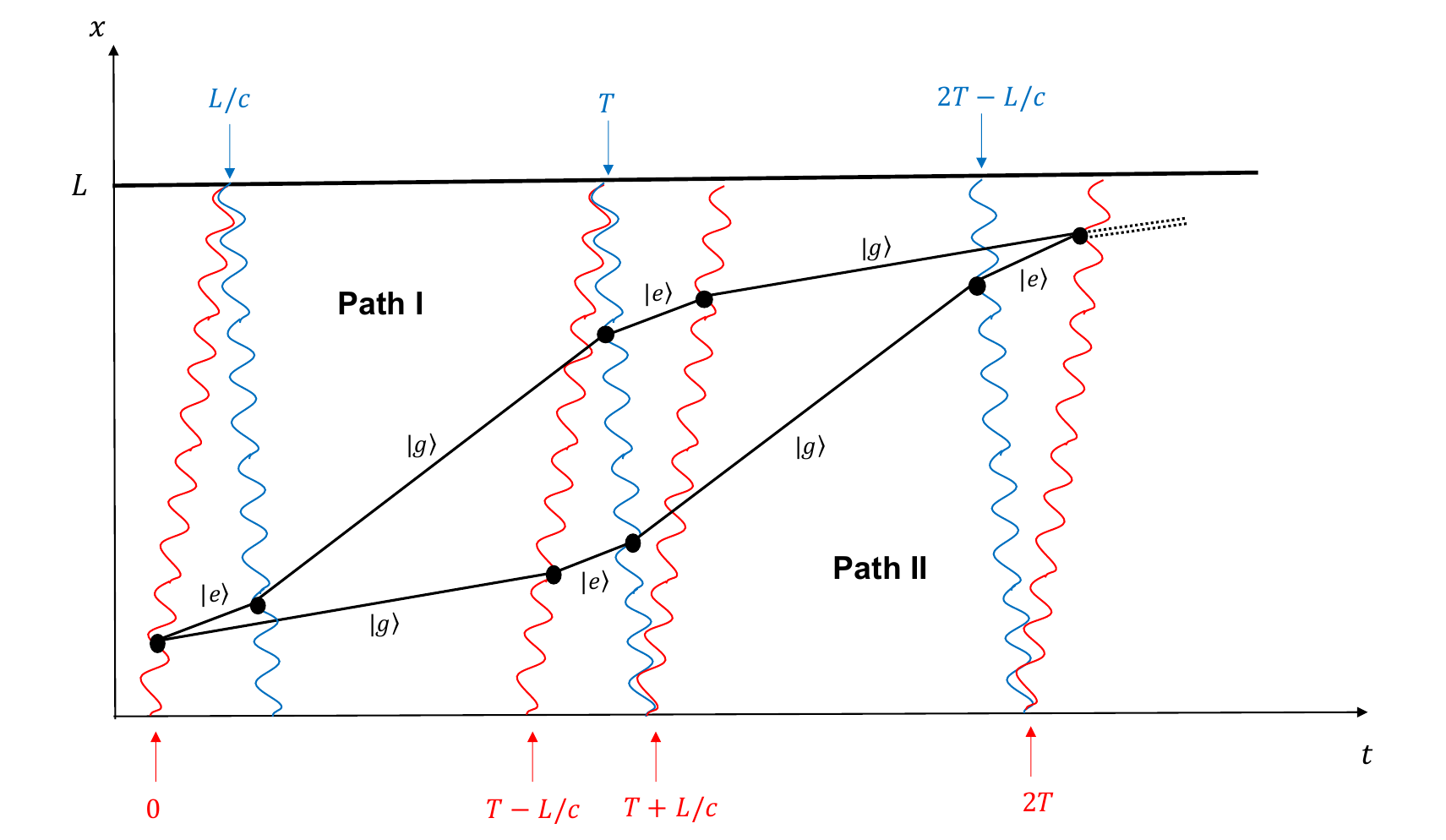}
    \caption{Space-time diagram of a single photon transition, as proposed in \cite{Graham13} with $n=2$ (using the convention of \cite{Arvanitaki18}, see text.) Two lasers are used, one located at $x=0$, with emission (in red) towards the second laser located at $x=L$ (with blue emission). The frequencies of the lasers are selected carefully to interact only with the wavepacket of interest (see text). The black dots indicate the location of matter-wave interactions.}
    \label{fig:gradio}
\end{figure}
The first single photon transition AI scheme considered consists in gradiometers, i.e a setup where two atom interferometers are stacked at different altitudes, for the study of e.g gravity gradients. We are interested in the setup initially proposed by \cite{Graham13} and then studied in e.g \cite{Arvanitaki18,Badurina22}. Practically, we consider two ensembles of atoms $A$ (one for each interferometer) located respectively at $x_1$ and $x_2$, all initially in the state $|g,\hbar \vec k\rangle$ and two lasers, one at coordinate $x=0$ with effective wavevector $\vec k_1$ and the other at coordinate $x=L$ with effective wavevector $\vec k_2$.  In those single photon interactions configurations, $k_1 \approx k_2 = k = \omega^0_A/c$. 

At an initial time $t_0$, the first laser sends a beam which interacts with both atom ensembles and which corresponds to a $\pi/2$ pulse. Then, the second laser beam sends a $\pi$ Doppler shifted laser pulse in order to interact only with the excited state wavepackets, whose motion induces a change in transition frequency, and to convert entirely this wavepacket to the ground state. After this sequence, all wavepackets are in the ground state, however one of them has gained momentum  $2\hbar k$. Following the convention of \cite{Arvanitaki18, Badurina22}, at the end of the sequence, the fast wavepackets have received a Large Momentum Transfer (LMT) photon kick of order\footnote{Note that in the original proposal by \cite{Graham13}, the LMT has a slightly different definition.} 2. If another pair of $\pi$ pulses (with the first one from the bottom laser and the second one from the top laser in the same way as before) is sent to the faster half of the atom, one makes a large momentum transfer (LMT) beam splitter of order 4. More generally, if $m$ pairs of $\pi$ pulses are sent, the order of the LMT beam splitter is $2(m+1)$ and the faster wavepacket has gained total momentum $2(m+1)\hbar k$. In other words, the order $n$ of the LMT is defined as $n=2(m+1)$.

Later at time $t=T$, a  sequence of state inversion similar to the one used in two-photon transitions interferometers is performed, but this time, with three different $\pi$ pulses, the first one coming from the bottom laser, the second one from the top laser and the last one from the bottom laser again. However, in order to slow down the faster wavepacket, $m$ pairs of $\pi$ pulses are added before this sequence, such that it loses $2(m+1)\hbar k$ momentum. Symmetrically, $m$ other pairs of LMT pulses are added after the state inversion to accelerate the other wavepacket, such that it gains $2(m+1)\hbar k$ momentum.

Finally at time $t=2T$, a sequence of pulses opposite to the one sent at the initial $t=0$   is sent to the wavepackets, i.e $m$ LMT pairs of $\pi$ pulses are sent to the wavepackets before the final $\pi-\pi/2$ pulses used for recombination. This whole sequence is depicted in Fig.~\ref{fig:gradio} with $n=2$ (i.e $m=0$) and only one interferometer.

\begin{figure}
    \centering
    \includegraphics[scale=0.3]{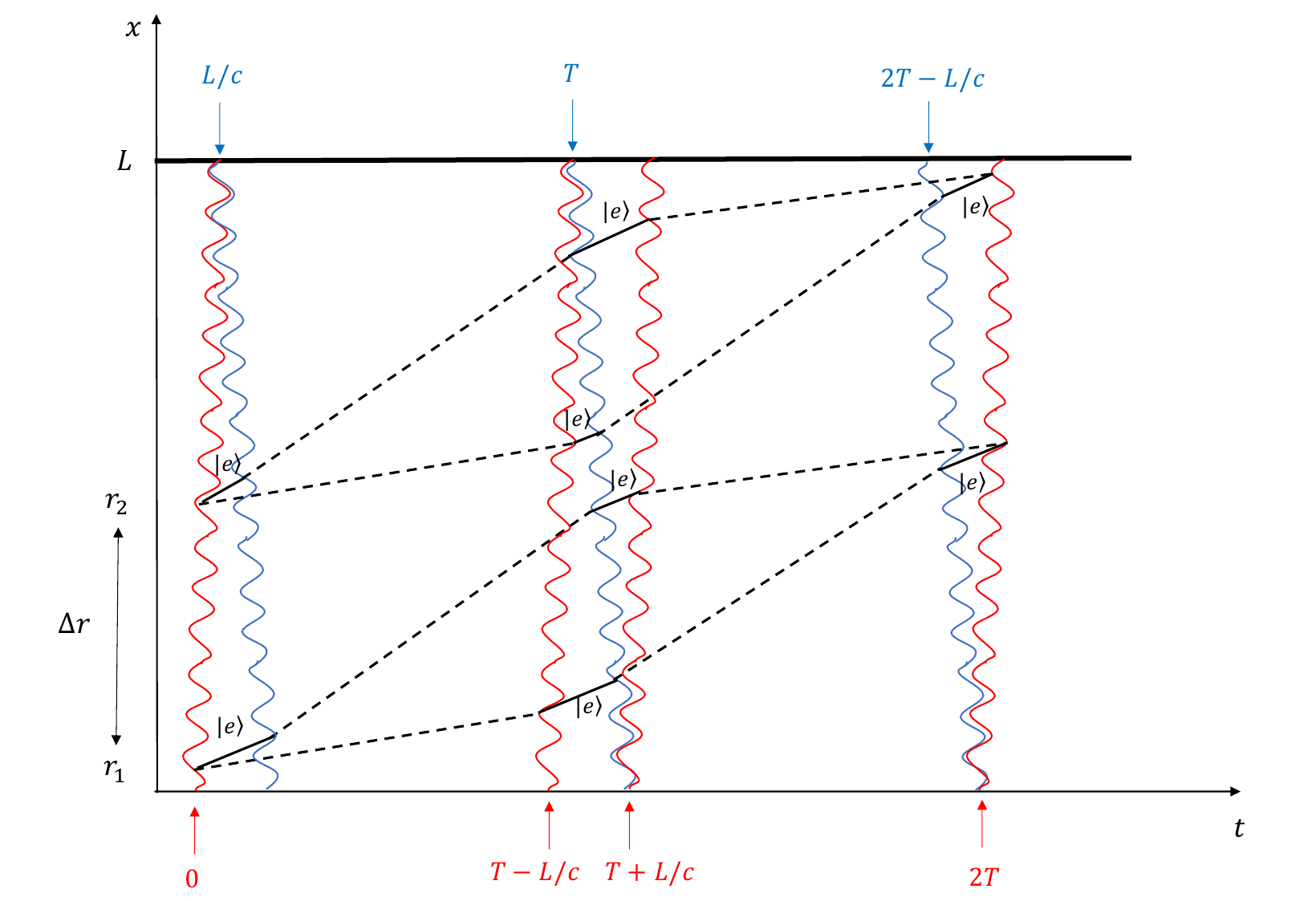}
    \caption{Space-time diagram of a gradiometer, where two interferometers as shown in Fig.~\ref{fig:gradio} are stacked at different altitudes. In this figure, trajectories of the atom in its ground state are represented by dashed lines, while full lines represent paths where the atom is in its excited state. }
    \label{fig:gradio_multiple}
\end{figure}

In gradiometers, two such interferometers are stacked at different altitudes, separated by a distance $\Delta r$, and the same laser beam is used for the laser-atom interaction in both interferometers, which is depicted in Fig.~\ref{fig:gradio_multiple}.
The main advantage of this setup is that laser phase noise is entirely cancelled when measuring the differential phase shift between the two interferometers.

Assuming $\omega L/c \ll 1$ and $n L/c \ll T$ where the LMT kick is of order $n$, and $L$ is the baseline separation between the two lasers, which we assume for simplicity to be the distance between the two interferometers, the differential phase shift can be computed following the exact same methodology as in Appendix~\ref{ap:MZ_phase} and reads 
\begin{align}
|\Delta \Phi^\mathrm{Grad}_\mathrm{A}| \approx \frac{4n\omega^0_A \Delta r Q^A_\omega}{c}\sin^2\left(\frac{\omega T}{2}\right) \, ,
    \label{eq:phase_gradio}
\end{align}
which is the phase shift amplitude derived in \cite{Graham13,Badurina22}.

\paragraph{\label{sec:new_proposal_pres}Single photon transition isotope differential AI (SPID)}
\begin{figure}
    \centering
    \includegraphics[scale=0.45]{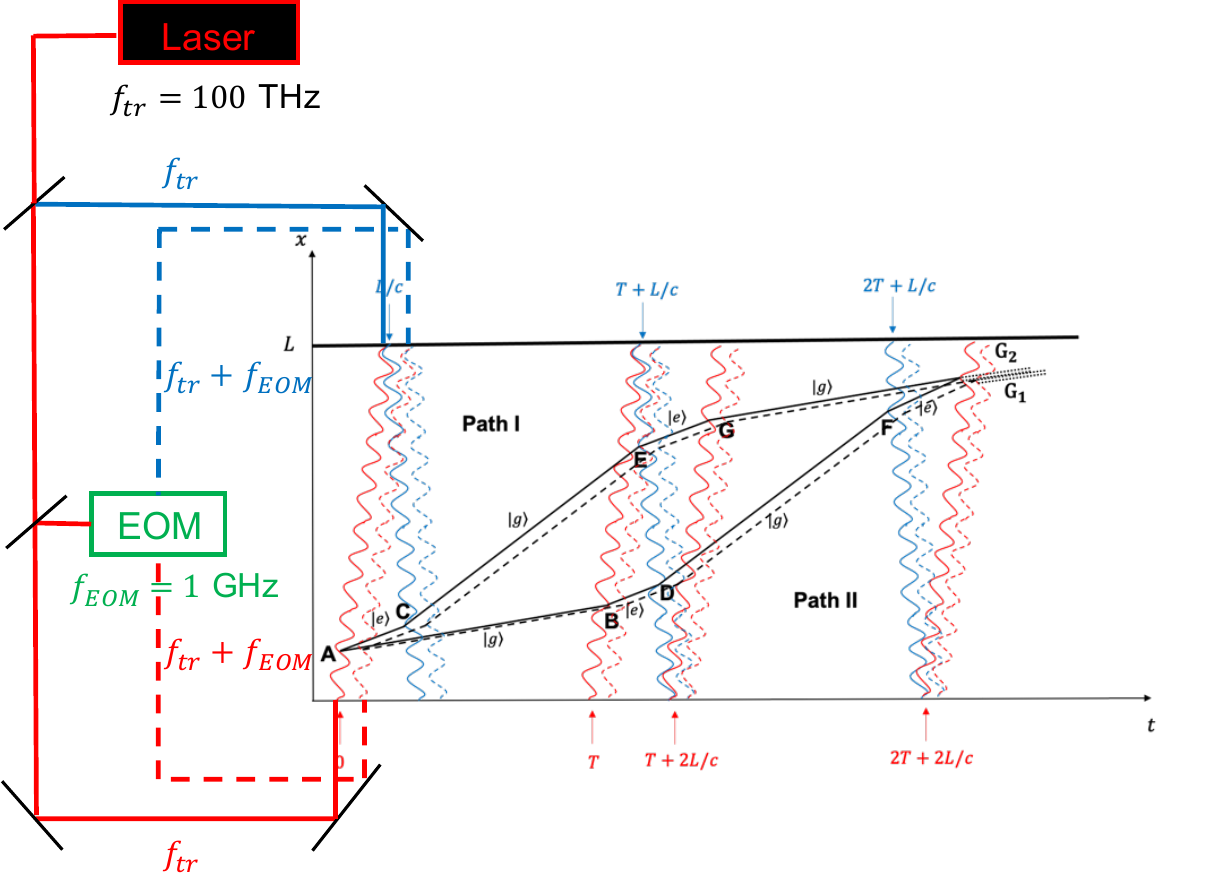}
    \caption{Simplified setup for the SPID experiment. For the sake of simplicity, we assume two different isotopes with respective transition frequencies $f_1=f_\mathrm{tr}=10^{14}$ Hz, $f_2 =f_\mathrm{tr}+f_\mathrm{EOM}=(1+10^{-5})\times10^{14}$ Hz. The laser is locked on the transition frequency of isotope 1 and essentially split in two: the first output is used for the isotope 1 AI (whose trajectory is shown in full lines), while the second output enters an EOM to shift its frequency by $f_\mathrm{EOM}$ to account for the isotope shift in order to be used with the second isotope (whose trajectory inside the AI is shown in dashed lines).}
    \label{fig:isotope_AI}
\end{figure}

In this section, we focus on another interferometric scheme, which is a variation of the interferometric sequence presented in Sec.~\ref{gradio_general}. We will show that this experimental setup is more sensitive to oscillations in atom rest mass and transition frequency compared to regular gradiometers. This type of setup has already been proposed, without any detail, for ULDM detection in MAGIS-100 \cite{Abe21}. The goal of this section is to show the expected signals of such a setup, in order to compare it directly with gradiometers, as the one expected to be used in AION-10 \cite{Badurina22}.

Contrary to usual gradiometers which only use one single species of atom, we consider two different atom isotopes, each of them undergoing individually the interferometric path described earlier in Sec.~\ref{gradio_general}. The setup is presented in Fig.~\ref{fig:isotope_AI}. The two interferometers overlap at the same elevation, so the experiment will test the universality of free fall between the two isotopes. This setup employs single photon transition (meaning we will consider only optical transitions) and measures the differential acceleration between two isotopes, so we will refer to this setup as Single Photon transition Isotope Differential (SPID). Typically, an optical transition has a frequency of the order of $\sim$10$^{14}$~Hz while the typical frequency shift between two isotopes is of order $\sim$10$^{9}$~Hz (i.e. typically 5-6 orders of magnitude smaller than the nominal transition frequency, see \cite{Takano17} for Sr). 
Then, a unique laser source can be used in this setup and separated in two different beams: one which is directly used in the laser-atom interactions inside the interferometer of the first isotope; and the other one whose frequency is shifted, e.g. using an electro-optic modulator (EOM), in order to interact with the second isotope. 

The calculation of the differential phase shift for such a setup can be computed following the exact same methodology as in Appendix~\ref{ap:MZ_phase}. If we assume the optical transition frequencies to be close, i.e $\omega^0_A \approx \omega^0_B \equiv \omega_0$ and the initial velocity to be much larger than the velocity kick $v_0 \gg \hbar k_\mathrm{eff}/m^0$ (see Appendix.~\ref{ap:acc_lab}), the leading order differential phase shift between the two interferometers reads
\begin{align}
|\Delta \Phi^\mathrm{SPID}_\mathrm{AB}| \approx \frac{4n\omega_0}{\omega c}\left|\left(v_\mathrm{lab}\hat e_{vL} \cdot \hat e_\mathrm{kick}+\tilde v_0 \hat e_{v0} \cdot \hat e_\mathrm{kick} \right)\right.\,\nonumber\\
\left.(Q^A_M-Q^B_M)\right|\sin^2\left(\frac{\omega T}{2}\right) \, ,
    \label{eq:phase_new_prop}
\end{align}
where, as previously, we set $v_0 \hat e_v= v_\mathrm{lab} \hat e_{vL} + \tilde v_0 \hat e_{v0}$. Here, we did not take into account the effect of an oscillating EOM frequency through $Q^\mathrm{EOM}_\omega \neq 0$. Nonetheless, its effect is suppressed by a factor $\omega L/v_0 \times \Delta \omega^0/\omega^0 \leq 10^{-7}$ compared to the leading term, where $\Delta \omega^0$ is the isotope shift. In such a case (and similarly for $L=0$), we recover the leading order differential Bragg phase shift derived in Eq.~\eqref{delta_phase_shift_Bragg} with $n\omega_0/c=k^A_\mathrm{eff}\sim k^B_\mathrm{eff}$, as expected.

The advantage of this setup compared to usual gradiometers can be immediately visualized by comparing Eq.~\eqref{eq:phase_new_prop} and Eq.~\eqref{eq:phase_gradio}: the SPID setup does not suffer from a small factor $\omega \Delta r/v_0$. More precisely, the ratio of amplitude of signals between this variation and usual gradiometers is roughly $v_0/(\omega\Delta r) \times (Q^A_M-Q^B_M)/Q^A_\omega \sim \left(1 \mathrm{\: rad.s^{-1}}/\omega\right)$ for the following values: $\Delta r \sim 5$ m, $v_0 \sim v_\mathrm{lab} \sim 10^{-3} c$, $\hat e_{vL} \cdot \hat e_\mathrm{kick} \sim \mathcal{O}(1)$ and mass and frequency charges presented in Table.~\ref{axionic_dilatonic_charge_table}). This implies that, at low angular frequency ($\omega < 1$ rad/s), the signal of the single photon transition isotope differential AI will be larger than the one in a gradiometer. The simple reason for this difference in signal amplitudes is that the gradiometer leading order phase shift in Eq.~\eqref{eq:phase_gradio} is proportional to the frequency charge $Q_\omega$, or in other words, it is proportional to the time in which the wavepackets are in their excited state. As can be noticed from Fig.~\ref{fig:gradio}, this happens for a limited time, of the order of $nL/c$. Conversely, the signal amplitude of the SPID variation is proportional to mass charges $Q_M$ whose effect is imprinted in the phase shift, whatever the internal state, i.e for a time $\sim T$.

In addition, as we shall see in Sec.~\ref{sec:charges_DM}, in some of the theoretical models considered, the frequency charge of optical transition is suppressed, making the gradiometer insensitive to such models, at leading order, while for all models of interest, the difference of mass charges of isotopes $Q^A_M-Q^B_M \neq 0$ (see Table.~\ref{axionic_dilatonic_charge_table}).

In the next sections, we show how these theoretical results Eqs.~\eqref{eq:delta_a_UFF}, \eqref{eq:delta_phase_shift},\eqref{eq:phase_gradio} and \eqref{eq:phase_new_prop} could constrain dark matter models which would produce such time oscillations, namely axions and dilatons.

\section{\label{sec:charges_DM}Mass and frequency charges for some DM candidates}

In the previous section, we have considered a particular phenomenological time oscillation of atomic rest mass and transition frequencies, see Eq.~\eqref{general_mass_freq_osc}. We have derived the observable signatures produced by such an oscillations on various types of observables.  In this section, we will focus on two particularly well studied and motivated DM candidates: the pseudo-scalar field axion and the scalar field dilaton. We will show that these DM candidates induce oscillations of atomic rest mass and transition frequency, and derive the expression of the mass and frequency charges appearing in Eq.~\eqref{general_mass_freq_osc} from the fundamental parameters of the theory.

\subsection{\label{axions}Axion}

The QCD axion was originally proposed to solve the strong CP problem \cite{Peccei_Quinn} by promoting the $\theta$-vacuum of QCD to a dynamical field \cite{Marsh16}. A new U(1) symmetry is introduced and spontaneously broken at a given energy scale $f_a$. The axion $a$ is the pseudo Goldstone boson of mass $m_a$ of this broken symmetry. It is defined as
\begin{align}
    \theta = \sqrt{\hbar c}\frac{a}{f_a} \, .
    \label{theta_axion}
\end{align}
The QCD axion mass is inversely proportional to $f_a$. A particle similar to the QCD axion, with the exception that it does not solve the strong CP problem, i.e whose mass is not proportional to $1/f_a$ is called an Axion Like Particle (ALP).

A scalar ALP field follows the usual Klein-Gordon equation in an expanding universe, and when $m_ac^2/\hbar \gg H$, the field oscillates at its Compton frequency $\omega_a$ and behaves as Cold Dark Matter \cite{Marsh16}.

Let us consider an ALP in the galactic centered reference frame 
\begin{equation}
	a=a_0 \cos (\omega_a t+\phi_a) \, ,
 \label{axion_field}
\end{equation}
with $a_0,\omega_a,\phi_a$ respectively the amplitude, frequency and phase of the field.
In this equation, the field $a$ has units of $\sqrt{\mathrm{J/m}}$.

Similarly to \cite{Kim22}, we consider a model where the ALP and gluons couple as follows
\begin{equation}\label{eq:int_axion}
	\mathcal L_\mathrm{int} = \frac{\sqrt{\hbar c} g^2_s}{32 \pi^2} \frac{a}{f_a}\tilde G^{\mu\nu}G_{\mu\nu}\, ,
\end{equation}
in SI units and where $G_{\mu\nu}$ is the gluon strength tensor field, $\tilde G^{\mu\nu}$ its dual and $g_s$ is the dimensionless strong coupling.

If the ALP is identified as DM, its amplitude of oscillation $a_0$ is fixed by the local DM density $\rho_\mathrm{DM}$ i.e.\footnote{The scalar field $a$ energy density is $\rho = \langle \frac{\dot{a}^2}{2c^2}+\frac{\omega^2_a a^2}{2c^2} \rangle = \frac{\omega^2_a a^2_0}{2c^2}$ where $\langle \rangle$ represents the average over several field oscillations.}
\begin{equation}
	a_0 = \frac{\sqrt{2\rho_\mathrm{DM}}c}{\omega_a} \, .
 \label{axion_density}
\end{equation} 

\subsubsection{Axionic mass charge}

In \cite{Kim22}, it is shown that the interaction Lagrangian from Eq.~(\ref{eq:int_axion}) induces a dependency of the mass of pions to the axion field, which implies a dependency of the mass of nucleons and atomic binding energy on the axion field. As a consequence, the mass of any atom will also depend on the axion field and its coupling strength with gluons. In this sense, we can define the dimensionless axionic mass charge of the atom A as 
\begin{align}
    [Q^A_M]_a = \frac{\partial \ln m_A}{\partial \left(\theta^2\right)} \, ,
    \label{axionic_charge}
\end{align}
such that the mass of the atom A has a small time dependency
\begin{subequations}
\begin{align}
    dm_A(t) &= m^0_A [Q^A_M]_a d\theta^2(t) \, ,
\end{align}
with $m^0_A$ the nominal mass of the atom. We can use Eqs.~\eqref{theta_axion}, \eqref{axion_field} and \eqref{axion_density} to derive the full expression of the mass of the atom A, i.e
\begin{align}
&m_A(t)= m^0_A(1+[Q^A_M]_a \theta^2(t)) \,\\
&\equiv m^0_A\left(1+\frac{\hbar c^3 \rho_\mathrm{DM}[Q^A_M]_a}{f^2_a \omega^2_a}\cos(2\omega_at+2\phi_a)\right) \, \label{mass_osc_axion} ,
\end{align}
\end{subequations}
at first order in axionic mass charge and where we made a reparametrization of the nominal mass of the atom at the last line, i.e $m^0_A \rightarrow m^0_A\left(1+\hbar c^3 \rho_\mathrm{DM}[Q^A_M]_a/f^2_a \omega^2_a\right)$. Note that, through this reparametrization, we neglect the time constant amplitude $\propto (a/f_a)^2$, which could still be detected through its variation due to finite DM coherence time, see \cite{Flambaum23}. The time dependency of atomic masses induced by an ALP can therefore be parametrized using the parametrization used in Sec.~\ref{sec:signatures} (see Eq.~\eqref{general_mass_osc})  by using the changes of variables $\omega \rightarrow 2\omega_a$ and $Q^A_M \rightarrow \hbar c^3 \rho_\mathrm{DM}[Q^A_M]_a/f^2_a \omega^2_a$.

Eq.~\eqref{axionic_charge} describes how much the mass of an atom depends on the axion field. In the following, we derive the expression of the axionic charge of an atom. To do so, we will combine the work of \cite{Kim22} who derived how the pion mass depends on the axion field with the explicit calculation of the dependency of the nucleons rest mass and the binding energy on the pion mass from \cite{damour:2010zr}.

The rest mass of an atom, with charge number Z and neutron number N, can be parameterized as 
\begin{subequations}
\begin{align}
m_{\mathrm{atom}} &= m_{\mathrm{const.}}+E_\mathrm{bind}\,\\
&=Z(m_p+m_e)+Nm_n+E_\mathrm{bind}\, ,
\end{align}
\end{subequations}
where $m_p, m_e, m_n$ are respectively the rest masses of the proton, the electron and the neutron, and where $m_{\mathrm{const.}}, E_\mathrm{bind}$ represent respectively the rest mass of the particle constituents of the atom (proton, neutron, electron) and the nuclear binding energy.

In \cite{Kim22}, it is shown the pion mass is $\theta$ dependent and influence the nucleon mass $m_\mathrm{N}$ through  
\begin{subequations}\label{pions_axions_dep}
	\begin{align}
		\frac{\partial \ln m_N}{\partial \ln m_\pi^2} &\approx 0.06 \, ,\\
		 \frac{\partial \ln m_\pi^2}{\partial \left(\theta^2\right)} & = -\frac{m_u m_d}{2(m_u+m_d)^2}=-0.109 \, \\
   \Rightarrow \frac{\partial \ln m_N}{\partial \left(\theta^2\right)} &\approx -0.065 \, ,
	\end{align}
\end{subequations}
such that, for an atom made of ($N+Z$) nucleons, the contribution of the nucleons rest mass to the atom rest mass to the axionic charge is given by 
\begin{equation}
	[Q^\mathrm{atom}_M]_a\Big|_\mathrm{const.} = \frac{\partial \ln m_N}{\partial \left(\theta^2\right)}=Q_N\approx -0.065 \, ,
\label{axionic_charge_rest}
\end{equation}
meaning that it is independent of the number of nucleons inside the atom, or in other words of the atomic species. If we consider only the impact of the axion field on the rest mass of nucleons inside the atoms, the macroscopic observable ~\eqref{eq:delta_a_UFF} would be independent of the atomic species. In other word, at this level of approximation (i.e considering only the rest mass of nucleons), following Eq.~\eqref{a_vUFF}, all atom will undergo an additional acceleration, proportional to the universal axionic charge, implying e.g. a null differential acceleration Eq.~\eqref{eq:delta_a_UFF}. However, following Eq.~\eqref{phase_shift_exact_AI}, this nucleon rest mass contribution would induce a non-zero phase shift in AI, but not relative to a violation of the UFF, as explained earlier and in agreement with the macroscopic measurement.

Let us now focus on the contribution of the binding energy of the nuclei to the axionic mass charge and show that it is composition dependent.

As computed in \cite{damour:2010zr}, the binding energy of the nuclei depends to first order on the mass of the pions and therefore, following Eq.~\eqref{pions_axions_dep}, on the ALP.  We will now use the results from Section IV of \cite{damour:2010zr} to infer the analytical expression of the dependency of the binding energy to the ALP and to the mass number $A$ and the charge number $Z$. 

Three different interactions contribute to the binding energy \cite{damour:2010zr}: the central force $E_{\mathrm{central}}$ coming from the isospin symmetric central nuclear force, the asymmetry energy $E_{\mathrm{asym}}$, i.e the residual energy from the asymmetry between neutrons and protons inside the nucleus and the Coulomb force $E_{\mathrm{Coulomb}}$ depending on how tightly the nucleons are packed together, leading to\footnote{The pairing energy is subdominant for all atoms compared to the other interactions \cite{damour:2010zr}, therefore we neglect it.}
\begin{align}
E_\mathrm{bind} &= E_{\mathrm{central}}+E_{\mathrm{asym}}\frac{(A-2Z)^2}{A}+\,\nonumber \\
&E_{\mathrm{Coulomb}}\frac{Z(Z-1)}{A^{1/3}}\, .
\end{align}
In Appendix.~\ref{ap:axionic_mass}, we review how each of these contribution depends on the mass of the pions and therefore, using Eq.~\eqref{pions_axions_dep} on the axion field, to derive the non-universal axionic mass charge of an atom due to the binding energy, which reads 
\begin{align}
[Q^\mathrm{atom}_M]_a &\approx - 0.070 + 10^{-3} \times \left(\frac{3.98}{A^{1/3}}+\right.\,\nonumber \\
&\left. 2.22\frac{(A-2Z)^2}{A^2}+ 0.015\frac{Z(Z-1)}{A^{4/3}}\right) \, .
\label{axionic_mass_charge}
\end{align}

\subsubsection{\label{axion_freq_charge}Axionic frequency charge}

In this section, we derive the axionic frequency charge (see Eq.~\ref{general_freq_osc}) for various transitions frequencies. We will consider two different types of atomic transitions, which do not depend on the same physical parameters: (i) the hyperfine transitions and (ii) the optical ones.

Similarly to the previous section, we define the axionic frequency charge  of the atom as
\begin{align}
    [Q^\mathrm{atom}_\omega]_a &\equiv \frac{\partial \ln \omega_\mathrm{atom}}{\partial (\theta)^2} \, ,
\end{align}
and derive the full expression of the frequency of the atom A as
\begin{align}
    \omega_\mathrm{atom}(t) \equiv \omega^0_\mathrm{atom}\left(1+\frac{\hbar c^3 \rho_\mathrm{DM}[Q^\mathrm{atom}_\omega]_a}{f^2_a \omega^2_a}\cos(2\omega_at+2\phi_a)\right) \, .
    \label{freq_osc_axion}
\end{align}
Again, one can notice that this time dependency can be parametrized using Eq.~\eqref{general_freq_osc} by the changes of variables $\omega \rightarrow 2\omega_a$ and $Q^A_\omega \rightarrow \hbar c^3 \rho_\mathrm{DM}[Q^\mathrm{atom}_\omega]_a/f^2_a \omega^2_a$.

We will first consider hyperfine atomic transitions ($\omega^H_\mathrm{atom}$) which are impacted by the axion-gluon coupling from Eq.~\eqref{eq:int_axion}  as \cite{Kim22}
\begin{align}\label{hyp_freq_dep_g}
    \frac{\partial \ln \omega^H_\mathrm{atom}}{\partial (\theta)^2} &= \left(\frac{\partial \ln g}{\partial \ln m^2_\pi}-\frac{\partial \ln m_p}{\partial \ln m^2_\pi}\right)\frac{\partial \ln m^2_\pi}{\partial (\theta)^2} \, ,
\end{align}
where $g$ is the nucleon g-factor whose dependence on pion mass is given by \cite{Kim22}
\begin{subequations}\label{g_dep_pion}
    \begin{align}
        \frac{\partial \ln g}{\partial \ln m^2_\pi} &= K_n \frac{\partial \ln g_n}{\partial \ln m^2_\pi} + K_p \frac{\partial \ln g_p}{\partial \ln m^2_\pi} -0.17 K_b \, ,
    \end{align}
with 
    \begin{align}
        \frac{\partial \ln g_n}{\partial \ln m^2_\pi} &\approx -0.25\, , \\
        \frac{\partial \ln g_p}{\partial \ln m^2_\pi} &\approx -0.17 \, ,
    \end{align}
\end{subequations}
where $g_{n}$ ($g_{p}$) respectively the neutron (proton) gyromagnetic factors and $K_n$, $K_p$ and $K_b$, coefficients computed from chiral perturbation theory (nuclear shell model). These coefficients are not measurable, but they are related to observable parameters, namely, the sensitivity coefficient of the nuclear magnetic moment to light quarks masses $\kappa_q$, to strange quarks masses $\kappa_s$ and to light quarks masses over QCD energy scale $\kappa$ through \cite{Flambaum06}
\begin{subequations}\label{kappa_K_relation}
    \begin{align}
        \kappa_q &= -0.118 K_n -0.087 K_p \, , \\
        \kappa_s &= 0.0013 K_n -0.013 K_p \, , \\
        \kappa &=-0.12 K_n -0.10 K_p -0.11 K_b \, .
    \end{align}
\end{subequations}
Using Eq.~\eqref{pions_axions_dep}, \eqref{hyp_freq_dep_g}, \eqref{g_dep_pion} and \eqref{kappa_K_relation}, we define the dimensionless axionic frequency charge of the atom as
\begin{align}
    [Q^\mathrm{atom}_\omega]_a  &\approx \left(-16.8 \kappa -5.69 \kappa_q +25.1\kappa_s + 0.65\right) \times 10^{-2} \, .
    \label{axionic_freq_charge}
\end{align}
In particular, Eq.~\eqref{axionic_freq_charge} is relevant  for $^{87}$Rb hyperfine transition, whose associated $\kappa$ parameters are $\kappa = -0.016, \kappa_q = -0.046, \kappa_s = -0.010$ \cite{Flambaum06}.

Let us now consider optical transitions whose frequency does depend neither on the nucleon g-factor nor on the proton mass at lowest order. This implies that the axionic frequency charge at lowest order is 0. Higher order contributions would lead to non-zero axionic frequency charge, in particular if one considers its dependence on the fine structure constant $\alpha$ which arises at loop level \cite{Beadle23, Kim24} or to the nuclear charge radius \cite{Zhao24} and are therefore highly suppressed. We now compute the axionic optical frequency charge of an atom A, considering both of these effects for some atomic optical transition. Using the dependence of optical transition on the fine structure constant $\alpha$, the gluon-photon coupling at loop level leads to a charge 
\begin{subequations}\label{axionic_freq_charge_2nd_order}
\begin{align}
    [Q^A_\omega]_a &= c_{F_2}\frac{(2+\epsilon) \alpha}{4\pi^2} \, ,
\end{align}
where $c_{F_2}$ is the parameter encoding the explicit symmetry-breaking, generating the one-loop coupling. While for QCD axion, this parameter is $\mathcal{O}(1)$, it is expected to be at most $10^{-2}$ in the ALP case \cite{Beadle23}. Therefore, the corresponding axionic charge is, e.g. for $^{171}$Yb,
\begin{align}
    [Q^{Yb}_\omega]_a & \sim 1.2 \times 10^{-8} \, .
\end{align}
The second contribution, arising from the coupling of the ALP to the nuclear charge radius has the form \cite{Zhao24}
\begin{align}
     [Q^A_\omega]_a &= \frac{\partial \ln \omega_A}{\partial \ln \langle r^2_N \rangle}\frac{\partial \ln \langle r^2_N \rangle}{\partial \ln m^2_\pi}\frac{\partial \ln m^2_\pi}{\partial (\theta)^2} = \frac{\beta F \langle r^2_N \rangle}{\omega_A} \frac{\partial \ln m^2_\pi}{\partial (\theta)^2} \, ,
\end{align}
where $\langle r^2_N \rangle$ is the mean squared charge nuclear radius, $F$ is the difference in field shift factor between excited and ground state, and $\beta = \partial \ln \langle r^2_N \rangle/\partial \ln m^2_\pi \equiv -0.2$ \cite{Zhao24}. For $^{171}$Yb optical transition, $F = -2\pi \times 10.955$ GHz/fm$^2$ \cite{Schelfhout21}, $r_N = 5.2906$ fm \cite{Angeli13}, which leads to 
\begin{align}
    [Q^{Yb}_\omega]_a &\sim -1.21 \times 10^{-5} \, .
\end{align}
\end{subequations}
To compute the nuclear charge radius contributions of other optical transitions, we use the fact that the field shift $F \langle r_N^2 \rangle$ scales approximately as $Z^2/A^{1/3}$ \cite{Blaum13}.
Note that, as expected, those two contributions are at least two orders of magnitude, smaller than the axionic charges of hyperfine transitions (see Table.~\ref{axionic_dilatonic_charge_table}).

\subsection{Dilaton scalar field}

Another example of scalar dark matter candidate is the dilaton \cite{arvanitaki:2015qy,Flambaum15,stadnik:2015yu,Flambaum16}. In the most general case, this field is non universally coupled to matter,  which makes some ``constants'' of Nature to evolve with the field \cite{damour94,Damour02,damour:2010zr,arvanitaki:2015qy,Flambaum15,stadnik:2015yu,Flambaum16,hees18}.
In this paper, we will only consider linear coupling\footnote{Quadratic coupling has also been widely studied, e.g in \cite{Flambaum15,Flambaum16,hees18}.} between the dimensionless dilatonic field $\phi$ and several SM fields, namely the matter fermions, the gluon and the photon fields through the following Lagrangian \cite{damour:2010zr}
\begin{align}
    \label{dilaton_lag}
    \mathcal{L}_\mathrm{int} &= \phi \left[\frac{d_e}{4\mu_0}F_{\mu\nu}F^{\mu\nu} -\frac{d_g\beta_3}{2g_s}G_{\mu\nu}G^{\mu\nu} - \right. \, \nonumber \\
    &\left. \sum_{i=e,u,d} (d_{m_i}+\gamma_{m_i}d_g)m_i\bar \Psi_i \Psi_i\right] \, ,
\end{align}
where $\mu_0$ is the magnetic permeability, $F_{\mu\nu}$ the electromagnetic strength tensor, $\beta_3$ the running coupling function of the QCD gauge coupling and $\Psi_i$ the different fermionic fields. All the $d_i$ are dimensionless couplings between the dilaton and the SM fields which will, amongst others, modify the mass of  atoms through the variation of several ``constants'' of Nature, namely the fine structure constant $\alpha$, the electron mass $m_e$, the  quark masses and the QCD energy scale $\Lambda_\mathrm{QCD}$ \cite{damour:2010zr}. More generally, a ``constant'' $X$ is modified by Eq.~\eqref{dilaton_lag} as \cite{damour:2010zr}
\begin{align}
    \label{variation_const}
    X \rightarrow X\left(1+ d_X \phi\right) \, .
\end{align}
At cosmological scales, the dilaton behaves as a classical oscillating field 
\begin{align}
    \phi = \phi_0 \cos(\omega_\phi t +\Phi) \, \label{dilaton_field},
\end{align}
whose amplitude $\phi_0$ is directly related to the DM energy density, i.e \cite{arvanitaki:2015qy,Flambaum15,hees18}
\begin{align}
\label{dilaton_energy_density}
\phi_0=\frac{\sqrt{8\pi G \rho_\mathrm{DM}}}{\omega_\phi c} \, ,
\end{align}
where $G$ is the Newton's gravitational constant.

\subsubsection{Dilatonic mass charges}

The dependency of the constants $X$ on the scalar field $\phi$, see Eq.~\eqref{variation_const}, induces variations of the rest mass of an atom which is characterized by \cite{damour:2010zr}
\begin{align}
    [Q^\mathrm{atom}_M]_\phi &= \frac{\partial \ln m_\mathrm{atom}}{\partial \phi}\, . \label{eq:dilaton_mass_charge}
\end{align}
In \cite{damour:2010zr}, it has been shown that the $[Q^\mathrm{atom}_M]_\phi$ coupling function depends on the dilatonic charges $Q_{M,X}$ and on the dilaton/matter coupling coefficients $d_X$ as\footnote{Note that following \cite{damour:2010zr}, we neglect the contribution of the QED trace anomaly that would add another small contribution \cite{nitti:2012ut}.}
\begin{align}
    \label{partial_dil_mass_charge}
    [Q^\mathrm{atom}_M]_\phi &= \left(Q^\mathrm{atom}_{M,m_e} (d_{m_e}-d_g)+ Q^\mathrm{atom}_{M,e} d_e+\right.\,\nonumber \\
    &\left.Q^\mathrm{atom}_{M,\hat m}(d_{\hat m}-d_g)+Q^\mathrm{atom}_{M,\delta m}(d_{\delta m}-d_g)\right)\, ,
\end{align}
with the expressions of the dilatonic charges given by
\begin{subequations}\label{dilatonic_mass_charge}
\begin{align}
Q_{M,\hat{m}} &= 0.093 - \frac{0.036}{A^{1/3}} - 0.02\frac{(A-2Z)^2}{A^2} -\,\\ 
&1.4 \times 10^{-4} \frac{Z(Z-1)}{A^{4/3}} \, , \nonumber \\
Q_{M,\delta m} &= 0.0017\frac{A-2Z}{A} \, , \\
Q_{M,m_e} &= 5.5\times 10^{-4} \frac{Z}{A}\, , \\
Q_{M,e} &= \left(-1.4 + 8.2 \frac{Z}{A} + 7.7\frac{Z(Z-1)}{A^{4/3}}\right)\times 10^{-4} \, ,
\end{align}
\end{subequations}
where $A$ and $Z$ represent respectively the mass and atomic numbers of the atom, $\hat m$ is the mean of the up and down quark masses and $\delta m$ their difference.

\subsubsection{Dilatonic frequency charges}
In a similar way, the dilatonic frequency charge is defined as 
\begin{align}\label{eq:dilaton_freq_charge}
    [Q^\mathrm{atom}_\omega]_\phi &= \frac{\partial \ln \omega_\mathrm{atom}}{\partial \phi}\, .
\end{align}
Similarly to the mass charge, the frequency charge depends on  dilatonic charges $Q_{\omega,X}$ and on the dilaton/matter coupling coefficients $d_X$ (see e.g. \cite{hees18})
\begin{align}
\label{partial_dil_freq_charge}
    [Q^\mathrm{atom}_\omega]_\phi &= \left(Q^\mathrm{atom}_{\omega, m_e}\left(d_{m_e}-d_g\right)+Q^\mathrm{atom}_{\omega, e} d_e + \right. \,\nonumber \\
    &\left.Q^\mathrm{atom}_{\omega,\hat m}\left(d_{\hat m}-d_g\right)+Q^\mathrm{atom}_{\omega,\delta m}\left(d_{\delta m}-d_g\right)\right)\, .
\end{align}

In case of hyperfine transitions, we have $f^{hyp}_\mathrm{atom} \propto \alpha^{k_\alpha}(m_e/m_p)(m_q/\Lambda_\mathrm{QCD})^{k_q}$ \cite{Flambaum06}, where $k_\alpha,k_q$ represent respectively the sensitivity coefficients of the hyperfine transition to the fine structure constant and to the ratio of the light quark masses $m_q$ to the QCD mass scale $\Lambda_\mathrm{QCD}$ ratio. Subsequently, the corresponding  dilatonic frequency charges are 
\begin{subequations}\label{dilatonic_hyp_freq_charge}
\begin{align}
Q^\mathrm{hyp}_{\omega, m_e} &= 1 \, , \\
Q^\mathrm{hyp}_{\omega,e} &=  k_\alpha \, , \\
Q^\mathrm{hyp}_{\omega, \hat m} &=-0.048+k_q \, , \\
Q^\mathrm{hyp}_{\omega, \delta m} &= 0.0017+k_q \, ,
\end{align}
\end{subequations}
where we used the dependency of the proton mass to the light quark masses $\partial \ln m_p/\partial \ln \hat m = 0.048$, $\partial \ln m_p/\partial \ln \delta m = -0.0017$ \cite{damour:2010zr}.

The frequencies of optical transition depend mainly on the electron mass and on the fine structure constant $f^\mathrm{opt}_\mathrm{atom} \propto (m_e m_N) \alpha^{2+\epsilon_\mathrm{atom}}/(m_e+m_N)$ \cite{Arvanitaki15,Kim22} (where $m_N$ is the total nucleus mass and $\epsilon_\mathrm{atom}$ accounts for the relativistic correction of the dependence of $f_\mathrm{atom}$ to $\alpha$). Therefore, the  frequency dilatonic charges write
\begin{subequations}\label{dilatonic_freq_charge}
\begin{align}
Q^\mathrm{opt}_{\omega, m_e} &= 1 \, , \\
Q^\mathrm{opt}_{\omega,e} &= 2+\epsilon_\mathrm{atom} \, , \\
Q^\mathrm{opt}_{\omega, \hat m} &\approx \frac{2.6 \times 10^{-5}}{A} \, , \\
Q^\mathrm{opt}_{\omega, \delta m} &\approx 9.0 \times 10^{-7}\frac{A-2Z}{A^2} \, ,
\end{align}
\end{subequations}
where for the third charge $Q^\mathrm{opt}_{\omega, \hat m}$, we assumed $m_n=m_p$.
In conclusion, similarly to Sec.~\ref{axions}, the oscillating mass and transition frequency of the atom A in the dilatonic framework can be written as 
\begin{subequations}
    \begin{align}
        m_\mathrm{atom}(t) &\equiv m^0_\mathrm{atom}\left(1+ \frac{\sqrt{8\pi G \rho_\mathrm{DM}}[Q^\mathrm{atom}_M]_\phi}{\omega_\phi c}\cos(\omega_\phi t +\Phi)\right) \label{mass_osc_dilaton}\, , \\
        \omega_\mathrm{atom}(t) &\equiv \omega^0_\mathrm{atom}\left(1+ \frac{\sqrt{8\pi G \rho_\mathrm{DM}}[Q^\mathrm{atom}_\omega]_\phi}{\omega_\phi c}\cos(\omega_\phi t +\Phi)\right) \label{freq_osc_dilaton}\, ,
    \end{align}
\end{subequations}
where we have used Eqs.~(\ref{dilaton_field}-\ref{eq:dilaton_mass_charge}) and \eqref{eq:dilaton_freq_charge}  and where both mass and frequency charges are respectively given by Eqs.~\eqref{partial_dil_mass_charge} and \eqref{partial_dil_freq_charge}. These time dependencies can easily be mapped into the parmeterization from Eq.~\eqref{general_mass_freq_osc} used in Sec.~\ref{sec:signatures} by the simple changes of variables $\omega \rightarrow \omega_\phi$ and $Q^A_M$ (resp. $Q^A_\omega$) $\rightarrow \left(\sqrt{8 \pi G\rho_\mathrm{DM}}/\omega_\phi c\right)[Q^A_M]_\phi$ (resp. $[Q^A_\omega]_\phi$).

\section{Description of existing and future experiments}\label{sec:experiments}

The previous sections were dedicated to a theoretical derivation of the signatures induced by some models of ULDM on several types of experiments. In this section, we will present several existing or future experiments, based either on the classical measurement of differential acceleration between macroscopic bodies or on atom interferometry. The main goal of this section is to present an overview of the experiments, their main characteristic and also provide an estimate of their noise level and integration time. These will be needed in the future section to estimate the sensitivity of the experiments to the various axionic and dilatonic couplings.

The experiments considered in this paper are the following:
\begin{enumerate}
    \item the MICROSCOPE space experiment \cite{Microscope17, Microscope22}.
    \item the AI experiment operating at Stanford \cite{Stanford20}.
    \item the AION-10 experiment \cite{Badurina20}.
    \item the SPID variation presented in Sec.~\ref{sec:new_proposal_pres} 
\end{enumerate}
The first two are existing experiments with available data, while AION-10 is a proposed experimental scheme with futuristic experimental parameters. We will derive the noise level of the SPID variation. Since we aim at comparing this variation with the gradiometer one of AION-10, we will also assume the same experimental parameters as AION-10.

\subsubsection{MICROSCOPE}

 MICROSCOPE  is a space mission that was launched to test the UFF in space \cite{Microscope17}. The payload contains (amongst other) two concentric test masses, one made of Platinum and the other one made of Titanium. By closely monitoring the relative displacement between the two test masses along the symmetry axis, one can measure the differential acceleration between them and estimate the E\"otvos $\eta$ parameter. The final result of MICROSCOPE is \cite{Microscope22}
\begin{subequations}
\begin{equation}
    \eta= (-1.5 \pm 2.7)\times 10^{-15}\, ,
\end{equation}
at a 1$\sigma$ confidence-level.

The full MICROSCOPE dataset used to constrain the UFF is made of 17 sessions \cite{Rodrigues22}. The noise PSD of the differential acceleration of the session 404 is given by \cite{Pihan}
\begin{align}
    S_a(f) =  2.2\times 10^{-24} f^{-1} + 2.3 \times 10^{-17} f^{4} \: \mathrm{\left(m/s^2\right)^2/Hz}\, ,
    \label{PSD_MICROSCOPE}
\end{align}
\end{subequations}
for frequencies between $10^{-5}$ Hz and 0.3 Hz. The $f^{-1}$ slope noise comes from thermal effects of the gold wire connecting the test masses to the cage, while the high frequency noise in $f^4$ is the second derivative of the position measurement white noise \cite{Microscope17}.

The UFF signal is modulated by spinning the satellite at a frequency chosen to minimize the noise, $f_\mathrm{EP} \sim 3$ mHz. Note that $f_\mathrm{EP}$ is in reality a linear combination of the spin frequency $f_\mathrm{spin}$ and orbital frequency $f_\mathrm{orb}$, with $f_\mathrm{spin} \gg f_\mathrm{orb} \sim 1.5 \times 10^{-4}$ Hz \cite{Microscope17} for session 404. Three different spin frequencies exist depending on the session \cite{Berge23}, such that one should take both contributions into account for the determination of $f_\mathrm{EP}$ for other sessions. 

The bucket frequency of this acceleration noise corresponds approximately to $f_\mathrm{bucket} \sim 30$ mHz \cite{Pihan,Microscope17}. 

To take into account the contribution of the full experiment, i.e the 17 different sessions, we make the hypothesis that all of them have the same level of acceleration noise Eq.~\eqref{PSD_MICROSCOPE} while we consider the orbital period to be constant $T_\mathrm{single \: orbit} \approx 5946$ s \cite{Rodrigues22}. This hypothesis is sufficient for the rough sensitivity analysis we do here, but will have to be revisited when doing a complete MICROSCOPE data analysis in search for DM candidates.

Using the total number of orbits of the experiment, i.e $N_\mathrm{orbits}= 1362$ \cite{Rodrigues22}, we find that the total integration time is $T_\mathrm{int} = T_\mathrm{single \: orbit} \times N_\mathrm{orbits} \equiv 8.1 \times 10^6$ s.

As an experiment measuring the differential acceleration between two test masses, the corresponding theoretical signal is given by Eq.~\eqref{eq:delta_a_UFF}.

\subsubsection{Atom interferometry in the Stanford Tower}\label{sec:stanford}

The most stringent constrain on the E\"otv\"os parameter $\eta$ obtained from atom interferometry experiment is achieved using a Bragg atom interferometry experiment in the Stanford Tower. In  this experiment, the relative acceleration of freely falling clouds of two isotopes of Rubidium ($^{85}$Rb and $^{87}$Rb) is measured \cite{Stanford20}. The differential interferometric phase measurement between these two inteferometers leads to a constraint on $\eta$ given by \cite{Stanford20}
\begin{subequations}
\begin{align} \label{equ:eta_Stanford}
\eta = (1.6 \pm 5.2)\times 10^{-12}\, .
\end{align}
Both atoms are launched upwards inside the Stanford Tower for an interferometric sequence of total duration $2T=1.91$ s, corresponding to the total time for the atoms to fall back \cite{Stanford20}.

Double diffraction interferometry is performed using two lasers  both resonant with the $|5^{2}S_{1/2}\rangle \rightarrow | 5^{2}P_{3/2}\rangle$ $^{87}$Rb transition, i.e. their wavevectors are $k_1 \approx k_2 \approx 2\pi/\lambda$, with $\lambda = 780$ nm. Three different momentum splittings have been used: $\{4\hbar k, 8\hbar k, 12 \hbar k\}$. All are in agreement with no EP violation \cite{Stanford20}. In the following, we will consider the highest momentum transfer since this will enhance the signal searched for, which corresponds to using an effective wavevector $k_\mathrm{eff}=24\pi/\lambda$ in Eq.~\eqref{phase_shift_exact_AI_Bragg}.

The resolution per shot on the differential acceleration is $\sigma_{\Delta a} = 1.4 \times10^{-11}g$ \cite{Stanford20}, with $g = 9.81$ m/$s^2$ the Earth gravitational acceleration on ground. We assume white noise (see below) with a corresponding acceleration noise PSD
\begin{align}
    S_{\Delta a} = \frac{2\sigma^2_{\Delta a}}{f_s}
    \approx \left(7.7 \times 10^{-11}\ g\right)^2 \ \mathrm{Hz}^{-1} \, ,
\end{align}
with $f_s$ the sampling frequency of the experiment, defined as $f_s = 1/T_\mathrm{cycle}$ where $T_\mathrm{cycle} = 15$ s is the duration of one interferometric sequence, including atom preparation, launch and free fall \cite{Stanford20}. 
The raw measurement in this experiment is actually a differential phase shift $\Delta \Phi$ which is related to the differential acceleration between the atoms $\Delta a$ through $\Delta \Phi = k_\mathrm{eff}T^2 \Delta a$ \cite{Storey}. Therefore, we can infer the  ASD of the phase shift using
\begin{align}\label{phase_acceleration_noise}
    \sqrt{S_\Phi} = k_\mathrm{eff}T^2\sqrt{S_{\Delta a}} \, .
\end{align}


The final uncertainty of this differential acceleration measurement is mostly limited by electromagnetic effects, coming from the Bragg lasers and non homogeneous magnetic field \cite{Stanford20}. Knowing the final uncertainty of the differential measurement \eqref{equ:eta_Stanford} we derive the total ``effective'' experiment time $T_\mathrm{int}$ under our white noise assumption i.e. assuming that individual experimental cycles are uncorrelated. Then, the number of cycles $N$ can be derived as
\begin{align}
    N = \frac{1}{g^2}\left(\frac{\sigma_{\Delta a}}{\sigma_\eta}\right)^2 \approx 77 \, ,
\end{align}
and
\begin{align}
    T_\mathrm{int} = N\times T_\mathrm{cycle} \approx 1148 \mathrm{\: s} \, .
\end{align}
\end{subequations}





\subsubsection{AION-10}

The Atom Interferometer Observatory and Network (AION) is an experimental program to search for ULDM and gravitational waves in the $10^{-1}-10$ Hz range using atom interferometry \cite{Badurina20}. AION-10 is a 10 meter-long single-photon atom gradiometer instrument which will use $^{87}\mathrm{Sr}$ atoms and that will be built in Oxford \cite{Badurina22}. Contrary to the other experiments previously introduced, AION-10 will operate in a distant future, assuming a much better control on noise than current experiments. Following \cite{Badurina22}, we will use the following experimental parameters ultimately envisaged for AION-10:
\begin{subequations}\label{AION_exp_param}
\begin{align}
    T_\mathrm{int} &= 10^8 \mathrm{\: s} \, , \\
    T &= 0.74 \mathrm{\: s} \, , \\
    S_\Phi(f) &= 10^{-8} \mathrm{\: rad^2/Hz} \, , \\
    n &= 1000 \, , \\
    \omega_\mathrm{Sr} &= 2.697 \times 10^{15} \mathrm{\: rad/s} \, , \\
    \Delta r &= 4.86 \mathrm{\: m} \, , \\
    L &= 10 \mathrm{\: m} \, ,
\end{align}
\end{subequations}
respectively the interrogation time, the free evolution time, the gradiometer phase noise PSD, the number of LMT kicks, the optical transition frequency used, the gradiometer separation, and the total size.

\subsubsection{\label{New_param}SPID}

The SPID experiment described in Sec.~\ref{sec:new_proposal_pres} consists of two colocated interferometers with an interferometric sequence similar to the one used in gradiometers, as AION-10 but using two different isotopes. 

We consider this setup variation for two different reasons. First, MAGIS-100 \cite{Abe21} will operate a similar mode for the search of ULDM and we want to derive the expected sensitivity of such a large scale experiment to axion and dilaton signals. More specifically, MAGIS-100 will run a Bragg interferometer (i.e with two photons transitions instead of single-photon transition, as in SPID). However, as mentioned in the text after Eq.~\eqref{eq:phase_new_prop}, the leading order signal of both setup is the same. The main difference between the two setup is that for SPID, the wavepackets spend some time in their excited state (which they do not in Bragg configuration), but as shown in Eq.~\eqref{eq:phase_new_prop}, the additional phase shift is next-to-leading order, and thus negligible. In addition, we argue that phase noise levels will be equivalent in both setups (see next paragraphs). Therefore, the optimal choice between the two setup is a matter of practicality (see end of this subsection for a small discussion). The second reason for the SPID consideration is that we would like to compare it with the gradiometer for the same experimental parameters to assess which one is the best to search for axion and dilaton DM. 

For this reason, we consider first the experimental parameters from one running mode of MAGIS 100. They are listed in \cite{Abe21}, see their Fig. 3. Their noise levels are much lower than current experiments, similar to AION-10. MAGIS-100 will use two isotopes of Strontium ($^{87}$Sr and $^{88}$Sr) in a 100-meter baseline. The transition under consideration in \cite{Abe21} is $|5^1 S_0\rangle \rightarrow |5^3 P_1\rangle$ with frequency $\omega_\mathrm{Sr} = 2.73 \times 10^{15}$ rad/s. In terms of acceleration noise level and order of LMT, we consider the upgraded parameters, i.e 
\begin{subequations}
    \begin{align}
        \sqrt{S_a}(f) &= 6 \times 10^{-17} \mathrm{g/}\sqrt{\mathrm{Hz}} \, , \\
        n &= 1000 \, .
    \end{align}
In addition, for a $L =100$ m high tower, the free-fall time of atoms is given by $2T=\sqrt{8L/g}$, which implies $T\sim 4.5$ s. Following Eq.~\eqref{phase_acceleration_noise}, the phase noise PSD is 
\begin{align}
    S_\Phi(f) &= \left(\frac{n \omega_\mathrm{Sr} T^2}{c}\right)^2 S_a(f) \sim 10^{-8} \: \mathrm{rad^2/Hz}  \, ,
\end{align}
\end{subequations}
i.e a similar phase noise PSD as AION-10. Finally, the full integration time of the experiment corresponds to 1 year of observation, $T_\mathrm{int} \approx 3.16 \times 10^7$ s.

The second reason why we consider the SPID setup is to directly compare the expected sensitivity on axion and dilaton couplings of two experiments with the exact same experimental parameters but operating a gradiometer on one hand and the SPID variation on the other hand. For this matter, we will compare the sensitivity of the current version of AION-10, i.e a gradiometer, with a SPID setup using the same noise levels. In this case, in addition to the phase noise, we also include the EOM noise (which we show to be negligible in the following), which is used to shift the frequency of the laser to account for the isotope shift ($\sim 1$~GHz) as discussed in Sec.~\ref{sec:new_proposal_pres} and depicted in Fig.~\ref{fig:isotope_AI}. An alternative scheme would be to phase-lock two lasers with a $\sim 1$~GHz frequency offset. In either case, this setup generates an additional frequency fluctuation $\sigma_f(t)$ in the second beam, which comes from the EOM internal noise and from the phase noise of the GHz reference frequency. Whilst the raw EOM phase noise PSD can be approximated as $S^\mathrm{GHz}_\Phi(f) \approx 10^{-13} \left(f/\mathrm{Hz}\right)^{-2}\, \mathrm{rad^2/Hz}$ \cite{Barke15,Hartnett2012,Xie2017a}, it enters only differentially in the SPID setup at times separated typically by $L/c$. The resulting PSD is altered by a factor $2\left(1-\cos\left(2 \pi f L/c\right)\right) \approx (2\pi f L/c)^2$, since $2\pi f L/c \ll 1$ for all frequencies $f$ of interest. Since this noise is uncorrelated with the gradiometer phase noise, the total phase noise PSD of the SPID setup is simply
\begin{align}
S^\mathrm{SPID}_\Phi(f)&=S^\mathrm{GHz}_\Phi(f)\left(\frac{2\pi f L}{c}\right)^2 + S^\mathrm{AION}_\Phi(f) \,\nonumber \\
&\approx S^\mathrm{AION}_\Phi(f) \, ,
\end{align}
For the remaining experimental parameters, such as the total time of experiment, the flight time of atoms and the number of LMT kicks, we will assume the same values as AION-10, see Eq.~\eqref{AION_exp_param}.
We will also consider the use of four different pairs of isotopes of alkaline-Earth neutral atoms commonly used as optical clocks \cite{Ludlow15}, and we will assess the impact of the choice of the atoms on the ULDM  sensitivity: (i) two bosonic isotopes of Strontium $^{86}\mathrm{Sr}-^{88}\mathrm{Sr}$ based on the   $5s^2\:^1S_0 \rightarrow 5s5p\:^3P_1$ transition whose frequency is $\omega_\mathrm{Sr}= 2.73 \times 10^{15}$ rad/s \cite{LeTargat06,Ludlow15}; (ii) two bosonic isotopes of Calcium $^{40}\mathrm{Ca}-^{44}\mathrm{Ca}$ using the $4s4s\:^1S_0 \rightarrow 4s4p\:^3P_1$ transition with frequency $\omega_\mathrm{Ca}= 2.87 \times 10^{15}$ rad/s \cite{Sterr04} ;  (iii) one mixture of fermionic and bosonic isotope of Ytterbium $^{171}\mathrm{Yb}-^{176}\mathrm{Yb}$ using the  $6s^2\:^1S_0 \rightarrow 6s6p\:^3P_1$ transition of  frequency $\omega_\mathrm{Yb}= 3.39 \times 10^{15}$ rad/s \cite{Atkinson19,Jones23} and (iv) two bosonic isotopes of Mercury, $^{196}\mathrm{Hg}-^{202}\mathrm{Hg}$ using the  $6s^2\:^1S_0 \rightarrow 6s6p\:^3P_1$ transition of frequency $\omega_\mathrm{Hg}= 7.45 \times 10^{15}$ rad/s \cite{Witkowski19}. All these parameters are summarized in Tab.~\ref{tab:alk_isotope_freq}. Note that in these choices of isotopes and transitions, we ignore the small lifetime of the various excited states (of hundreds of $ns$ to hundreds of $\mu s$), which could limit the number of atoms detected at the end of the sequence due to spontaneous emission\footnote{This limitation could be overcome for optical transitions due to the large Rabi frequency, implying a very short $\pi$ pulse duration \cite{Rudolph20}.}.
\begin{table}
\begin{tabular}{ccc}
\hline
\hline
Isotopes & Transition & Frequency (rad/s)\\
\hline
\hline
$^{86}$Sr, $^{88}$Sr & $5s^2\:^1S_0 \rightarrow 5s5p\:^3P_1$ & $2.73 \times 10^{15}$ \\
$^{40}$Ca, $^{44}$Ca & $4s4s\:^1S_0 \rightarrow 4s4p\:^3P_1$ & $2.87 \times 10^{15}$ \\
$^{171}$Yb, $^{176}$Yb & $6s^2\:^1S_0 \rightarrow 6s6p\:^3P_1$ & $3.39 \times 10^{15}$ \\
$^{196}$Hg, $^{202}$Hg & $6s^2\:^1S_0 \rightarrow 6s6p\:^3P_1$ & $7.45 \times 10^{15}$ \\
\hline
\end{tabular}
\caption{Various isotope pairs used in the SPID setup, for direct comparison with AION-10}
\label{tab:alk_isotope_freq}
\end{table}

\section{\label{sec:sens_experiments} Sensitivity of various experiments to axion and dilaton fields}

In this section, we will derive the sensitivity of the experiments listed in the previous section on the various axion and dilaton couplings discussed in Sec.~\ref{sec:charges_DM}. These estimates are obtained by comparing the theoretical signal presented in Sec.~\ref{sec:signatures} to the published experimental noise presented in Sec.~\ref{sec:experiments} (or predictions regarding AION-10 and MAGIS-100). The expected sensitivity of MAGIS-100 to another DM candidate, namely to a $U(1)$ $B-L$ field has already been demonstrated in \cite{Abe21}, but we show that the expected sensitivity of SPID to axion and dilaton couplings is also very competitive. We insist that the results obtained in this section needs to be interpreted as sensitivity analysis or expected experimental reach, while actual constraints would require careful statistical searches for the signal, with a full analysis of statistical and systematic uncertainties.

\subsection{Experimental features impacting the signal\label{sec:exp_features}}
In this section, we will review the various experimental parameters that impact the theoretical signal presented in Sec.~\ref{sec:signatures}.
 
First, the observable signatures induced by an axion or dilaton field depend on the properties of the atoms used in the experiment. More precisely, they depend directly on the axionic or dilatonic charges that have been introduced in Sec.~\ref{sec:charges_DM}. In Tab. ~\ref{axionic_dilatonic_charge_table}, we present all non-universal axionic and dilatonic charges of all atomic species used in the set of experiments presented in Sec.~\ref{sec:experiments}. The axionic charges are derived from Eqs.~\eqref{axionic_mass_charge} and \eqref{axionic_freq_charge} for hyperfine atomic transitions. The dilatonic charges are derived from Eqs.~\eqref{dilatonic_mass_charge}, \eqref{dilatonic_hyp_freq_charge} and \eqref{dilatonic_freq_charge}. Note that, in the experiments considered in Sec.~\ref{sec:experiments}, $^{85}$Rb is only used in a two-photon transitions Bragg-type interferometer which is independent of the frequency charge at leading order, see Eq.~(\ref{phase_shift_exact_AI_Bragg}) and comments after. Similarly, $^{195}$Pt and $^{48}$Ti are only used in a classical test of the UFF which is also independent of the frequency charge (at leading order, see Eq.~(\ref{a_vUFF})). For these reasons, the frequency charges of these atoms are not provided in Tab.~\ref{axionic_dilatonic_charge_table}.

Second, as can be noticed from Eqs.~(\ref{eq:delta_a_UFF}), (\ref{phase_shift_exact_AI}) and (\ref{eq:phase_new_prop}), some observables depend on the initial velocity of the bodies.  Note that the method used to perform the calculations in Sec.~\ref{phase_AI_general} is only valid for Lagrangians that are at most quadratic in the position and velocity \cite{Storey}.  In the case of massive axion and dilaton, this is the case only in the galactocentric frame, which we assume to be the scalar field rest frame. In any other frame (in particular in a lab-centered reference frame), the field will behave as $\cos\left(\omega t - \vec k\cdot\vec x\right)$ leading to a Lagrangian not at most quadratic in the position.  In the galactocentric frame, the atoms and laser have an initial velocity corresponding to the velocity of the Sun in the DM halo, i.e 
\begin{align}\label{vel_DM}
    \vec v_0 =  v_\mathrm{lab} \hat e_{vL} + \tilde v_0 \hat e_{v0} \approx v_\mathrm{DM} \hat e_\mathrm{DM} \approx 10^{-3}c \: \hat e_\mathrm{DM} \, .
\end{align}
The Sun velocity in the galactic halo points towards $\alpha$ Cygni, the biggest star of the Cygnus constellation \cite{Miuchi20}, which corresponds to a right ascension $\alpha_\mathrm{DM}=310.36\degree \mathrm{E}$ and declination $\delta_\mathrm{DM}=45.28\degree \mathrm{N}$ \cite{vanLeeuwen07}.

The observable signatures depend also on the orientation of the experimental setup with respect to this initial velocity. Indeed, for classical tests of the universality of free fall between two macroscopic bodies, one measures $\Delta \vec a\propto \vec v_0$ (see Eq.\eqref{eq:delta_a_UFF}) projected onto the sensitive axis of the instrument. Therefore, the signal depends on the projection of the velocity of the bodies in the galactocentric frame with the sensitive axis of the experiment. Similarly, all the AI experiments depend on the scalar product between the initial galactocentric velocity of the atomic clouds with the direction of the velocity kick undergone in the interferometric schemes, i.e. on $\hat e_v \cdot \hat e_\mathrm{kick}$, see Eqs.~\eqref{phase_shift_exact_AI},~\eqref{eq:delta_phase_shift} and \eqref{eq:phase_new_prop}.

All the AI-based experiments operate at constant location, \textit{loc}, on Earth with longitude $\lambda_\mathrm{loc}$ and latitude $\phi_\mathrm{loc}$. We assume the velocity kick to be directed vertically, i.e. in the Earth's reference frame $\hat e_\mathrm{kick}=\left(\cos (\lambda_\mathrm{loc})\cos (\phi_\mathrm{loc}),\sin (\lambda_\mathrm{loc})\cos (\phi_\mathrm{loc}),\sin (\phi_\mathrm{loc})\right)$. As first approximation, we consider the declination as equivalent to the terrestrial latitude, i.e $\delta_\mathrm{DM}\approx \phi_\mathrm{DM}$, such that the dot product $\hat e_v \cdot \hat e_\mathrm{kick}$  is simply given by $\cos(\phi_\mathrm{loc})\cos(\phi_\mathrm{DM})\cos(\lambda_\mathrm{loc}-\lambda_\mathrm{DM}(t))+\sin(\phi_\mathrm{loc})\sin(\phi_\mathrm{DM})$. $\lambda_\mathrm{DM}(t)$ is the longitude of $\alpha$ Cygni at the time $t$ of the experiment. Indeed, while $\alpha_\mathrm{DM}$ is fixed, the former follows the Earth rotation with frequency $\omega_E \sim 7 \times 10^{-5}$ Hz and is therefore time dependent
\begin{subequations}
\begin{align}
    \lambda_\mathrm{DM}(t)&=\omega_E t+\varphi \, ,
\end{align}
where the phase $\varphi$ corresponds to the longitude of $\alpha$ Cygni at the origin of time reference considered. 
For short experiments with an experimental time  much smaller than a day (i.e. relevant for Stanford's experiment, see Sec~\ref{sec:stanford}), the dot product is roughly constant and depends therefore on the exact time of the day when the experiment was conducted. In order to infer a sensitivity estimate, we will only consider the mean value of the dot product which is given by $\sin(\phi_\mathrm{loc})\sin(\phi_\mathrm{DM})$, i.e 
\begin{align}\label{dot_product_DM_axis_1}
    \hat e_\mathrm{DM} \cdot \hat e_\mathrm{kick}\Big|_\mathrm{Stanford} &\approx 0.43 \, .
\end{align}
For longer time experiments, like the SPID variation, the dot product evolves with time, such that the signal evolves as 
\begin{align}
    s(t) &\propto \hat e_\mathrm{DM} \cdot \hat e_\mathrm{kick} \cos(\omega_\mathrm{DM}t +\phi_0)\,\\
    &= [\cos(\phi_\mathrm{loc})\cos(\phi_\mathrm{DM})\cos(\omega_E t +\varphi)+\,\nonumber\\
    &\sin(\phi_\mathrm{loc})\sin(\phi_\mathrm{DM})] \cos(\omega_\mathrm{DM}t +\phi_0) , 
\end{align}
i.e the Earth rotation modulates the signal at frequency $f_e=\omega_E/2\pi$. We will make two approximations for our estimates. For all the DM frequencies of interest $\omega_\mathrm{DM}$, the Earth rotation is a slowly varying function, because $\omega_E < \omega_\mathrm{DM}$. Therefore, we will assume that the signal will manifest itself by a single peak at frequency $f_\mathrm{DM}= \omega_\mathrm{DM}/2\pi$ in Fourier space. In addition, we will be interested only by the maximum value of the dot product, which is only a function of the different latitudes.
\begin{align}
    \hat e_\mathrm{DM} \cdot \hat e_\mathrm{kick} = \mathrm{Max}\left(|\cos(\phi_\mathrm{loc}\pm \phi_\mathrm{DM})|\right) \, .
\end{align}
For the various locations under consideration in this paper, we have
\begin{align}\label{dot_product_DM_axis_2}
\hat e_\mathrm{DM} \cdot \hat e_\mathrm{kick}\Big|_\mathrm{Oxford} &\approx 0.99 \, , \\
\hat e_\mathrm{DM} \cdot \hat e_\mathrm{kick}\Big|_\mathrm{Fermilab} &\approx 1.00 \, ,
\end{align}
where we assume the SPID variation operates at the same location as AION-10, i.e Oxford \cite{Badurina20} for consistent comparison. 

For MICROSCOPE, the axis of measurement is alongside the test masses cylinders' longitudinal symmetry axis \cite{Microscope17}. The orbital motion of the satellite around Earth is sun-synchronous, which means that the orientation of the orbital plane evolves with time with an annual period. As it was mentioned in Sec.~\ref{sec:experiments}, the measurements are distributed on 17 different sessions, each of them lasting $T_\mathrm{int}/17 \sim 5$ days on average. For such durations, we can assume the orbital plane to be fixed during each session. In addition, the satellite spins around an axis that is orthogonal to the axis of measurement with angular frequency $\omega_\mathrm{spin}$. In total, this means the dot product can be written as 
\begin{align}
    \hat e_\mathrm{DM} \cdot \hat e_\mathrm{meas.}(t)\Big|_\mathrm{\mu SCOPE} &= A(t) \cos(\omega_\mathrm{spin} t + \psi) \, ,
\end{align}
where $A(t)$ depends on the orientation of the orbital plane, i.e is fixed for one session but changes from one session to another, and with $\psi$ an irrelevant phase.
Using MICROSCOPE's publicly available data, we estimated numerically the coefficient $A(t)$ for every session and we find that its value oscillates between 0.71 and 1. For our estimates, we will consider its mean value, i.e $|\hat e_\mathrm{DM} \cdot \hat e_\mathrm{meas.}| \approx 0.85$. In conclusion, the signal presented in Eq.~\eqref{eq:delta_a_UFF} is modulated by the factor
\begin{align}
    \hat e_\mathrm{DM} \cdot \hat e_\mathrm{meas.}(t)\Big|_\mathrm{\mu SCOPE} &\approx 0.85 \cos(\omega_\mathrm{spin} t + \psi) \label{dot_product_DM_MICRO} \, .
\end{align}
\end{subequations}
Thus, the observable signal in the MICROSCOPE experiment is oscillating at the combination of the DM frequency and the spin frequency, i.e. has harmonics at the two frequencies $\omega_\mathrm{DM} \pm \omega_\mathrm{spin}$. Therefore, for DM frequencies in Eq.~\eqref{eq:delta_a_UFF} such that $\omega_\mathrm{DM} \ll \omega_\mathrm{spin}$, we will consider that the signal in MICROSCOPE oscillates at $\omega_\mathrm{spin}$, whereas if $\omega_\mathrm{DM} \gg \omega_\mathrm{spin}$, we will make our estimate with the signal oscillating at $\omega_\mathrm{DM}$. As explained in Sec.~\ref{sec:experiments}, $\omega_\mathrm{spin}$ depends on the measurement session since three different spinning frequencies have been used during the full mission. They differ by a factor 5 roughly \cite{Berge23}. For our rough sensitivity estimates, we use the data from session 404 as a basis, therefore we will assume a constant $\omega_\mathrm{spin} \sim 18.4$~mrad/s.


\begin{table*}
\centering
\begin{tblr}{
    vlines,
    colspec={cccccccccccc}
}
\hline
\SetCell[c=2,r=1]{}  & & \SetCell[c=2]{} Axionic charges & & \SetCell[c=7]{} Dilatonic charges \\
\hline\hline
Experiment & Species & $Q_M$ & $Q_\omega$  & $Q_{M,e}$ & $Q_{M,m_e}$  & $Q_{M,\hat m}$ & $Q_{M,\delta m}$ & $Q_{\omega, e}$  & $Q_{\omega, \hat m}$ & $Q_{\omega, \delta m} $\\
 & & [$\times 10^{-3}$] & [$\times 10^{-5}$] & [$\times 10^{-3}$] & [$\times 10^{-4}$] & [$\times 10^{-3}$]& [$\times 10^{-4}$] &&[$\times 10^{-4}$]&[$\times 10^{-2}$]
 \\
\hline\hline
\SetCell[r=2]{} $\mu$SCOPE & $^{195}$Pt \cite{Microscope18} & -69.065 & $-$ & 4.278 & 2.20 & 85.25 & 3.40 & $-$ &$-$ &$-$\\
& $^{48}$Ti \cite{Microscope18} & -68.770 & $-$ & 2.282 & 2.53 & 82.58 & 1.38 & $-$& $-$ & $-$\\\hline
\SetCell[r=2]{} Stanford  & $^{87}$Rb & -68.920 & 930 \cite{Flambaum06} & 2.869 & 2.34 & 83.95 & 2.54 & 2.34 \cite{Guena12} & -670 \cite{Guena12} & -1.73 \cite{Guena12} \\
& $^{85}$Rb & -68.924 & $-$ & 2.961 & 2.39 & 83.98 & 2.20 & $-$ &$-$ & $-$ \\\hline
& $^{40}$Ca & -68.715 & -0.188 & 2.409 & 2.75 & 82.08 & 0 & 2.02 \cite{Angstmann04} & 0.007 & 0 \\
& $^{44}$Ca & -68.738 & -0.182 & 2.116 & 2.50 & 82.29 & 1.55 & 2.02 & 0.006 & $\mathcal{O}(10^{-7})$ \\
&$^{86}$Sr & -68.933 & -0.554 & 3.074 & 2.43 & 84.06 & 1.98 & 2.06 & 0.003 & $\mathcal{O}(10^{-7})$\\
\SetCell[r=1]{} AION &$^{87}$Sr & -68.932 & -0.552 & 3.027 & 2.40 & 84.05 & 2.15 &  2.06 \cite{Bloom14} & 0.003 & $\mathcal{O}(10^{-7})$\\
\SetCell[r=1]{} MAGIS &$^{88}$Sr & -68.930 & -0.550 & 2.980 & 2.38 & 84.03 & 2.32 & 2.06 & 0.003 &$\mathcal{O}(10^{-7})$ \\
\SetCell[r=1]{} SPID &$^{171}$Yb & -69.054 & -1.206 & 4.114 & 2.25 & 85.14 & 3.08 & 2.31 \cite{Hinkley13} & 0.002 & $\mathcal{O}(10^{-7})$ \\
&$^{176}$Yb & -69.043 & -1.194 & 3.957 & 2.19 & 85.05 & 3.48 & 2.31 & 0.002 & $\mathcal{O}(10^{-7})$\\
&$^{196}$Hg & -69.077 & -0.684 & 4.469 & 2.24 & 85.35 & 3.12 &  2.81 \cite{Angstmann04} & 0.001 &$\mathcal{O}(10^{-7})$ \\
&$^{202}$Hg & -69.066 & -0.677 & 4.291 & 2.18 & 85.25 & 3.53 &  2.81 & 0.001 & $\mathcal{O}(10^{-7})$ \\\hline
\SetCell[r=1]{} All AI & SiO$_2$ & -68.442 & $-$ & 1.607 & 2.75 & 79.62 & 0.03 & $-$ & $-$ & $-$\\\hline
\end{tblr}
\caption{Axionic/dilatonic charges for species of atoms of interest (we assume Earth is made of SiO$_2$, composed of 47\% Si and 53\% O). The transitions are hyperfine for $^{87}$Rb and optical for the rest. Axionic charges are derived from Eqs.~\eqref{axionic_mass_charge} and \eqref{axionic_freq_charge}, while dilatonic charges are derived from Eqs.~\eqref{dilatonic_mass_charge},\eqref{dilatonic_hyp_freq_charge} and \eqref{dilatonic_freq_charge}. The leading order value of $Q_{\omega,m_e}$ is universal for all atomic transitions, see Eqs.~\eqref{dilatonic_hyp_freq_charge} and \eqref{dilatonic_freq_charge}, so it is not provided. For isotopes, we assume the same $Q_{\omega,e}$ value (see text).}
\label{axionic_dilatonic_charge_table}
\end{table*}

In addition, the DM has a velocity distribution in the galaxy, which implies that the scalar field oscillation will have a coherence time $\tau(\omega)$ given by \cite{Derevianko}
\begin{align}
    \tau(\omega) \approx \frac{2\pi c^2}{\sigma^2_v \omega} \approx \frac{2\pi \times 4\times 10^6}{\omega}\, ,
\end{align}
where $\sigma_v \approx 5\times 10^{-4}c$ is the DM galactic velocity dispersion \cite{Evans19}. As we shall see in the next paragraphs, the coherence time will affect the sensitivity of the experiment.

We will now present the general principle used to derive the sensitivity curves.  The sensitivity of a given experiment to a general DM-SM coupling $\chi$ at a given DM Compton frequency $\omega_\mathrm{DM}$ depends on: (i) the time of integration of the  experiment, (ii) possibly the coherence time of the DM field, (iii) the PSD of the noise at $\omega_\mathrm{DM}$ and (iv) on the expected amplitude of the theoretical signal produced by the DM candidate. In the following, we will assume a constant threshold SNR value of 1 for all experiments under consideration.  Let us write very generically the signal searched for $s(t)$ as 
\begin{align}\label{eq:s(t)}
    s(t)= \chi \left[X_s\right] \cos \left(\omega_\mathrm{DM} t+\varphi \right) \, ,
\end{align}
where we factorize the coupling from the rest of the signal.
If the time of integration is much shorter than the coherence time, i.e $T_\mathrm{int} \ll \tau(\omega_\mathrm{DM})$, this equation is fully valid and the experiment sensitivity on the coupling is 
\begin{subequations}\label{coupling_constraint_general_DM}
\begin{align}
    \chi &= \frac{\sqrt{\mathrm{SNR}}}{[X_s]}\sqrt{\frac{S_{s}}{T_{\mathrm{int}}}} \, .
\end{align}
Note that in this regime, where $T_\mathrm{int} \ll \tau(\omega_\mathrm{DM})$, a correction factor to the sensitivity arises due to the stochastic nature of the amplitude of the field \cite{Centers21}. In our case where we consider a 68\% detection threshold (i.e. SNR = 1), this correction factor induces a loss in signal of $\sim 1.5$.

On the other hand, when the integration time is much longer than the coherence time of the field, i.e $T_\mathrm{int} \gg \tau(\omega_\mathrm{DM})$, this means that the signal searched for and parametrized by Eq.~(\ref{eq:s(t)}) is no longer coherent, i.e. it should in principle be modeled as a sum of several stochastic harmonics, see \cite{foster:2018aa}. Another method to analyse the data  is to cut the dataset in fragments with duration smaller than $\tau(\omega_\mathrm{DM})$ and search for a coherent signal in each of these blocks of data. In such a case, the experimental sensitivity to the coupling is reduced and becomes \cite{Budker14}
\begin{align}
    \chi &= \frac{\sqrt{\mathrm{SNR}}}{[X_s]}\sqrt{\frac{S_s}{\sqrt{T_{\mathrm{int}}\tau(\omega_\mathrm{DM})}}} \, .
\end{align}
\end{subequations}

Finally, we will assume the local DM energy density to be $\rho_\mathrm{DM} = 0.4$ GeV/cm$^3$ \cite{McMillan11}.

\subsection{Sensitivity of the experiments to the axion-gluon coupling}

First of all, let us remind that all signals considered in this paper are quadratic in the coupling $1/f_a$.

\begin{figure*}
\centering
\includegraphics[width=0.8\textwidth]{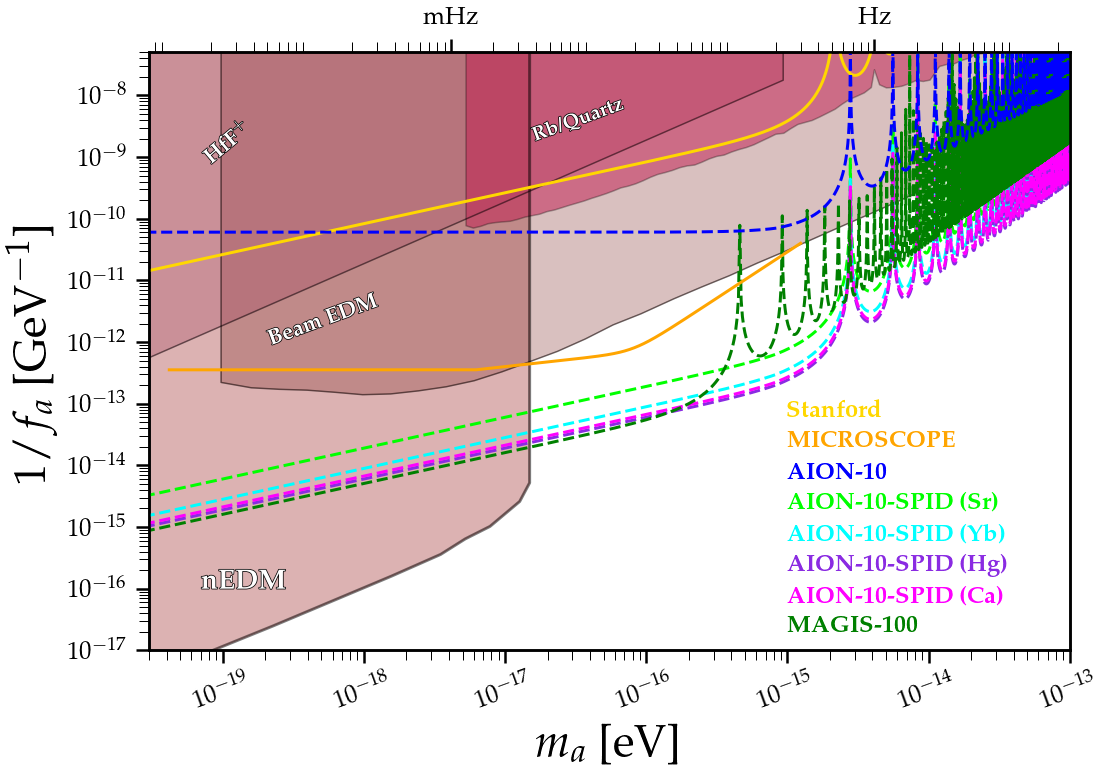}
\caption{Current lab constraints on $1/f_a$ axion coupling from \cite{AxionLimits} (the constraint from \cite{Rb_quartz} has been rescaled for consistent value of local DM energy density), including new sensitivity estimate resulting from this work, i.e from Stanford 
and MAGIS-100 atom interferometry experiments and EP classical test from MICROSCOPE. The expected sensitivity of the AION-10-SPID-like experiment is shown in four different colors, each using four different pairs of isotopes, denoted "AION-10-SPID" (see text).}
\label{contraints_fa}
\end{figure*}

Several laboratory experiments \cite{Beam_EDM,HfF,Rb_quartz,nEDM} already constrain the $f_a$ coupling of the axion, which are all shown by the red-brown full lines in Fig.~\ref{contraints_fa}.

\subsubsection{\label{Delta_a_axions} MICROSCOPE}

Using Eqs.~\eqref{eq:delta_a_UFF} and \eqref{mass_osc_axion}, we can express the amplitude of the differential acceleration  between two test-masses A and B as
\begin{widetext}
\begin{align}
|\Delta \vec a(t)| &= \frac{2\hbar c^3 \rho_\mathrm{DM} v_\mathrm{DM}}{f^2_a\omega_a}\left|\left([Q^A_M]_a-[Q^B_M]_a\right)\hat e_\mathrm{DM} \cdot \hat e_\mathrm{meas.}(t)\Big|_\mathrm{\mu SCOPE}\right| 
\equiv \frac{1}{f^2_a}[X_a] \, .
\label{Amp_delta_acc_axion}
\end{align}
\end{widetext}

In Fig.~\ref{contraints_fa}, the expected sensitivity of MICROSCOPE is shown by the orange full line. As a reminder, all the experimental parameters we used to obtain this curve are explicited in Sec.~\ref{sec:experiments} and \ref{sec:exp_features}. This curve presents two breaking point frequencies. The first one $f\sim 1$ mHz corresponds to half of $f_\mathrm{spin} \sim 3$ mHz. As we discussed in the previous section, we consider two different frequency regimes, depending on whether the signal frequency is higher or lower than $f_\mathrm{spin}$. In the axion case, the signal Eq.~\eqref{Amp_delta_acc_axion} oscillates at twice the axion field frequency, therefore the breaking point is $f_\mathrm{spin}/2$. The second breaking point arising at $f \sim 15$ mHz corresponds to half of the bucket frequency of the acceleration noise PSD, for the same reason as above. Note that there is no breaking point frequency associated with the coherence time of the field because it would arise at a frequency larger than the bandwidth of the noise PSD, at around 500 mHz.
As it can be seen from this curve, a complete re-analysis of MICROSCOPE's data would enable to constrain a new region of the parameter space, over approximately two orders of magnitude in mass, compared to existing laboratory experiments. 

\subsubsection{\label{AI_axions} Atom interferometry}

For the atom interferometry experiments, i.e Stanford Tower, AION-10, MAGIS-100, and the SPID variation, we use Eqs.~\eqref{eq:delta_phase_shift}, \eqref{eq:phase_gradio} and \eqref{eq:phase_new_prop} to express respectively the amplitude of the phase shift between atomic species A and B in differential two-photon transition AI (Stanford experiment), gradiometers (AION-10) and SPID with both AION-10 and MAGIS-100 experimental parameters. The Stanford experiment is a Bragg-transition differential AI involving 2 isotopes, we will therefore assume $k^A_\mathrm{eff} = k^B_\mathrm{eff} \equiv k^\mathrm{AB}_\mathrm{eff}$. Then, the phase shifts read 
\begin{widetext}
\begin{subequations}\label{Amp_phase_shift_axion}
\begin{align}
\Delta \Phi^\mathrm{Stanford}_\mathrm{AB} &\approx \frac{2 \hbar c^3 \rho_\mathrm{DM}v_\mathrm{DM}k^\mathrm{AB}_\mathrm{eff}}{\omega^3_a f^2_a}\Big|\left([Q^A_M]_a-[Q^B_M]_a\right)\hat e_\mathrm{DM} \cdot \hat e_\mathrm{kick}\Big|_\mathrm{Stanford}\Big|\sin^2(\omega_aT) \equiv \frac{1}{f^2_a}[X_\Phi]_\mathrm{Stanford} \label{Amp_phase_shift_MZ_axion} \, , \\
\Delta \Phi^\mathrm{AION-10}_\mathrm{A} &\approx \frac{4 n \omega^0_A \Delta r \hbar c^2 \rho_\mathrm{DM} [Q^A_\omega]_a}{\omega^2_a f^2_a}\sin^2(\omega_aT) \equiv \frac{1}{f^2_a}[X_\Phi]_\mathrm{AION-10}\, , \\
\Delta \Phi^\mathrm{MAGIS-100}_\mathrm{AB} &\approx \frac{2n \omega_0 \hbar c^2 \rho_\mathrm{DM} v_\mathrm{DM}}{\omega^3_a f^2_a}\left|\left([Q^A_M]_a-[Q^B_M]_a\right)\hat e_\mathrm{DM} \cdot \hat e_\mathrm{kick}\Big|_\mathrm{Fermilab}\right|\sin^2(\omega_aT) \equiv \frac{1}{f^2_a}[X_\Phi]_\mathrm{MAGIS-100}\, , \\
\Delta \Phi^\mathrm{AION-10-SPID}_\mathrm{AB} &\approx \frac{2n \omega_0 \hbar c^2 \rho_\mathrm{DM} v_\mathrm{DM}}{\omega^3_a f^2_a}\left|\left([Q^A_M]_a-[Q^B_M]_a\right)\hat e_\mathrm{DM} \cdot \hat e_\mathrm{kick}\Big|_\mathrm{Oxford}\right|\sin^2(\omega_aT) \equiv \frac{1}{f^2_a}[X_\Phi]_\mathrm{AION-10-SPID}\, ,
\end{align}
\end{subequations}
\end{widetext}
at first order in the axionic charges and where we again factorized the axion-gluon coupling from the rest.  

In Fig.~\ref{contraints_fa}, we present the sensitivity of the Stanford Tower \cite{Stanford20}, 
MAGIS-100 experiment \cite{Abe21} and the SPID AI setup with AION-10 experimental parameters, denoted ``AION-10-SPID''.

One can notice that MICROSCOPE is approximately two to three orders of magnitude more sensitive than Stanford. This is consistent considering that the signal is quadratic in $1/f_a$, that MICROSCOPE constrains the E\"otv\"os parameter $\eta$ three orders of magnitude better than the Stanford experiment and that the difference in axionic mass charges of species used in the experiments is 2 orders of magnitude larger for MICROSCOPE. 

Contrary to the gradiometer setup of AION-10 which is almost insensitive to the axion-gluon coupling (at leading order, the sensitivity is proportional to the axionic frequency charge of the optical transition of Sr, which is 0), the SPID variation or AION-10 would have the largest sensitivity to this coupling compared to existing laboratory experiments. Such an experiments would improve the current lab constraint \cite{Beam_EDM} by 2 orders of magnitude over a  mass range of 4 orders of magnitude  ($10^{-17}-10^{-13}$ eV). In this mass range, the MAGIS-100 experiment, which uses the same setup has also a very interesting sensitivity on the coupling, which is comparable to the one of AION-10 in the SPID setup. 

\subsection{Sensitivity of the experiments to the dilaton couplings}

The calculations are similar to the ones from the previous section with one difference: the signal induced by a dilaton DM candidate is linear to the dilaton coupling coefficient $d_i$, while it was quadratic in the coupling for the axion.

\subsubsection{\label{Delta_a_dilatons} MICROSCOPE}
Using Eqs.~\eqref{eq:delta_a_UFF} and \eqref{mass_osc_dilaton}, we can express the amplitude of the acceleration difference between two test-masses A and B as 
\begin{widetext}
\begin{align}
 |\Delta \vec a(t)| &= \frac{\sqrt{8 \pi G\rho_\mathrm{DM}}v_\mathrm{DM}}{c}\left|\left([Q^A_M]_\phi-[Q^B_M]_\phi\right)\hat e_\mathrm{DM} \cdot \hat e_\mathrm{meas.}(t)\Big|_\mathrm{\mu SCOPE}\right| \equiv \sum_{i=e,m_e,\hat m, \delta m} d^*_i[X^i_a] \, ,
\label{Amp_delta_acc_dilaton}
\end{align}
\end{widetext}
where $d^*_i = d_i -d_g$ for $i=m_e,\hat m, \delta m$ or $d^*_i =d_e$ (following Eq.~\eqref{partial_dil_mass_charge}), are the couplings  we factorized from the rest of the signal and where each function $X^i_a$ has units of acceleration and contains the partial dilatonic mass charge $Q_{M,i}$ defined in Eq.~\eqref{partial_dil_mass_charge}. 

As it has already been discussed in \cite{hees18}, the sensitivity of MICROSCOPE from the oscillatory behavior of the field, descrbed by \eqref{Amp_delta_acc_dilaton} is not competitive compared to the Yukawa-type fifth-force generated by the Earth on the two test masses of MICROSCOPE (shown in Fig. \ref{full_dilaton_new}).

\subsubsection{\label{AI_dilatons}Atom Interferometry}


Regarding atom interferometry experiments, we use \eqref{eq:phase_gradio} and \eqref{eq:phase_new_prop} and we express respectively the amplitude of the phase shift between atomic species A and B in differential two-photon transition AI, gradiometers and the SPID setup, with respectively AION-10 and MAGIS-100 experimental parameters. Then, the different phase shifts read
\begin{widetext}
\begin{subequations}
\begin{align}
\Delta \Phi^\mathrm{Stanford}_\mathrm{AB} &\approx \frac{4\sqrt{8 \pi G \rho_\mathrm{DM}}v_\mathrm{DM}k^\mathrm{AB}_\mathrm{eff}}{\omega^2_\phi c}\Big|\left([Q^A_M]_\phi-[Q^B_M]_\phi\right) \hat e_\mathrm{DM} \cdot \hat e_\mathrm{kick}\Big|_\mathrm{Stanford}\Big|\sin^2\left(\frac{\omega_\phi T}{2}\right) \equiv d^*_i[X^i_\Phi]_\mathrm{Stanford}\, , \\
\Delta \Phi^\mathrm{AION-10}_\mathrm{A} &\approx \frac{4 n \omega^0_A \Delta r \sqrt{8 \pi G \rho_\mathrm{DM}}[Q^A_\omega]_\phi}{\omega_\phi c}\sin^2\left(\frac{\omega_\phi T}{2}\right) \equiv d^*_i [X^i_\Phi]_\mathrm{AION-10}\, , \\
\Delta \Phi^\mathrm{MAGIS-100}_\mathrm{AB} &\approx \frac{4 n \omega_0 \sqrt{8 \pi G\rho_\mathrm{DM}}v_\mathrm{DM}}{\omega^2_\phi c^2}\left|\left([Q^A_M]_\phi-[Q^B_M]_\phi\right)\hat e_\mathrm{DM} \cdot \hat e_\mathrm{kick}\Big|_\mathrm{Fermilab}\right|\sin^2\left(\frac{\omega_\phi T}{2}\right) \equiv d^*_i [X^i_\Phi]_\mathrm{MAGIS-100}\, , \\
\Delta \Phi^\mathrm{AION-10-SPID}_\mathrm{AB} &\approx \frac{4 n \omega_0 \sqrt{8 \pi G\rho_\mathrm{DM}}v_\mathrm{DM}}{\omega^2_\phi c^2}\left|\left([Q^A_M]_\phi-[Q^B_M]_\phi\right)\hat e_\mathrm{DM} \cdot \hat e_\mathrm{kick}\Big|_\mathrm{Oxford}\right|\sin^2\left(\frac{\omega_\phi T}{2}\right) \equiv d^*_i [X^i_\Phi]_\mathrm{AION-10-SPID}\, ,
\label{Amp_phase_shift_dilaton}
\end{align}
\end{subequations}
\end{widetext}
where we sum over the various coupling constants, i.e $i =\{e,m_e,\hat m, \delta m\}$ and where we again factorized the couplings from the rest of the signal containing partial dilatonic mass and/or frequency charges defined in Eqs.~\eqref{partial_dil_mass_charge},\eqref{partial_dil_freq_charge} depending on the experiment.

\begin{figure*}
\centering
\includegraphics[width=\textwidth]{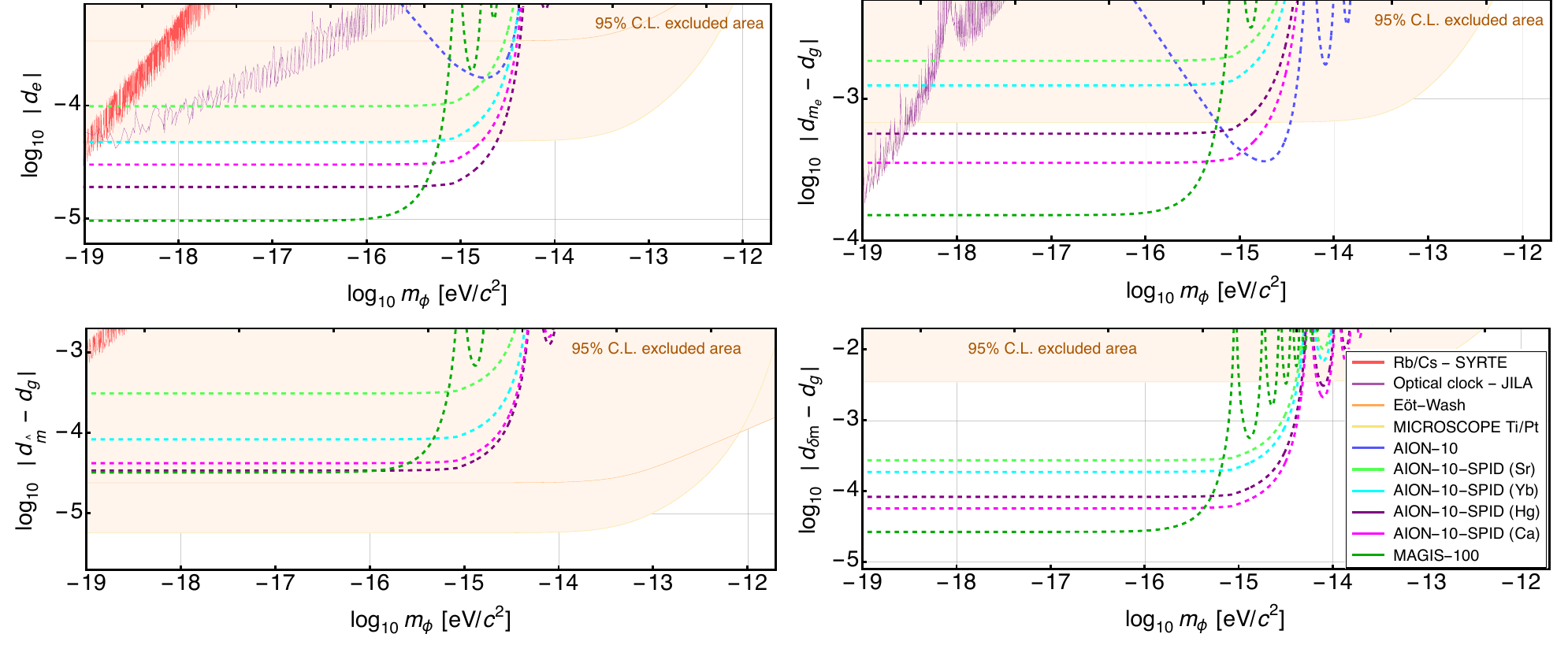}
\caption{Current constraints on all the dilatonic couplings of interest in this paper : $d_e$ (top left), $d_{m_e}-d_g$ (top right), $d_{\hat m}-d_g$ (bottom left), $d_{\delta m}-d_g$ (bottom right) from \cite{Microscope22,Wagner12,Hees16,Kennedy20}, with 95\% confidence level (shown in light orange background). The expected sensitivities of AION-10 are shown in dark blue dashed line while the expected sensitivity of AION-10 using the SPID variation, noted "AION-10-SPID" are respectively shown in green, light blue, purple and magenta dashed lines, depending on the isotope pair used. Finally, the expected sensitivity of MAGIS-100 is shown in dark green.}
\label{full_dilaton_new}
\end{figure*}

On the dilatonic side, laboratory constraints such as Torsion balances \cite{Wagner12}, MICROSCOPE\footnote{As discussed earlier, the best MICROSCOPE sensitivity comes from the static term of the field, i.e when considering a fifth force generated by Earth on the two test masses, while we only focused on the oscillatory term in this paper.} \cite{Microscope22} or hyperfine and optical clocks \cite{Hees16, Kennedy20} are currently the most stringent constraints on the different dilatonic couplings. 


In Fig.~\ref{full_dilaton_new}, we do not present the sensitivity of the Stanford tower experiment, as it is not competitive compared to the best existing constraints.
As it was discussed in \cite{Badurina22}, AION-10 would improve the current constraint on the $d_{m_e}-d_g$ coupling, over a mass range approximately between $7 \times 10^{-16}$ eV and $4\times 10^{-15}$ eV.

With the same experimental parameters as AION-10, operating the SPID setup would improve the current constraints on various dilatonic couplings. By respectively using Ca and Hg isotopes pairs, the constraint on $d_e$ would be improved by a factor $\sim 2$ and 3 respectively, over a mass range covering four orders of magnitude (approximately from $10^{-19}$ eV to $10^{-15}$ eV), compared to MICROSCOPE, the current best constraint in this mass range. Regarding the $d_{m_e}-d_g$ coupling, the use of Ca and Hg isotopes respectively would improve the best constraints, by a factor 2.5 and 1.5 respectively over the same mass range. Finally, all pairs of isotopes presented would improve the current best constraint on $d_{\delta m}-d_g$ by one order of magnitude, depending on the isotope pair used, over a mass range covering more than 4 orders of magnitude  (from $10^{-19}$ eV to $3 \times 10^{-15}$ eV). 

One can notice also that operating both gradiometers and SPID at AION-10 would give complementary sensitivities for the search of the $d_{m_e}-d_g$ coupling of the dilaton. Indeed, while the gradiometer would improve the current constrain in the $\sim 7 \times 10^{-16}-3\times 10^{-15}$ eV/$c^2$ mass range, the SPID setup's sensitivity at lower mass is better, as described in the last paragraph. In addition, SPID would be more sensitive than AION-10 to the $d_e, d_{\hat m}-d_g$ and $d_{\delta m}-d_g$ couplings in the full range of masses of interest (i.e lower than $10^{-14}$ eV/$c^2$). 

Regarding the sensitivity of MAGIS-100 operating the SPID setup, one can notice that it would overall give the best constraint at low masses.
One can notice that MAGIS-100 would reach better sensitivity on most couplings, compared to the AION-10-SPID variation. The reason is that for low masses, such that $\omega_\phi T \ll 1$, the signal is quadratic in the free fall time T, and MAGIS-100 using a much longer baseline than AION-10 (100m vs 10m), the free fall time is roughly 6 times longer for MAGIS-100, resulting in increased signal. Note that compared to the sensitivity curves on dilaton couplings presented in \cite{Abe21}, but using the gradiometer setup of MAGIS-100, the curve presented in this paper would constrain a larger DM mass range. Indeed, while the former has a peak sensitivity around a mass $10^{-15}$ eV and quickly loses sensitivity for lower masses, the SPID-like setup of MAGIS-100 would have a constant sensitivity of the same order of magnitude for 3 orders of magnitude of mass ($10^{-19}-10^{-16}$ eV).

\section{Conclusion and outlook}

In this paper, we derive the complete form of the expected signals in experiments involving differential acceleration between two test masses and in various AI setups in the general framework of atoms with time oscillating mass and transition frequency. We show that if the mass and frequency charges, which characterize respectively the amplitude of oscillation of the rest mass and transition frequency of the atom, are not universal among atoms, non zero signals in such experiments can be expected. If observed, such signals would constitute a violation of the equivalence principle. We study two particular models that lead to time varying mass and transition frequencies: axion like particles (ALP) and dilatons. 

We demonstrated a new way of constraining ALP dark matter coupling to gluons. In particular, we showed that the MICROSCOPE experiment has sufficient sensitivity to constrain this coupling at an unprecedented level, compared to already existing laboratory experiments. This, in our opinion, warrants a corresponding detailed analysis of MICROSCOPE data to search for a potential ALP signal. 

We also show the sensitivity to dilaton and ALP couplings of various AI setups. In particular, we derive the sensitivity of MAGIS-100-like setups \cite{Abe21}, denoted as SPID in this paper, to dilaton and axion couplings. Using AION-10 experimental parameters, we show that the SPID variation would be able to constrain both ALP coupling to gluons and dilaton couplings to photons, electrons and quarks $d_e, d_{m_e}-d_g, d_{\delta m}-d_g$ to remarkable levels compared to existing laboratory experiments and future experiments, including AION-10. Indeed, while AION-10 would not be sensitive to the axion-gluon at first order, the SPID experiment would improve the best laboratory constraints by roughly 2 orders of magnitude over 4 orders of magnitude mass range. Regarding the dilatonic couplings, the SPID experiment would largely surpass the expected sensitivity of AION-10 on $d_e, d_{\hat m}-d_g, d_{\delta m}-d_g$ couplings over the full mass range, while being more sensitive than AION-10 on the $d_{m_e}-d_g$ coupling at masses $m_\phi \leq 10^{-15}$ eV/$c^2$. We also show that the MAGIS-100 sensitivity to all of these couplings is very competitive for DM masses in the $10^{-19}-10^{-16}$ eV range due to its larger baseline.

Despite being extremely sensitive to various DM candidates-SM fields couplings, the SPID setup would also test the universality of free fall, with unprecedented level. Assuming MAGIS-100 experimental set of parameters described in Section.~\ref{New_param}, the free fall acceleration on ground $g=9.81$ m/s$^2$, a total cycle time $T_\mathrm{cycle} = 20$ s, to account for the atom preparation, free fall and measurement, the corresponding constraint on the E\"otv\"os parameter would reach 
\begin{align}
    \eta = \frac{1}{gk_\mathrm{eff} T^2}\sqrt{\frac{S^\mathrm{MAGIS-100}_\Phi}{2T_\mathrm{cycle}}}\approx 9 \times 10^{-18} \, ,
\end{align}
which would improve the current bound from MICROSCOPE \cite{Microscope22} by more than 2 orders of magnitude. However, this estimate is certainly over-optimistic as it does not take into account additional DC effects on the isotopes, such as gravity gradients, temperature gradients (blackbody radiation), wave-front aberrations, magnetic field gradients, etc...

\begin{acknowledgments}

The authors acknowledge Leonardo Badurina, Diego Blas, Robin Corgier, Daniel Derr and Albert Roura for helpful discussions. This work was supported by the Programme National GRAM of CNRS/INSU with INP and IN2P3 cofunded by CNES.

\end{acknowledgments}

\appendix

\section{\label{ap:MZ_phase}Full calculation of the two photon transition AI phase shift}

We consider an atom A whose nominal rest mass and transition frequency $m^0_A,\omega^0_A$ are perturbed such that they oscillate in phase as in Eq.~\eqref{general_mass_freq_osc}, i.e $m^0_A\left(1+Q^A_M\cos(\omega t + \phi_0)\right), \omega^0_A\left(1+Q^A_\omega\cos(\omega t + \phi_0)\right)$, where the charges $Q^A_M,Q^A_\omega$ represent the respective amplitude of oscillation of the rest mass and transition frequency. We show how this perturbation produces a phase at the end of a two-photon transition atom interferometer, namely we derive the results found in Eq.\eqref{phase_shift_exact_AI}. We will make the calculations assuming a Raman interferometer, i.e when the transition frequency of the atom is relevant, but note that the leading order phase of the Bragg interferometer can be recovered by setting $\omega^0_A=0$, as we will show it through the appendix.
We also make all calculations at first order in the perturbations $Q_M,Q_\omega$.

We describe the quantum state of the different atom wavepackets by a wavefunction $\Psi$, which we break down into two different wavefunctions $\Psi_\mathrm{I}$ and $\Psi_\mathrm{II}$, where I,II respectively the up/down paths in Fig.~\ref{Mach-Zehnder_osc}, depending on the classical path the atom followed. 
Considering free particles at the input of the interferometer, the atomic plane waves have the form \cite{Storey}
\begin{align}
\Psi_\mathrm{init}(t_0,\vec x_0) = \Psi_0 e^{i\Phi(t_0,\vec x_0)} \, ,
\label{wavefunction_init}
\end{align}
with the amplitude of the wavefunction $\Psi_0$, $\Phi(t_0,\vec x_0) = \vec k \cdot \vec x_0 - \omega t_0 - \varphi$ with $\omega, \vec k, \varphi$ respectively its angular frequency, wavevector and constant phase.
At the output of the interferometer, at a time $T_f \geq 2T$, considering all the different phase contributions listed previously, the atomic plane wave along the trajectory j read
\begin{align}
\Psi_\mathrm{j}(T_f,\vec x_f) = \Psi_\mathrm{init}(T_f,\vec x_f) \times  e^{i\Phi_\mathrm{j}} \, ,
\label{wavefunction_fin}
\end{align}
where $\Phi_\mathrm{j}=\Phi_\mathrm{sj}+\Phi_\mathrm{\ell j}+\Phi_\mathrm{uj}$ represents the trajectory dependent phase factor.
Then, at some detection time $T_\mathrm{d} \geq T_\mathrm{f}$, a detector measures by fluorescence the number of atoms on each quantum state, which is essentially the measurement of the probability that the two wavepackets are in the same quantum state, i.e
\begin{align}
  \int \Big|\Psi_\mathrm{I}(T_\mathrm{d},\vec x_\mathrm{d}) + \Psi_\mathrm{II}(T_\mathrm{d},\vec x_\mathrm{d}) \Big|^2 dS \, ,
  \label{overlap}
\end{align}
where the integral is taken over the detector area S. Plugging Eq.~\eqref{wavefunction_fin} into Eq.~\eqref{overlap} and neglecting loss of contrast due to decoherence, we find that the measurement result is proportional to (1 + $\cos \Delta \Phi$) where $\Delta \Phi = \Phi_\mathrm{I} - \Phi_\mathrm{II}$ is the phase difference between the two wave functions at $T_\mathrm{d}$.
For simplicity, we consider $T_\mathrm{d} \equiv 2T$.

\begin{widetext}
We first derive the perturbed equations of motion of the atom following by the perturbation to the acceleration Eq.~\eqref{a_vUFF}, to get the motion followed by the different wavepackets along the trajectories presented in Fig.~\ref{Mach-Zehnder_osc}. They read 
\begin{subequations}\label{EoM}
\begin{align}
&\vec v_A(t,t_0) \approx \vec v_0\left(1-Q^A_M\left(\cos(\omega  t+\phi_0)-\cos(\omega t_0+\phi_0)\right)\right) \, , \\
&\vec x_A(t,t_0) \approx  \vec x_0 + \vec v_0(t-t_0) -Q^A_M\frac{\vec v_0}{\omega}\left(\sin(\omega t+\phi_0)-\sin(\omega t_0+\phi_0)-\omega(t-t_0)\cos(\omega t_0+\phi_0)\right) \, ,
\end{align}
\end{subequations}    
at first order in the perturbation $Q^A_M$, where we discarded the frequency term from the acceleration, as it is negligible, and with $\vec x_0 = \vec x(t_0), \vec v_0=\vec v(t_0)$ respectively the initial position and velocity of the atom when entering the interferometer.

We start by the calculation of the position and velocity of the atom at the end of the trajectory portion 1 in Fig.~\ref{Mach-Zehnder_osc}. From Eq.\eqref{EoM}, the atom does not undergo any kick velocities from the laser pulses, hence after a time T, its equations of motion read
\begin{subequations}
    \begin{align}
        \vec x^{(1)}_A(T) &= \vec v_0\left(T-\frac{Q^A_M}{\omega}(\sin(\omega T+\phi_0)-\sin(\phi_0)- \omega T\cos(\phi_0))\right)\, , \\
       \vec v^{(1)}_A(T) &= \vec v_0\left(1-Q^A_M\left(\cos(\omega T+\phi_0)-\cos(\phi_0)\right)\right) \, .
    \end{align}
\end{subequations}
On the other hand, the wavepacket on portion 2 in Fig.~\ref{Mach-Zehnder_osc} has undergone a kick velocity with amplitude $v_\mathrm{kick,0}+\delta v_\mathrm{kick}(t=0)$, following Eq.~\eqref{kick_modif}, with 
\begin{align}
\delta \vec v_\mathrm{kick}(t=0) = \vec v_\mathrm{kick,0}\left(Q^L_\omega-Q^A_M\right)\cos(\phi_0) \, ,
\end{align}
with $\vec v_\mathrm{kick,0}$ along an arbitrary direction $\hat x$, hence
\begin{subequations}
\begin{align}
    &\vec x^{(2)}_A(T) = (\vec v_0+\vec v_\mathrm{kick,0})T - Q^A_M\frac{\sin(\omega T+\phi_0)-\sin(\phi_0)}{\omega}(\vec v_0 +\vec v_\mathrm{kick,0})+T(Q^A_M\vec v_{0}-Q^L_\omega \vec v_\mathrm{kick,0})\cos(\phi_0)\, , \\
    &\vec v^{(2)}_A(T) = \vec v_0+\vec v_\mathrm{kick,0} - Q^A_M\cos(\omega T+\phi_0)(\vec v_0 +\vec v_\mathrm{kick,0})+\left(Q^A_M\vec v_0 +Q^L_\omega \vec v_\mathrm{kick,0}\right)\cos(\phi_0) \, ,
    \end{align}
\end{subequations}
At time T, both wavepackets undergo a kick velocity with opposite direction, such that their momenta states are exchanged. Following Eq.~\eqref{kick_modif}, this means that at the end of portion 1, the atom undergoes a kick velocity of amplitude $v_\mathrm{kick,0}+\delta v_\mathrm{kick}(t=T)$, with
\begin{align}
    \delta \vec v_\mathrm{kick}(t=T)=\vec v_\mathrm{kick,0}\left(Q^L_\omega-Q^A_M\right)\cos(\omega T+\phi_0) \, ,
\end{align}
with same direction $\hat x$, hence, at the end of the portion 3, its coordinates read
\begin{subequations}
\begin{align}
    &\vec x^{(3)}_A(2T) = (2\vec v_0+\vec v_\mathrm{kick,0})T+\frac{Q^A_M}{\omega}\left(\vec v_\mathrm{kick,0}\sin(\omega T+\phi_0)+\vec v_0\sin(\phi_0)+2\vec v_0\omega T\cos(\phi_0)-(\vec v_0+\vec v_\mathrm{kick,0})\sin(2\omega T+\phi_0)\right)+\,\nonumber\\
    &Q^L_\omega \vec v_\mathrm{kick,0}T \cos(\omega T + \phi_0) \, , \\
    &\vec v^{(3)}_A(2T) = \vec v_0+\vec v_\mathrm{kick,0} - Q^A_M \left((\vec v_0+\vec v_\mathrm{kick,0})\cos(2\omega T+\phi_0)-\vec v_0\cos(\phi_0)\right)+Q^L_\omega \vec v_\mathrm{kick,0}\cos(\omega T + \phi_0)\, .
    \label{v3_2T}
\end{align}
\end{subequations}
The atom at the end of the portion 2 undergoes a kick velocity in the other direction, i.e of amplitude $v_\mathrm{kick,0}+\delta v_\mathrm{kick}(t=T)$, but with opposite direction (-$\hat x$), compared to the previous laser kicks, hence  
\begin{subequations}
\begin{align}
    &\vec x^{(4)}_A(2T) = (2\vec v_0+\vec v_\mathrm{kick,0})T-\frac{Q^A_M}{\omega}\left(\vec v_\mathrm{kick,0}\sin(\omega T+\phi_0)+\vec v_0\sin(2\omega T+\phi_0)-(\vec v_0+\vec v_\mathrm{kick,0})\sin(\phi_0)-2\vec v_0\omega T\cos(\phi_0)\right)-\, \nonumber \\
    &Q^L_\omega \vec v_\mathrm{kick,0} T \left(\cos(\omega T +\phi_0)-2\cos(\phi_0)\right)\, , \\
    &\vec v^{(4)}_A(2T) = \vec v_0\left(1 - Q^A_M\left(\cos(2\omega T+\phi_0)-\cos(\phi_0)\right)\right) - Q^L_\omega \vec v_\mathrm{kick,0}\left(\cos(\omega T + \phi_0)-\cos(\phi_0)\right)\, ,
\end{align}
\end{subequations}
at the end of portion 4.
\end{widetext}

\subsection{Propagation phase contribution}
The first component of phase shift is the one coming from the phase accumulated by atoms throughout the whole interferometric paths taking into account modified equations of motion and perturbed kicks. In the special case of quadratic lagrangian in the position and velocity of the atom at maximum, this phase is by the principle of least action the integral of the lagrangian over the path from initial point $i$ to final point $f$ $(t_i,x_i)\rightarrow (t_f,x_f)$
\begin{align}\label{eq:prop_phase_ap}
    \Phi_\mathrm{s} = \frac{1}{\hbar}\int_{t_i}^{t_f} L(x,\dot{x}) dt \, ,
\end{align}
where the lagrangian is defined in Eq.~\eqref{macro_lagrangian}. The internal energy term of the lagrangian being associated with the oscillation of the transition energy, it only contributes when the atom is on the excited state, i.e on paths 2 and 3 in Fig.~\ref{Mach-Zehnder_osc}.
Then, the phase accumulated by the atoms on the path I of the interferometer is 
\begin{align}
\Phi_\mathrm{sI} &=-\omega^0_A\int_0^{T}dt (1+Q^A_\omega \cos(\omega t +\phi_0))\left(1-\frac{|\vec v^{(2)}_A(t)|^2}{2c^2}\right)-\,\nonumber \\
&\frac{m^0_Ac^2}{\hbar}\left[\int_0^T dt \left(1+Q^A_M\cos(\omega t +\phi_0)\right)\left(1-\frac{|\vec v^{(2)}_A(t)|^2}{2c^2}\right)+\right.\,\nonumber\\
&\left.\int_T^{2T} dt\left(1+Q^A_M\cos(\omega t +\phi_0)\right)\left(1-\frac{|\vec v^{(4)}_A(t)|^2}{2c^2}\right)\right] \, ,
\end{align}
while the atom wavepacket on the path II accumulates a phase 
\begin{align}
\Phi_\mathrm{sII} &=-\omega^0_A\int_T^{2T}dt (1+Q^A_\omega \cos(\omega t +\phi_0))\left(1-\frac{|\vec v^{(3)}_A(t)|^2}{2c^2}\right)-\,\nonumber \\
&\frac{m^0_Ac^2}{\hbar} \left[\int_0^{T} dt \left(1+Q^A_M\cos(\omega t +\phi_0)\right)\left(1-\frac{|\vec v^{(1)}_A(t)|^2}{2c^2}\right)+\right.\,\nonumber\\
&\left.\int_T^{2T} dt\left(1+Q^A_M\cos(\omega t +\phi_0)\right)\left(1-\frac{|\vec v^{(3)}_A(t)|^2}{2c^2}\right)\right] \, ,
\end{align}
where $v^{(1)},v^{(2)},v^{(3)},v^{(4)}$ are respectively the atom velocities along portions 1, 2, 3 and 4 in Fig.~\ref{Mach-Zehnder_osc}, which can all be calculated explicitly using the first part of this appendix.

Then, the propagation phase shift between the two perturbed trajectories is simply 
\begin{widetext}
\begin{align}
\Phi_\mathrm{s} &= \Phi_\mathrm{sI}-\Phi_\mathrm{sII}=-\frac{4}{\omega}\left[k_\mathrm{eff}\left(v_{0}\hat e_v \cdot \hat e_\mathrm{kick}+\frac{\hbar k_\mathrm{eff}}{2m^0_A}\right)Q^A_M+\omega^0_AQ^A_\omega\right]\sin^2\left(\frac{\omega T}{2}\right)\sin(\omega T + \phi_0) +\, \nonumber \\
&4k_\mathrm{eff} T Q^L_\omega\left(v_0\hat e_v \cdot \hat e_\mathrm{kick} +\frac{\hbar k_\mathrm{eff}}{2m^0_A}\right)\sin\left(\frac{\omega T}{2}\right)\sin\left(\frac{\omega T}{2} + \phi_0\right) + \mathcal{O}\left(\left(\frac{v_0}{c}\right)^2\right) \, ,
\label{action_phase}
\end{align}
\end{widetext}
where we used $m^0_Av_\mathrm{kick}=\hbar k_\mathrm{eff}$ at zeroth order in the perturbation, following Eq.~\eqref{kick_modif} and where we defined $\vec v_0 = v_0 \hat e_v$ and $\vec v_\mathrm{kick} = v_\mathrm{kick}\hat e_\mathrm{kick}$. Note that the Bragg propagation phase is obtained by setting $\omega^0_A=0$. The terms $\propto (v^2_0/c^2)$ arise from the small contribution of the atoms velocity to the internal state kinetic energy. 

\subsection{Laser phase contribution}

We now consider the phase from light-matter interaction between the laser and the atoms.
In both Bragg and Raman schemes, we assume that the atoms are freely falling inside a "vacuum tower" where the laser are on the ground, located on a mount at initial position $x_G(0)=0$, while a retro-reflective mirror at initial position $x_M(0)=L$ is used to reflect the beams in order to create the counter-propagating scheme. In the Bragg case, only one laser beam is used and retro-reflected, while for the Raman case, two beams $L_1$, $L_2$ with respective frequencies $\omega_{L_1},\omega_{L_2}$ (such that $\omega_{L_1}-\omega_{L_2}=\omega^0_A$) are retro-reflected and the atomic wavepackets interact only with $L_1$ going up and $L_2$ going down. Considering that the whole tower on Earth is freely falling during the entire interferometric process, its own mass composition is affected by the oscillating mass behavior Eq.~\eqref{general_mass_freq_osc}, i.e $m_M = m^0_M\left(1+Q^M_M\cos(\omega t +\phi_0)\right)$, with $m^0_M, Q^M_M$ respectively its unperturbed mass and mass charge, implying it follows the same perturbed equations of motion as the atom Eq.\eqref{EoM}. Therefore, both retro-reflective mirror and mount on which laser rest upon are oscillating together.
At each space-time points of light-matter interaction is associated a phase, which contributes to the laser phase shift. These points are denoted $A, B, C, D_2$ in Fig.~\ref{Mach-Zehnder_osc}, since we assume the measurement of the interference pattern between the two wavepackets ending up on the internal ground state $|g\rangle$.

Keeping in mind the wavefunction expansion form $\Phi_L(t,\vec x) = \vec k_L \cdot \vec x - \omega_L t - \varphi$, with $\omega_L, \vec k_L$ the light angular frequency and wavevector respectively, and that $\vec k_L \cdot \vec x - \omega_L t = 0$ along a photon geodesic in a flat spacetime (which is the case at the surface of the Earth, see \cite{Badurina22}), the total light phase felt by the atom along a path $X$ is given by 
\begin{subequations}
\begin{align}
    \Phi_{\ell X} &=- \sum_{j=0}^n  \varphi_\mathrm{j}(t_i) \, \\
    &=- \sum_{j=0}^n s_j\left(\varphi_\mathrm{j}(t^\mathrm{down}_i) -\varphi_\mathrm{j}(t^\mathrm{up}_i)\right) \, ,
\end{align}
\end{subequations}
where we sum on the total number of points interaction of path X (occurring at times $\{0, T, 2T\}$), and where $t_i$ corresponds respectively to the time of laser at emission of the photon (the superscripts $up$ and $down$ refer to a photon coming from the up or down laser). At the second line, we took into account the fact that we have two-photon interactions, where the atom takes a photon from one of the laser and emit in the second one, and the $s_j = \pm 1$ parameter depends on the transition of the wavepacket at interaction j (it is $+1$ for a $|g\rangle \rightarrow |e\rangle $ transition and $-1$ for a $|g\rangle \rightarrow |e\rangle $ transition). Therefore, only the initial phase of the laser at the time of photon emission is needed. Since the laser frequency is locked on an atom ensemble whose frequency oscillates through Eq.~\eqref{general_mass_freq_osc}, the phase, as the integral of the frequency, is also time dependent, i.e 
\begin{align}\label{eq:init_phase_laser}
    \varphi_\mathrm{j}(t) &= \int^t_0 dt' \omega_L(t') \equiv \int^t_0 dt' \omega^0_L\left(1+Q^L_\omega \cos(\omega t' + \phi_0)\right) \, .
\end{align}
\begin{figure}
    \centering
    \includegraphics[width=0.3\textwidth]{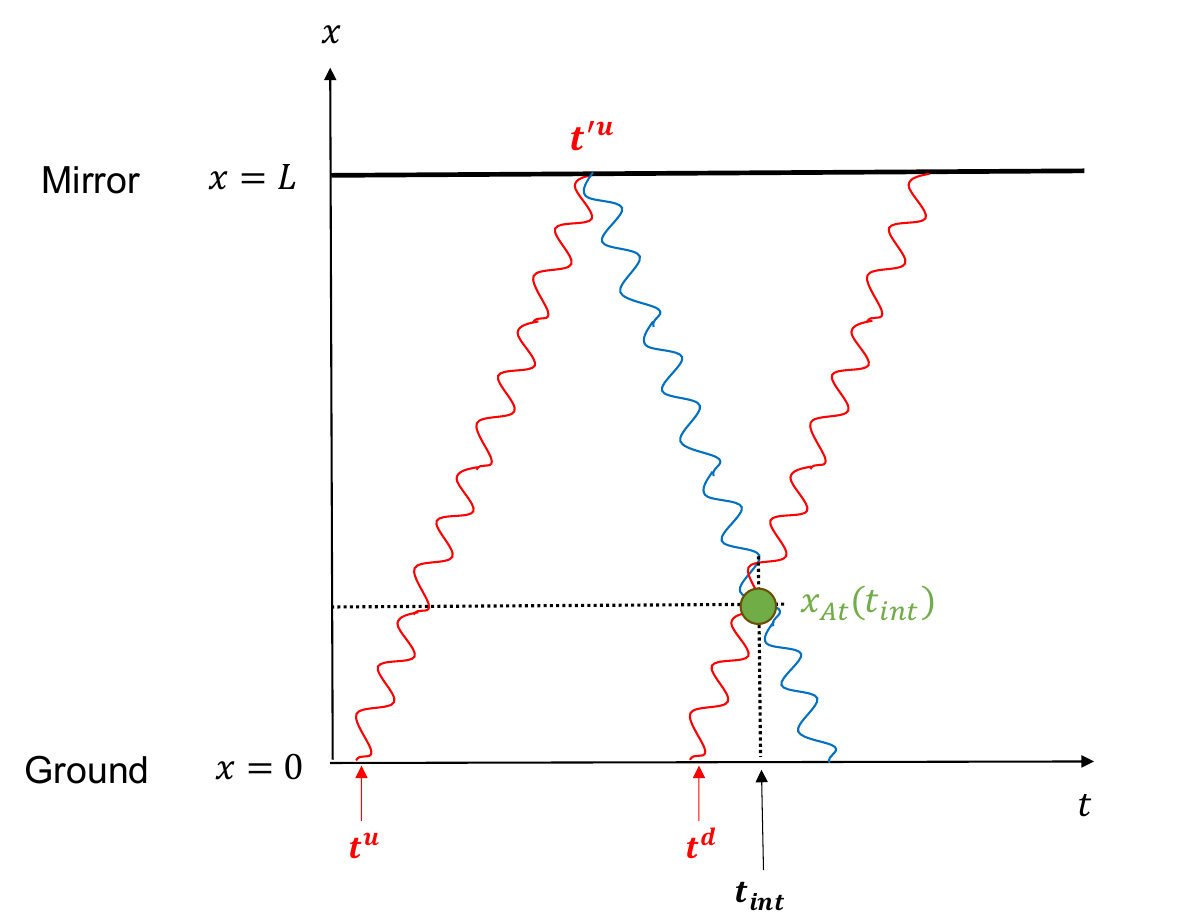}
    \caption{Simple scheme on the computation of the time of emission of photons from both laser.}
    \label{fig:laser_phase}
\end{figure}
To compute the laser phase, we need to know the different times $t_i$ of emission of photons. For a given interaction time $t_\mathrm{int}$, they are given by (see Fig.~\ref{fig:laser_phase}).
\begin{subequations}\label{eq:emission_times_laser}
    \begin{align}
        t^u &= t_\mathrm{int}-\frac{2x_M(t_\mathrm{int}-t^{'u})-x_\mathrm{At}(t_\mathrm{int})-x_G(t^{'u}-t^u)}{c} \, , \\
        t^d &= t_\mathrm{int} - \frac{x_\mathrm{At}(t_\mathrm{int})-x_G(t_\mathrm{int}-t^d)}{c} \, ,
    \end{align}
\end{subequations}
respectively for the beam retro-reflected (up) and not retro-reflected (down), where $x_\mathrm{M}$ is the mirror coordinate, $x_\mathrm{G}$ is the mount coordinate and $x_\mathrm{At}$ is the perturbed vertex of the atomic path. In order to solve Eq.~\eqref{eq:emission_times_laser}, the various $t^u,t^{'u},t^d$ on the right side are treated as unperturbed times.

Then, for Raman AI, the total laser phase is simply
\begin{widetext}\label{eq:laser_phase}
\begin{align}
   \Phi_\mathrm{\ell} &= \Phi_\mathrm{\ell I}-\Phi_\mathrm{\ell II} =\frac{4}{\omega}\left(k_\mathrm{eff}v_0\left(Q^A_M-Q^M_M\right)\hat e_v\cdot \hat e_\mathrm{kick}+\frac{\hbar k^2_\mathrm{eff}}{m^0_A}Q^A_M+\omega^0_A Q^L_\omega\right)\sin^2\left(\frac{\omega T}{2}\right)\sin(\omega T + \phi_0) +\, \nonumber \\
   &2Q^L_\omega \left(2L\left(k_\mathrm{eff} -\frac{\omega^0_A}{c}\right)\sin^2\left(\frac{\omega T}{2}\right)\cos(\omega T + \phi_0)-\frac{\hbar k^2_\mathrm{eff} T}{m^0_A}\sin\left(\frac{\omega T}{2}\right)\sin\left(\frac{\omega T}{2}+\phi_0\right)\right) \,  . \label{laser_phase_Raman} 
\end{align}
\end{widetext}
The Bragg laser phase is obtained by setting $\omega^0_A = 0$.

\subsection{Separation phase contribution}

Following Eq.~\eqref{overlap} and considering that, at the end of the classical paths at $t=T_d$, the wave packet of the path $i$=$\{$I,II$\}$ can be expressed as \cite{Wolf04}
\begin{align}
    \Psi_i = \Psi_0 e^{i\left(\Phi_{si}+\Phi_{\ell i}+\frac{p_i}{\hbar}(x-x_i)\right)} \, ,
\end{align}
where $\Phi_s,\Phi_\ell$ are respectively the propagation and laser phase contributions and where $p_i,x_i$ are respectively the momentum and position of the wavepacket just after the last $\pi/2$ pulse at $t=T_d$. Then, it can be shown easily that for small difference in momentum between the two wavepackets $\Delta \vec p = \vec p_\mathrm{I} - \vec p_\mathrm{II}$ (which we show in the following), the separation phase shift can be expressed as 
\begin{align}
\label{unclosed_path_gen}
    \Phi_u &= \Delta \vec p \cdot \vec x_\mathrm{det., COM}(2T)-\frac{\Delta (\vec p \cdot \vec x)}{\hbar} \, ,
\end{align}
where we compute the different of momenta $\vec p = \hbar \vec k$ and position $\vec x$ between $D_1$ and $D_2$ in Fig.~\ref{Mach-Zehnder_osc}, the two wavepackets in the same energy state, i.e $|g\rangle$ in our calculation. $\vec x_\mathrm{det., COM}$ represents the detector center of mass position at time $t=T_d=2T$, i.e in our case it is simply $\vec x_\mathrm{det., COM}(2T) = 2 \vec v_0 T$.

Since at the end of portion 3, the wavepacket is in the excited state, we must take into account an additional kick at time $t=2T$ of amplitude $v_\mathrm{kick,0}+\delta v_\mathrm{kick}(t=2T)$ and direction -$\hat x$ to this wavepacket, in order to put it back to the ground state, such that it state corresponds to the point $D_2$ in Fig.~\ref{Mach-Zehnder_osc}. Since, we assume the detection to be immediately after the kick, the position of the wavepacket will not be impacted by it. The final velocity of this wavepacket read
\begin{align} 
    \vec v^{(D_2)}_A(2T) &= \vec v_0\left((1 - Q^A_M\left(\cos(2\omega T+\phi_0)-\cos(\phi_0)\right)\right)- \, \nonumber\\
    &Q^L_\omega \vec v_\mathrm{kick,0}\left(\cos(2\omega T + \phi_0)-\cos(\omega T +\phi_0)\right) \, ,
    \label{v_D2}
\end{align}
resulting in a different velocity compared to the other wavepacket ($\vec v^{(4)}_A(2T)$). 
Then, the separation phase is a function of both differences of position and velocities at the end of the two paths 
\begin{subequations}
\begin{align}
    &\Delta \vec x(2T) = \vec x^{(4)}_A(2T)-\vec x^{(3)}_A(2T)\,\\
    &= 4\vec v_\mathrm{kick} \left(Q^L_\omega T-\frac{Q^A_M}{\omega}\sin\left(\frac{\omega T}{2}\right)\right)\sin\left(\frac{\omega T}{2}\right)\sin(\omega T +\phi_0) \,  , \\
    &\Delta \vec v(2T) = \vec v^{(4)}_A(2T)-\vec v^{(D_2)}_A(2T)\,\\
    &= -4\vec v_\mathrm{kick}Q^L_\omega\sin^2\left(\frac{\omega T}{2}\right)\cos(\omega T +\phi_0) \, ,
\end{align}
\end{subequations}
implying that the difference in velocities is small (of order $Q^L_\omega$), justifying the form of the separation phase Eq.~\eqref{unclosed_path_gen}.
Then, the separation phase shift is 
\begin{widetext}
\begin{align}
    \Phi_u &= \frac{m^0_A}{\hbar}\left(\Delta \vec v(2T)\cdot\vec x_\mathrm{det., COM}(2T) -\Delta \vec v(2T)\cdot \vec x(2T) -\vec v(2T) \cdot \Delta \vec x(2T)\right)\, \nonumber \\
    &=\frac{4k_\mathrm{eff}}{\omega}\left(v_0 \hat e_v \cdot \hat e_\mathrm{kick}\left[Q^A_M\sin\left(\frac{\omega T}{2}\right)\sin(\omega T + \phi_0)-\omega T Q^L_\omega\sin\left(\frac{\omega T}{2} + \phi_0\right)\right]+\right. \,\nonumber \\
    &\left.\frac{\hbar k_\mathrm{eff} \omega T }{m^0_A}Q^L_\omega\sin\left(\frac{\omega T}{2}\right)\cos(\omega T +\phi_0)\right)\sin\left(\frac{\omega T}{2}\right) \, ,
    \label{unclosed_path_shift}
\end{align}
\end{widetext}
where $\vec v(2T), \vec x(2T)$ respectively correspond to the unperturbed velocity and position of the wavepackets.

\subsection{Total phase shift}

Adding all contributions of phase shift Eqs.\eqref{unclosed_path_shift}, ~\eqref{laser_phase_Raman} and \eqref{action_phase}, one recovers Eq.~\eqref{phase_shift_exact_AI} for Raman and Bragg AI phases.

\section{\label{ap:acc_lab}A discussion regarding the various reference frames}

In the previous appendix and in the main text, the observable phase shift is computed in a reference frame where the perturbation to the rest mass/atomic frequency is proportional to $\cos \left(\omega t + \phi_0\right)$. In the case of a massive DM field, this frame of reference is the field rest frame, which is usually assumed to be a galactocentric reference frame. A simple Lorentz transformation can be used to show that the perturbation to the rest mass or atomic frequency is proportional to $\cos(\omega_\ell t_\ell - \vec k_\ell \cdot \vec x_\ell + \phi_0)$ in a lab-centric frame, moving at velocity $\vec v_\mathrm{lab}$ with respect to the galactocentric frame (with $v_\mathrm{lab} = v_\mathrm{DM} \sim 300$ km/s). The $\ell$ subscript denotes quantities expressed in the lab-centric frame. The reason why we performed the calculations in the galactocentric frame relies on the fact that the method presented in \cite{Storey}  is valid only for Lagrangians at most quadratic in the position and velocity, which is the case in the galactocentric case but not in the lab-centric one. For this reason, the quantities appearing in Eqs.~\eqref{phase_shift_exact_AI}, \eqref{eq:delta_phase_shift}, \eqref{eq:phase_gradio} and \eqref{eq:phase_new_prop} are quantities evaluated in the galactocentric frame. The first goal of this appendix is to justify why one can safely replace the values of $\omega$ , $k_\mathrm{eff}$ and $L$ appearing in these equations by their lab-centric counterpart. The result from Eq.~(\ref{phase_shift_exact_AI}) differs from the one obtained in \cite{Geraci16} by the fact that in \cite{Geraci16}, the velocity $v_0$ is the lab-centric initial velocity of the atoms while in the main text, it corresponds to its galactic counterpart. To strengthen our point, we will explicitly derive the classical equations of motion of the atoms in the lab-centric frame and show that the galactic velocity is indeed expected to appear in the phase shift. Finally, we will explain how a solution can be derived in the lab-centric frame (at first order in $v/c$) using the formalism from \cite{Storey} and show that this is also consistent with Eqs.~(\ref{phase_shift_exact_AI}).

First, the two reference frames are related to each other by a Lorentz transformation. This means that the angular frequency $\omega$ and wave vector $\vec k$ of the laser beam transform following
\begin{subequations}
    \begin{align}
        \omega_\ell &= \gamma \left(\omega - \vec v_\mathrm{lab} \cdot \vec k\right) = \omega \left(1 + \mathcal O\left(\frac{v_\mathrm{lab}}{c}\right) \right) \, , \\
        \vec k_\ell & = \vec k + \frac{1}{v_\mathrm{lab}^2}\left(\gamma -1 \right)\left(\vec v_\mathrm{lab} \cdot \vec k\right)\vec v_\mathrm{lab} - \frac{1}{c^2}\gamma\omega \vec v_\mathrm{lab} \nonumber \\
        &= \vec k\left(1  + \mathcal O\left(\frac{v_\mathrm{lab}}{c}\right)\right) \, ,
    \end{align}
\end{subequations}
where $\gamma=(1-v_\mathrm{lab}^2/c^2)^{-1/2}$ and $v_\mathrm{lab}/c = v_\mathrm{DM}/c\sim 10^{-3}$. For this reason, although Eqs.~(\ref{phase_shift_exact_AI}) are expressed in galactocentric reference frame, one can safely replace $\vec k_\mathrm{eff}$, $\omega_A^0$ by their lab-centric counterpart. This would lead to a correction three orders of magnitude smaller than the leading order term. A similar argument applies for the other quantities such as $L$, $T$ the interrogation time, etc. This demonstrates that one can safely use lab-centric quantities in Eqs.~(\ref{phase_shift_exact_AI}).

Let us now convince ourselves that a derivation directly performed in the lab-centric frame would also lead to a phase shift whose main term is also proportional to the lab velocity with respect to the galactic reference frame, in agreement with Eqs.~(\ref{phase_shift_exact_AI}). In a lab-centric frame, the Lagrangian becomes
\begin{align}
    L =  &-m^0_Ac^2\left(1-Q^A_M\left(\cos(\omega_\ell t_\ell -\vec k_\ell \cdot \vec x_\ell + \phi_0\right)\right)\nonumber \\
    &\qquad\qquad \times\left(1-\frac{v^2_{A,\ell}}{c^2}\right) \, ,\label{eq:langrangian_lab}
\end{align}
where $\vec v_{A,\ell}$ is the velocity of the atom in the lab frame (typically the launch velocity) and
\begin{equation}
    \vec k_\ell = -\frac{\omega_\ell \vec v_\mathrm{lab}}{c^2} \, ,
\end{equation}
is the de Broglie wavenumber of the massive DM scalar field in the lab frame. The classical equations of motion deriving from this Lagrangian are given by
\begin{equation}\label{eq:acc_uff}
    [\vec a_A]^\ell_{\overbar{\mathrm{UFF}}} = \left(\omega_\ell \vec v_{A,\ell} - \vec k_\ell c^2 \right)Q^A_M\sin\left(\omega_\ell t_\ell -\vec k_\ell \cdot \vec x_\ell + \phi_0\right) \, , 
\end{equation}
which can also be obtained by directly transforming Eq.~(\ref{a_vUFF}). In the lab frame, the second term in the acceleration is dominant since $v_\mathrm{lab} \gg v_{A,\ell}$. In addition, the AI schemes considered in Eqs.~(\ref{phase_shift_exact_AI}) act as accelerometers that provide, to first order, a phase shift $\Delta \Phi = k_\mathrm{eff} a T^2$ where $a$ is the local acceleration. For this reason, as an output to the perturbative acceleration form Eq.~(\ref{eq:acc_uff}), one expects that the leading order term for $\Delta \Phi$ is $\sim k_\mathrm{eff} k_\ell c^2 Q^A_M T^2= k_\mathrm{eff} \omega v_\mathrm{lab} Q^A_M T^2$, consistent for $\omega T\ll 1$ with the first term from Eqs.~(\ref{phase_shift_exact_AI}). This reasoning provides an argument showing that the velocity of the laboratory with respect to the galactocentric reference frame also appears when reasoning directly in the lab frame.
 
Formally, the method from \cite{Storey} cannot be used to compute the AI phase shift from the lab frame Lagrangian provided by Eq.~(\ref{eq:langrangian_lab}) since it is not at most quadratic in the position. Nevertheless, it is possible to perform an approximate calculation in the lab frame in the case where the de Broglie wavelength of the new field is much larger than the typical size of the experiment, i.e. if $\vec k_\ell\cdot \vec x_\ell \ll 1$ (which is the case for the experiments considered in this paper characterized by  $\vec k_\ell \cdot \vec x_\ell \lessapprox 5 \times 10^{-8} \ll 1$\footnote{We consider DM frequencies characterized by $mc^2 \leq 10^{-13}$ eV and size of experiments $L \leq 100$ m).}). In such a case, an expansion of the Lagrangian leads to 
\begin{align}
    L \approx& -m^0_Ac^2\left(1-\frac{v^2_{A,\ell}}{c^2}\right)\times \nonumber \\
    & \left(1-Q^A_M\left(\cos(\omega_\ell t_\ell + \phi_0)+\vec k_\ell \cdot \vec x_\ell \sin(\omega_\ell t_\ell+\phi_0)\right)\right) \, .
\end{align}
From this Lagrangian, one can derive the UFF violating acceleration in the laboratory frame, which reads
\begin{subequations}
\begin{align}\label{a_vUFF_lab}
    [\vec a_A]^\ell_{\overbar{\mathrm{UFF}}} &\approx \left(\omega_\ell \vec v_{A,\ell} - \vec k_\ell c^2 \right)Q^A_M\sin(\omega_\ell t_\ell + \phi_0) \, \\
    &= \omega_\ell \left(\vec v_{A,\ell} +\vec v_\mathrm{DM} \right)Q^A_M\sin(\omega_\ell t_\ell + \phi_0) \, .
\end{align}
\end{subequations}
This Lagrangian is now linear in the position and can be used to compute the phase shift using the method from \cite{Storey}. Note that there is an important difference compared to the calculation performed in \cite{Geraci16} where the second term has implicitly been neglected.

The equations of motion deriving from this Lagrangian consists in the ones from Eq.~(\ref{eq:acc_uff}) where one neglects the $\vec k_\ell\cdot \vec x_\ell$ within the sine function. The calculation of the phase shift directly in the lab frame follows exactly the equations presented in Appendix~\ref{ap:MZ_phase}. There are mainly two differences in the derived equations. First, the perturbed trajectory and velocity of the atom in the lab frame now read (taking $t_0 = 0$ immediately and $t_\ell \rightarrow t$)
\begin{subequations}
\begin{align}
&\vec v_{A,\ell}(t) \approx \vec v_{\ell,0} - \left(\vec v_{\ell,0}+\vec v_\mathrm{DM}\right)Q^A_M\left(\cos(\omega  t+\phi_0)-\cos(\phi_0)\right) \, , \\
&\vec x_{A,\ell}(t) \approx  \vec x_{\ell,0} + \vec v_{\ell,0} t -\frac{\vec v_{\ell,0}+\vec v_\mathrm{DM}}{\omega}Q^A_M\left(\sin(\omega t+\phi_0)-\right.\,\nonumber \\
&\left.\sin(\phi_0) - \omega t \cos(\phi_0)\right) \, ,
\end{align}
\end{subequations}
where $v_{\ell,0}$ is the initial velocity of the atom in the lab frame, which corresponds to the launch velocity. These equations replace Eqs.~(\ref{EoM}). Secondly, one needs to keep the $\vec k_\ell \cdot \vec x_\ell$ term in the Lagrangian when computing the propagation phase using Eq.~(\ref{eq:prop_phase_ap}). The full calculation following the method from Appendix~\ref{ap:MZ_phase} leads to a result that is consistent with Eq.~\eqref{phase_shift_exact_AI} to first order in $v_\mathrm{lab}/c$, in Bragg and Raman AI respectively, which demonstrates the equivalence between the two frames.

\section{\label{ap:axionic_mass}Axionic mass derivation}

In this appendix, we aim at deriving the expression of the axionic mass charge Eq.~\eqref{axionic_mass_charge}. The four contributions to the binding energy of the nuclei are the central force, the symmetry energy, the Coulomb force and the pairing energy. The dependence of the axion field to these four energies are as follows. 
\paragraph{\textbf{Central force}}\mbox{}
This interaction comes from the isospin symmetric central nuclear force, which is the dominant contribution in the binding of heavy nuclei \cite{damour:2010zr}. 
The dominant interactions are an attractive scalar $\eta_S$ and a repulsive vector $\eta_V$, and \cite{damour:2010zr} shows that the former is more sensitive to the pion mass, implying in our case
\begin{subequations}
\begin{align}
\frac{\partial E_\mathrm{Central}}{\partial (\theta^2)}\approx \frac{\partial E_\mathrm{Central}}{\partial \eta_S} \frac{\partial \eta_S}{\partial \ln m^2_\pi}\frac{\partial \ln m^2_\pi}{\partial (\theta^2)} \, ,
\end{align}
with \cite{Introductory_Nucl_Phys, damour:2010zr}
    \begin{align}
        E_\mathrm{Central} &\approx -(120A-97A^{2/3})\eta_S\, ,\\
        \frac{\partial \eta_S}{\partial \ln m^2_\pi} &= -0.35 \mathrm{ \ MeV} \, .
    \end{align}
\end{subequations}

\paragraph{\textbf{Asymmetry energy}}\mbox{}
The residual energy from the asymmetry between neutrons and protons inside the nucleus contains two components : 1) from the Pauli exclusion principle, requiring that when there are more neutrons than protons, the extra neutrons are in the higher energy states than the protons; and 2) from the nuclear force, which is more attractive for a neutron and a proton than with a pair of neutrons or a pair of protons. 
The asymmetry energy depends mainly on the scalar coupling strength between the nucleons $G_S$ \cite{damour:2010zr} implying 
\begin{subequations}
\begin{align}
\frac{\partial  E_{\mathrm{Asym}}}{\partial (\theta^2)} =\frac{\partial E_{\mathrm{Asym}}}{\partial G_S}\frac{\partial G_S}{\partial \ln m^2_\pi}\frac{\partial \ln m^2_\pi}{\partial (\theta^2)} \, ,
\end{align}
with \cite{damour:2010zr, Introductory_Nucl_Phys}
\begin{align}
    \frac{\partial E_{\mathrm{Asym}}}{\partial G_S}\frac{\partial G_S}{\partial \ln m^2_\pi} &= -19 \mathrm{ \ MeV} \, .
\end{align}
\end{subequations}

\paragraph{\textbf{Coulomb force}}\mbox{}
The Coulomb energy has a dependency on the strong interaction coupling terms since it depends on how tightly the nucleons are packed together. It can be shown that this contribution depends on the same scalar coupling as the asymmetry energy, hence
\begin{subequations}
\begin{align}
\frac{\partial E_{\mathrm{Coulomb}}}{\partial (\theta^2)} =\frac{\partial E_{\mathrm{Coulomb}}}{\partial G_S}\frac{\partial G_S}{\partial \ln m^2_\pi}\frac{\partial \ln m^2_\pi}{\partial (\theta^2)} \, ,
\end{align}
with \cite{damour:2010zr,Introductory_Nucl_Phys}
\begin{align}
    \frac{\partial E_{\mathrm{Coulomb}}}{\partial G_S}\frac{\partial G_S}{\partial \ln m^2_\pi} &=-0.13 \mathrm{ \ MeV} \, .
\end{align}
\end{subequations}

\paragraph{\textbf{Pairing energy}}\mbox{}
The pairing interaction contributes to binding energy and its numerical value is \cite{Introductory_Nucl_Phys}
\begin{equation}
    E_{\mathrm{Pairing}} = 12 \mathrm{ \ MeV} \, .
\end{equation}
\cite{damour:2010zr} shows that this contribution is subdominant for all atoms, due to its dependency to mass number A, compared to the other interactions, hence we will not consider the pairing energy in the calculation of the axionic charges of atoms.

Then, the axionic mass charge is given by

\begin{widetext}
\begin{align}
[Q^\mathrm{atom}_M]_a&=\frac{1}{m_\mathrm{atom}}\frac{\partial m_\mathrm{atom}}{\partial (\theta^2)}\,\nonumber\\
    &= \frac{1}{m_\mathrm{atom}}\left(\frac{\partial m_\mathrm{rest \ mass}}{\partial (\theta^2)}+\frac{\partial E_\mathrm{bind}}{\partial (\theta^2)}\right)\,\nonumber\\
&\approx -0.065 \frac{m_\mathrm{rest \ mass}}{m_\mathrm{atom}}+\frac{1 \mathrm{\: MeV}}{m_\mathrm{atom}}\left(-4.578 A +3.701 A^{2/3} + 2.071 \frac{(A-2Z)^2}{A} +0.014 \frac{Z(Z-1)}{A^{1/3}}\right)\,\nonumber\\
&\approx -0.065+F_A\Big(-4.92+\frac{3.98}{A^{1/3}}+2.22\frac{(A-2Z)^2}{A^2}+0.015\frac{Z(Z-1)}{A^{4/3}}\Big)\times 10^{-3}\, \nonumber \\
&\approx -0.070+F_A\Big(\frac{3.98}{A^{1/3}}+2.22\frac{(A-2Z)^2}{A^2}+0.015\frac{Z(Z-1)}{A^{4/3}}\Big)\times 10^{-3} \, .
\end{align}
\end{widetext}
At the last line, we considered $m_\mathrm{rest \ mass}/m_\mathrm{atom} \approx 1 + \langle E_\mathrm{bind}\rangle/m_\mathrm{amu} c^2$ with $\langle E_\mathrm{bind}\rangle \sim 8$ MeV, the average binding energy per nucleon and $m_\mathrm{amu}$ the atomic mass unit \cite{damour:2010zr}. Also, we factorized $F_A=Am_\mathrm{amu}/m_\mathrm{atom}$ which is of order unity at first order for all species of atoms (the relative error is $\mathcal{O}(10^{-3})$)\cite{damour:2010zr}.

\bibliographystyle{apsrev4-1}
\bibliography{AI}

\begin{thebibliography}{83}%
\makeatletter
\providecommand \@ifxundefined [1]{%
 \@ifx{#1\undefined}
}%
\providecommand \@ifnum [1]{%
 \ifnum #1\expandafter \@firstoftwo
 \else \expandafter \@secondoftwo
 \fi
}%
\providecommand \@ifx [1]{%
 \ifx #1\expandafter \@firstoftwo
 \else \expandafter \@secondoftwo
 \fi
}%
\providecommand \natexlab [1]{#1}%
\providecommand \enquote  [1]{``#1''}%
\providecommand \bibnamefont  [1]{#1}%
\providecommand \bibfnamefont [1]{#1}%
\providecommand \citenamefont [1]{#1}%
\providecommand \href@noop [0]{\@secondoftwo}%
\providecommand \href [0]{\begingroup \@sanitize@url \@href}%
\providecommand \@href[1]{\@@startlink{#1}\@@href}%
\providecommand \@@href[1]{\endgroup#1\@@endlink}%
\providecommand \@sanitize@url [0]{\catcode `\\12\catcode `\$12\catcode
  `\&12\catcode `\#12\catcode `\^12\catcode `\_12\catcode `\%12\relax}%
\providecommand \@@startlink[1]{}%
\providecommand \@@endlink[0]{}%
\providecommand \url  [0]{\begingroup\@sanitize@url \@url }%
\providecommand \@url [1]{\endgroup\@href {#1}{\urlprefix }}%
\providecommand \urlprefix  [0]{URL }%
\providecommand \Eprint [0]{\href }%
\providecommand \doibase [0]{http://dx.doi.org/}%
\providecommand \selectlanguage [0]{\@gobble}%
\providecommand \bibinfo  [0]{\@secondoftwo}%
\providecommand \bibfield  [0]{\@secondoftwo}%
\providecommand \translation [1]{[#1]}%
\providecommand \BibitemOpen [0]{}%
\providecommand \bibitemStop [0]{}%
\providecommand \bibitemNoStop [0]{.\EOS\space}%
\providecommand \EOS [0]{\spacefactor3000\relax}%
\providecommand \BibitemShut  [1]{\csname bibitem#1\endcsname}%
\let\auto@bib@innerbib\@empty
\bibitem [{\citenamefont {Weinberg}(1972)}]{Weinberg}%
  \BibitemOpen
  \bibfield  {author} {\bibinfo {author} {\bibfnamefont {S.}~\bibnamefont
  {Weinberg}},\ }\href@noop {} {\emph {\bibinfo {title} {Gravitation and
  Cosmology: Principles and Applications of the General Theory of
  Relativity}}}\ (\bibinfo  {publisher} {Wiley},\ \bibinfo {address} {New
  York,},\ \bibinfo {year} {1972})\BibitemShut {NoStop}%
\bibitem [{\citenamefont {Wagner}\ \emph {et~al.}(2012)\citenamefont {Wagner},
  \citenamefont {Schlamminger}, \citenamefont {Gundlach},\ and\ \citenamefont
  {Adelberger}}]{Wagner12}%
  \BibitemOpen
  \bibfield  {author} {\bibinfo {author} {\bibfnamefont {T.~A.}\ \bibnamefont
  {Wagner}}, \bibinfo {author} {\bibfnamefont {S.}~\bibnamefont
  {Schlamminger}}, \bibinfo {author} {\bibfnamefont {J.~H.}\ \bibnamefont
  {Gundlach}}, \ and\ \bibinfo {author} {\bibfnamefont {E.~G.}\ \bibnamefont
  {Adelberger}},\ }\href {\doibase 10.1088/0264-9381/29/18/184002} {\bibfield
  {journal} {\bibinfo  {journal} {Classical and Quantum Gravity}\ }\textbf
  {\bibinfo {volume} {29}},\ \bibinfo {pages} {184002} (\bibinfo {year}
  {2012})}\BibitemShut {NoStop}%
\bibitem [{\citenamefont {Touboul}\ \emph {et~al.}(2022)\citenamefont
  {Touboul}, \citenamefont {M\'etris}, \citenamefont {Rodrigues}, \citenamefont
  {Berg\'e}, \citenamefont {Robert}, \citenamefont {Baghi}, \citenamefont
  {Andr\'e}, \citenamefont {Bedouet}, \citenamefont {Boulanger}, \citenamefont
  {Bremer}, \citenamefont {Carle}, \citenamefont {Chhun}, \citenamefont
  {Christophe}, \citenamefont {Cipolla}, \citenamefont {Damour}, \citenamefont
  {Danto}, \citenamefont {Demange}, \citenamefont {Dittus}, \citenamefont
  {Dhuicque}, \citenamefont {Fayet}, \citenamefont {Foulon}, \citenamefont
  {Guidotti}, \citenamefont {Hagedorn}, \citenamefont {Hardy}, \citenamefont
  {Huynh}, \citenamefont {Kayser}, \citenamefont {Lala}, \citenamefont
  {L\"ammerzahl}, \citenamefont {Lebat}, \citenamefont {Liorzou}, \citenamefont
  {List}, \citenamefont {L\"offler}, \citenamefont {Panet}, \citenamefont
  {Pernot-Borr\`as}, \citenamefont {Perraud}, \citenamefont {Pires},
  \citenamefont {Pouilloux}, \citenamefont {Prieur}, \citenamefont {Rebray},
  \citenamefont {Reynaud}, \citenamefont {Rievers}, \citenamefont {Selig},
  \citenamefont {Serron}, \citenamefont {Sumner}, \citenamefont {Tanguy},
  \citenamefont {Torresi},\ and\ \citenamefont {Visser}}]{Microscope22}%
  \BibitemOpen
  \bibfield  {author} {\bibinfo {author} {\bibfnamefont {P.}~\bibnamefont
  {Touboul}}, \bibinfo {author} {\bibfnamefont {G.}~\bibnamefont {M\'etris}},
  \bibinfo {author} {\bibfnamefont {M.}~\bibnamefont {Rodrigues}}, \bibinfo
  {author} {\bibfnamefont {J.}~\bibnamefont {Berg\'e}}, \bibinfo {author}
  {\bibfnamefont {A.}~\bibnamefont {Robert}}, \bibinfo {author} {\bibfnamefont
  {Q.}~\bibnamefont {Baghi}}, \bibinfo {author} {\bibfnamefont
  {Y.}~\bibnamefont {Andr\'e}}, \bibinfo {author} {\bibfnamefont
  {J.}~\bibnamefont {Bedouet}}, \bibinfo {author} {\bibfnamefont
  {D.}~\bibnamefont {Boulanger}}, \bibinfo {author} {\bibfnamefont
  {S.}~\bibnamefont {Bremer}}, \bibinfo {author} {\bibfnamefont
  {P.}~\bibnamefont {Carle}}, \bibinfo {author} {\bibfnamefont
  {R.}~\bibnamefont {Chhun}}, \bibinfo {author} {\bibfnamefont
  {B.}~\bibnamefont {Christophe}}, \bibinfo {author} {\bibfnamefont
  {V.}~\bibnamefont {Cipolla}}, \bibinfo {author} {\bibfnamefont
  {T.}~\bibnamefont {Damour}}, \bibinfo {author} {\bibfnamefont
  {P.}~\bibnamefont {Danto}}, \bibinfo {author} {\bibfnamefont
  {L.}~\bibnamefont {Demange}}, \bibinfo {author} {\bibfnamefont
  {H.}~\bibnamefont {Dittus}}, \bibinfo {author} {\bibfnamefont
  {O.}~\bibnamefont {Dhuicque}}, \bibinfo {author} {\bibfnamefont
  {P.}~\bibnamefont {Fayet}}, \bibinfo {author} {\bibfnamefont
  {B.}~\bibnamefont {Foulon}}, \bibinfo {author} {\bibfnamefont {P.-Y.}\
  \bibnamefont {Guidotti}}, \bibinfo {author} {\bibfnamefont {D.}~\bibnamefont
  {Hagedorn}}, \bibinfo {author} {\bibfnamefont {E.}~\bibnamefont {Hardy}},
  \bibinfo {author} {\bibfnamefont {P.-A.}\ \bibnamefont {Huynh}}, \bibinfo
  {author} {\bibfnamefont {P.}~\bibnamefont {Kayser}}, \bibinfo {author}
  {\bibfnamefont {S.}~\bibnamefont {Lala}}, \bibinfo {author} {\bibfnamefont
  {C.}~\bibnamefont {L\"ammerzahl}}, \bibinfo {author} {\bibfnamefont
  {V.}~\bibnamefont {Lebat}}, \bibinfo {author} {\bibfnamefont
  {F.}~\bibnamefont {Liorzou}}, \bibinfo {author} {\bibfnamefont
  {M.}~\bibnamefont {List}}, \bibinfo {author} {\bibfnamefont {F.}~\bibnamefont
  {L\"offler}}, \bibinfo {author} {\bibfnamefont {I.}~\bibnamefont {Panet}},
  \bibinfo {author} {\bibfnamefont {M.}~\bibnamefont {Pernot-Borr\`as}},
  \bibinfo {author} {\bibfnamefont {L.}~\bibnamefont {Perraud}}, \bibinfo
  {author} {\bibfnamefont {S.}~\bibnamefont {Pires}}, \bibinfo {author}
  {\bibfnamefont {B.}~\bibnamefont {Pouilloux}}, \bibinfo {author}
  {\bibfnamefont {P.}~\bibnamefont {Prieur}}, \bibinfo {author} {\bibfnamefont
  {A.}~\bibnamefont {Rebray}}, \bibinfo {author} {\bibfnamefont
  {S.}~\bibnamefont {Reynaud}}, \bibinfo {author} {\bibfnamefont
  {B.}~\bibnamefont {Rievers}}, \bibinfo {author} {\bibfnamefont
  {H.}~\bibnamefont {Selig}}, \bibinfo {author} {\bibfnamefont
  {L.}~\bibnamefont {Serron}}, \bibinfo {author} {\bibfnamefont
  {T.}~\bibnamefont {Sumner}}, \bibinfo {author} {\bibfnamefont
  {N.}~\bibnamefont {Tanguy}}, \bibinfo {author} {\bibfnamefont
  {P.}~\bibnamefont {Torresi}}, \ and\ \bibinfo {author} {\bibfnamefont
  {P.}~\bibnamefont {Visser}} (\bibinfo {collaboration} {MICROSCOPE
  Collaboration}),\ }\href {\doibase 10.1103/PhysRevLett.129.121102} {\bibfield
   {journal} {\bibinfo  {journal} {Phys. Rev. Lett.}\ }\textbf {\bibinfo
  {volume} {129}},\ \bibinfo {pages} {121102} (\bibinfo {year}
  {2022})}\BibitemShut {NoStop}%
\bibitem [{\citenamefont {Damour}\ and\ \citenamefont
  {Polyakov}(1994)}]{damour94}%
  \BibitemOpen
  \bibfield  {author} {\bibinfo {author} {\bibfnamefont {T.}~\bibnamefont
  {Damour}}\ and\ \bibinfo {author} {\bibfnamefont {A.}~\bibnamefont
  {Polyakov}},\ }\href {\doibase https://doi.org/10.1016/0550-3213(94)90143-0}
  {\bibfield  {journal} {\bibinfo  {journal} {Nuclear Physics B}\ }\textbf
  {\bibinfo {volume} {423}},\ \bibinfo {pages} {532} (\bibinfo {year}
  {1994})}\BibitemShut {NoStop}%
\bibitem [{\citenamefont {Fayet}(2018)}]{Fayet18}%
  \BibitemOpen
  \bibfield  {author} {\bibinfo {author} {\bibfnamefont {P.}~\bibnamefont
  {Fayet}},\ }\href {\doibase 10.1103/PhysRevD.97.055039} {\bibfield  {journal}
  {\bibinfo  {journal} {Phys. Rev. D}\ }\textbf {\bibinfo {volume} {97}},\
  \bibinfo {pages} {055039} (\bibinfo {year} {2018})}\BibitemShut {NoStop}%
\bibitem [{\citenamefont {Fayet}(2019)}]{Fayet2019}%
  \BibitemOpen
  \bibfield  {author} {\bibinfo {author} {\bibfnamefont {P.}~\bibnamefont
  {Fayet}},\ }\href {\doibase 10.1103/PhysRevD.99.055043} {\bibfield  {journal}
  {\bibinfo  {journal} {Phys. Rev. D}\ }\textbf {\bibinfo {volume} {99}},\
  \bibinfo {pages} {055043} (\bibinfo {year} {2019})}\BibitemShut {NoStop}%
\bibitem [{\citenamefont {Will}(2014)}]{Will14}%
  \BibitemOpen
  \bibfield  {author} {\bibinfo {author} {\bibfnamefont {C.~M.}\ \bibnamefont
  {Will}},\ }\href {\doibase 10.12942/lrr-2014-4} {\bibfield  {journal}
  {\bibinfo  {journal} {Living Reviews in Relativity}\ }\textbf {\bibinfo
  {volume} {17}} (\bibinfo {year} {2014}),\ 10.12942/lrr-2014-4}\BibitemShut
  {NoStop}%
\bibitem [{\citenamefont {Uzan}(2011)}]{Uzan11}%
  \BibitemOpen
  \bibfield  {author} {\bibinfo {author} {\bibfnamefont {J.-P.}\ \bibnamefont
  {Uzan}},\ }\href {\doibase 10.12942/lrr-2011-2} {\bibfield  {journal}
  {\bibinfo  {journal} {Living Reviews in Relativity}\ }\textbf {\bibinfo
  {volume} {14}},\ \bibinfo {pages} {2} (\bibinfo {year} {2011})}\BibitemShut
  {NoStop}%
\bibitem [{\citenamefont {Damour}\ and\ \citenamefont
  {Donoghue}(2010)}]{damour:2010zr}%
  \BibitemOpen
  \bibfield  {author} {\bibinfo {author} {\bibfnamefont {T.}~\bibnamefont
  {Damour}}\ and\ \bibinfo {author} {\bibfnamefont {J.~F.}\ \bibnamefont
  {Donoghue}},\ }\href {\doibase 10.1103/PhysRevD.82.084033} {\bibfield
  {journal} {\bibinfo  {journal} {Phys. Rev. D}\ }\textbf {\bibinfo {volume}
  {82}},\ \bibinfo {pages} {084033} (\bibinfo {year} {2010})}\BibitemShut
  {NoStop}%
\bibitem [{\citenamefont {{Damour}}\ and\ \citenamefont
  {{Donoghue}}(2010)}]{damour:2010ve}%
  \BibitemOpen
  \bibfield  {author} {\bibinfo {author} {\bibfnamefont {T.}~\bibnamefont
  {{Damour}}}\ and\ \bibinfo {author} {\bibfnamefont {J.~F.}\ \bibnamefont
  {{Donoghue}}},\ }\href {\doibase 10.1088/0264-9381/27/20/202001} {\bibfield
  {journal} {\bibinfo  {journal} {Classical and Quantum Gravity}\ }\textbf
  {\bibinfo {volume} {27}},\ \bibinfo {pages} {202001} (\bibinfo {year}
  {2010})},\ \Eprint {http://arxiv.org/abs/1007.2790} {arXiv:1007.2790 [gr-qc]}
  \BibitemShut {NoStop}%
\bibitem [{\citenamefont {Kim}\ and\ \citenamefont {Perez}(2024)}]{Kim22}%
  \BibitemOpen
  \bibfield  {author} {\bibinfo {author} {\bibfnamefont {H.}~\bibnamefont
  {Kim}}\ and\ \bibinfo {author} {\bibfnamefont {G.}~\bibnamefont {Perez}},\
  }\href {\doibase 10.1103/PhysRevD.109.015005} {\bibfield  {journal} {\bibinfo
   {journal} {Phys. Rev. D}\ }\textbf {\bibinfo {volume} {109}},\ \bibinfo
  {pages} {015005} (\bibinfo {year} {2024})}\BibitemShut {NoStop}%
\bibitem [{\citenamefont {Marsh}(2016)}]{Marsh16}%
  \BibitemOpen
  \bibfield  {author} {\bibinfo {author} {\bibfnamefont {D.~J.}\ \bibnamefont
  {Marsh}},\ }\href {\doibase 10.1016/j.physrep.2016.06.005} {\bibfield
  {journal} {\bibinfo  {journal} {Physics Reports}\ }\textbf {\bibinfo {volume}
  {643}},\ \bibinfo {pages} {1–79} (\bibinfo {year} {2016})}\BibitemShut
  {NoStop}%
\bibitem [{\citenamefont {Graham}\ \emph {et~al.}(2016)\citenamefont {Graham},
  \citenamefont {Kaplan}, \citenamefont {Mardon}, \citenamefont {Rajendran},\
  and\ \citenamefont {Terrano}}]{Graham16}%
  \BibitemOpen
  \bibfield  {author} {\bibinfo {author} {\bibfnamefont {P.~W.}\ \bibnamefont
  {Graham}}, \bibinfo {author} {\bibfnamefont {D.~E.}\ \bibnamefont {Kaplan}},
  \bibinfo {author} {\bibfnamefont {J.}~\bibnamefont {Mardon}}, \bibinfo
  {author} {\bibfnamefont {S.}~\bibnamefont {Rajendran}}, \ and\ \bibinfo
  {author} {\bibfnamefont {W.~A.}\ \bibnamefont {Terrano}},\ }\href {\doibase
  10.1103/PhysRevD.93.075029} {\bibfield  {journal} {\bibinfo  {journal} {Phys.
  Rev. D}\ }\textbf {\bibinfo {volume} {93}},\ \bibinfo {pages} {075029}
  (\bibinfo {year} {2016})}\BibitemShut {NoStop}%
\bibitem [{\citenamefont {Geraci}\ and\ \citenamefont
  {Derevianko}(2016)}]{Geraci16}%
  \BibitemOpen
  \bibfield  {author} {\bibinfo {author} {\bibfnamefont {A.~A.}\ \bibnamefont
  {Geraci}}\ and\ \bibinfo {author} {\bibfnamefont {A.}~\bibnamefont
  {Derevianko}},\ }\href {\doibase 10.1103/PhysRevLett.117.261301} {\bibfield
  {journal} {\bibinfo  {journal} {Phys. Rev. Lett.}\ }\textbf {\bibinfo
  {volume} {117}},\ \bibinfo {pages} {261301} (\bibinfo {year}
  {2016})}\BibitemShut {NoStop}%
\bibitem [{\citenamefont {Graham}\ \emph {et~al.}(2013)\citenamefont {Graham},
  \citenamefont {Hogan}, \citenamefont {Kasevich},\ and\ \citenamefont
  {Rajendran}}]{Graham13}%
  \BibitemOpen
  \bibfield  {author} {\bibinfo {author} {\bibfnamefont {P.~W.}\ \bibnamefont
  {Graham}}, \bibinfo {author} {\bibfnamefont {J.~M.}\ \bibnamefont {Hogan}},
  \bibinfo {author} {\bibfnamefont {M.~A.}\ \bibnamefont {Kasevich}}, \ and\
  \bibinfo {author} {\bibfnamefont {S.}~\bibnamefont {Rajendran}},\ }\href
  {\doibase 10.1103/PhysRevLett.110.171102} {\bibfield  {journal} {\bibinfo
  {journal} {Phys. Rev. Lett.}\ }\textbf {\bibinfo {volume} {110}},\ \bibinfo
  {pages} {171102} (\bibinfo {year} {2013})}\BibitemShut {NoStop}%
\bibitem [{\citenamefont {Badurina}\ \emph {et~al.}(2020)\citenamefont
  {Badurina}, \citenamefont {Bentine}, \citenamefont {Blas}, \citenamefont
  {Bongs}, \citenamefont {Bortoletto}, \citenamefont {Bowcock}, \citenamefont
  {Bridges}, \citenamefont {Bowden}, \citenamefont {Buchmueller}, \citenamefont
  {Burrage}, \citenamefont {Coleman}, \citenamefont {Elertas}, \citenamefont
  {Ellis}, \citenamefont {Foot}, \citenamefont {Gibson}, \citenamefont
  {Haehnelt}, \citenamefont {Harte}, \citenamefont {Hedges}, \citenamefont
  {Hobson}, \citenamefont {Holynski}, \citenamefont {Jones}, \citenamefont
  {Langlois}, \citenamefont {Lellouch}, \citenamefont {Lewicki}, \citenamefont
  {Maiolino}, \citenamefont {Majewski}, \citenamefont {Malik}, \citenamefont
  {March-Russell}, \citenamefont {McCabe}, \citenamefont {Newbold},
  \citenamefont {Sauer}, \citenamefont {Schneider}, \citenamefont {Shipsey},
  \citenamefont {Singh}, \citenamefont {Uchida}, \citenamefont {Valenzuela},
  \citenamefont {van~der Grinten}, \citenamefont {Vaskonen}, \citenamefont
  {Vossebeld}, \citenamefont {Weatherill},\ and\ \citenamefont
  {Wilmut}}]{Badurina20}%
  \BibitemOpen
  \bibfield  {author} {\bibinfo {author} {\bibfnamefont {L.}~\bibnamefont
  {Badurina}}, \bibinfo {author} {\bibfnamefont {E.}~\bibnamefont {Bentine}},
  \bibinfo {author} {\bibfnamefont {D.}~\bibnamefont {Blas}}, \bibinfo {author}
  {\bibfnamefont {K.}~\bibnamefont {Bongs}}, \bibinfo {author} {\bibfnamefont
  {D.}~\bibnamefont {Bortoletto}}, \bibinfo {author} {\bibfnamefont
  {T.}~\bibnamefont {Bowcock}}, \bibinfo {author} {\bibfnamefont
  {K.}~\bibnamefont {Bridges}}, \bibinfo {author} {\bibfnamefont
  {W.}~\bibnamefont {Bowden}}, \bibinfo {author} {\bibfnamefont
  {O.}~\bibnamefont {Buchmueller}}, \bibinfo {author} {\bibfnamefont
  {C.}~\bibnamefont {Burrage}}, \bibinfo {author} {\bibfnamefont
  {J.}~\bibnamefont {Coleman}}, \bibinfo {author} {\bibfnamefont
  {G.}~\bibnamefont {Elertas}}, \bibinfo {author} {\bibfnamefont
  {J.}~\bibnamefont {Ellis}}, \bibinfo {author} {\bibfnamefont
  {C.}~\bibnamefont {Foot}}, \bibinfo {author} {\bibfnamefont {V.}~\bibnamefont
  {Gibson}}, \bibinfo {author} {\bibfnamefont {M.}~\bibnamefont {Haehnelt}},
  \bibinfo {author} {\bibfnamefont {T.}~\bibnamefont {Harte}}, \bibinfo
  {author} {\bibfnamefont {S.}~\bibnamefont {Hedges}}, \bibinfo {author}
  {\bibfnamefont {R.}~\bibnamefont {Hobson}}, \bibinfo {author} {\bibfnamefont
  {M.}~\bibnamefont {Holynski}}, \bibinfo {author} {\bibfnamefont
  {T.}~\bibnamefont {Jones}}, \bibinfo {author} {\bibfnamefont
  {M.}~\bibnamefont {Langlois}}, \bibinfo {author} {\bibfnamefont
  {S.}~\bibnamefont {Lellouch}}, \bibinfo {author} {\bibfnamefont
  {M.}~\bibnamefont {Lewicki}}, \bibinfo {author} {\bibfnamefont
  {R.}~\bibnamefont {Maiolino}}, \bibinfo {author} {\bibfnamefont
  {P.}~\bibnamefont {Majewski}}, \bibinfo {author} {\bibfnamefont
  {S.}~\bibnamefont {Malik}}, \bibinfo {author} {\bibfnamefont
  {J.}~\bibnamefont {March-Russell}}, \bibinfo {author} {\bibfnamefont
  {C.}~\bibnamefont {McCabe}}, \bibinfo {author} {\bibfnamefont
  {D.}~\bibnamefont {Newbold}}, \bibinfo {author} {\bibfnamefont
  {B.}~\bibnamefont {Sauer}}, \bibinfo {author} {\bibfnamefont
  {U.}~\bibnamefont {Schneider}}, \bibinfo {author} {\bibfnamefont
  {I.}~\bibnamefont {Shipsey}}, \bibinfo {author} {\bibfnamefont
  {Y.}~\bibnamefont {Singh}}, \bibinfo {author} {\bibfnamefont
  {M.}~\bibnamefont {Uchida}}, \bibinfo {author} {\bibfnamefont
  {T.}~\bibnamefont {Valenzuela}}, \bibinfo {author} {\bibfnamefont
  {M.}~\bibnamefont {van~der Grinten}}, \bibinfo {author} {\bibfnamefont
  {V.}~\bibnamefont {Vaskonen}}, \bibinfo {author} {\bibfnamefont
  {J.}~\bibnamefont {Vossebeld}}, \bibinfo {author} {\bibfnamefont
  {D.}~\bibnamefont {Weatherill}}, \ and\ \bibinfo {author} {\bibfnamefont
  {I.}~\bibnamefont {Wilmut}},\ }\href {\doibase 10.1088/1475-7516/2020/05/011}
  {\bibfield  {journal} {\bibinfo  {journal} {Journal of Cosmology and
  Astroparticle Physics}\ }\textbf {\bibinfo {volume} {2020}},\ \bibinfo
  {pages} {011} (\bibinfo {year} {2020})}\BibitemShut {NoStop}%
\bibitem [{\citenamefont {Badurina}\ \emph {et~al.}(2022)\citenamefont
  {Badurina}, \citenamefont {Blas},\ and\ \citenamefont {McCabe}}]{Badurina22}%
  \BibitemOpen
  \bibfield  {author} {\bibinfo {author} {\bibfnamefont {L.}~\bibnamefont
  {Badurina}}, \bibinfo {author} {\bibfnamefont {D.}~\bibnamefont {Blas}}, \
  and\ \bibinfo {author} {\bibfnamefont {C.}~\bibnamefont {McCabe}},\ }\href
  {\doibase 10.1103/PhysRevD.105.023006} {\bibfield  {journal} {\bibinfo
  {journal} {Phys. Rev. D}\ }\textbf {\bibinfo {volume} {105}},\ \bibinfo
  {pages} {023006} (\bibinfo {year} {2022})}\BibitemShut {NoStop}%
\bibitem [{\citenamefont {Arvanitaki}\ \emph {et~al.}(2018)\citenamefont
  {Arvanitaki}, \citenamefont {Graham}, \citenamefont {Hogan}, \citenamefont
  {Rajendran},\ and\ \citenamefont {Van~Tilburg}}]{Arvanitaki18}%
  \BibitemOpen
  \bibfield  {author} {\bibinfo {author} {\bibfnamefont {A.}~\bibnamefont
  {Arvanitaki}}, \bibinfo {author} {\bibfnamefont {P.~W.}\ \bibnamefont
  {Graham}}, \bibinfo {author} {\bibfnamefont {J.~M.}\ \bibnamefont {Hogan}},
  \bibinfo {author} {\bibfnamefont {S.}~\bibnamefont {Rajendran}}, \ and\
  \bibinfo {author} {\bibfnamefont {K.}~\bibnamefont {Van~Tilburg}},\ }\href
  {\doibase 10.1103/PhysRevD.97.075020} {\bibfield  {journal} {\bibinfo
  {journal} {Phys. Rev. D}\ }\textbf {\bibinfo {volume} {97}},\ \bibinfo
  {pages} {075020} (\bibinfo {year} {2018})}\BibitemShut {NoStop}%
\bibitem [{\citenamefont {Abe}\ \emph {et~al.}(2021)\citenamefont {Abe},
  \citenamefont {Adamson}, \citenamefont {Borcean}, \citenamefont {Bortoletto},
  \citenamefont {Bridges}, \citenamefont {Carman}, \citenamefont
  {Chattopadhyay}, \citenamefont {Coleman}, \citenamefont {Curfman},
  \citenamefont {DeRose}, \citenamefont {Deshpande}, \citenamefont
  {Dimopoulos}, \citenamefont {Foot}, \citenamefont {Frisch}, \citenamefont
  {Garber}, \citenamefont {Geer}, \citenamefont {Gibson}, \citenamefont
  {Glick}, \citenamefont {Graham}, \citenamefont {Hahn}, \citenamefont
  {Harnik}, \citenamefont {Hawkins}, \citenamefont {Hindley}, \citenamefont
  {Hogan}, \citenamefont {Jiang}, \citenamefont {Kasevich}, \citenamefont
  {Kellett}, \citenamefont {Kiburg}, \citenamefont {Kovachy}, \citenamefont
  {Lykken}, \citenamefont {March-Russell}, \citenamefont {Mitchell},
  \citenamefont {Murphy}, \citenamefont {Nantel}, \citenamefont {Nobrega},
  \citenamefont {Plunkett}, \citenamefont {Rajendran}, \citenamefont {Rudolph},
  \citenamefont {Sachdeva}, \citenamefont {Safdari}, \citenamefont {Santucci},
  \citenamefont {Schwartzman}, \citenamefont {Shipsey}, \citenamefont {Swan},
  \citenamefont {Valerio}, \citenamefont {Vasonis}, \citenamefont {Wang},\ and\
  \citenamefont {Wilkason}}]{Abe21}%
  \BibitemOpen
  \bibfield  {author} {\bibinfo {author} {\bibfnamefont {M.}~\bibnamefont
  {Abe}}, \bibinfo {author} {\bibfnamefont {P.}~\bibnamefont {Adamson}},
  \bibinfo {author} {\bibfnamefont {M.}~\bibnamefont {Borcean}}, \bibinfo
  {author} {\bibfnamefont {D.}~\bibnamefont {Bortoletto}}, \bibinfo {author}
  {\bibfnamefont {K.}~\bibnamefont {Bridges}}, \bibinfo {author} {\bibfnamefont
  {S.~P.}\ \bibnamefont {Carman}}, \bibinfo {author} {\bibfnamefont
  {S.}~\bibnamefont {Chattopadhyay}}, \bibinfo {author} {\bibfnamefont
  {J.}~\bibnamefont {Coleman}}, \bibinfo {author} {\bibfnamefont {N.~M.}\
  \bibnamefont {Curfman}}, \bibinfo {author} {\bibfnamefont {K.}~\bibnamefont
  {DeRose}}, \bibinfo {author} {\bibfnamefont {T.}~\bibnamefont {Deshpande}},
  \bibinfo {author} {\bibfnamefont {S.}~\bibnamefont {Dimopoulos}}, \bibinfo
  {author} {\bibfnamefont {C.~J.}\ \bibnamefont {Foot}}, \bibinfo {author}
  {\bibfnamefont {J.~C.}\ \bibnamefont {Frisch}}, \bibinfo {author}
  {\bibfnamefont {B.~E.}\ \bibnamefont {Garber}}, \bibinfo {author}
  {\bibfnamefont {S.}~\bibnamefont {Geer}}, \bibinfo {author} {\bibfnamefont
  {V.}~\bibnamefont {Gibson}}, \bibinfo {author} {\bibfnamefont
  {J.}~\bibnamefont {Glick}}, \bibinfo {author} {\bibfnamefont {P.~W.}\
  \bibnamefont {Graham}}, \bibinfo {author} {\bibfnamefont {S.~R.}\
  \bibnamefont {Hahn}}, \bibinfo {author} {\bibfnamefont {R.}~\bibnamefont
  {Harnik}}, \bibinfo {author} {\bibfnamefont {L.}~\bibnamefont {Hawkins}},
  \bibinfo {author} {\bibfnamefont {S.}~\bibnamefont {Hindley}}, \bibinfo
  {author} {\bibfnamefont {J.~M.}\ \bibnamefont {Hogan}}, \bibinfo {author}
  {\bibfnamefont {Y.}~\bibnamefont {Jiang}}, \bibinfo {author} {\bibfnamefont
  {M.~A.}\ \bibnamefont {Kasevich}}, \bibinfo {author} {\bibfnamefont {R.~J.}\
  \bibnamefont {Kellett}}, \bibinfo {author} {\bibfnamefont {M.}~\bibnamefont
  {Kiburg}}, \bibinfo {author} {\bibfnamefont {T.}~\bibnamefont {Kovachy}},
  \bibinfo {author} {\bibfnamefont {J.~D.}\ \bibnamefont {Lykken}}, \bibinfo
  {author} {\bibfnamefont {J.}~\bibnamefont {March-Russell}}, \bibinfo {author}
  {\bibfnamefont {J.}~\bibnamefont {Mitchell}}, \bibinfo {author}
  {\bibfnamefont {M.}~\bibnamefont {Murphy}}, \bibinfo {author} {\bibfnamefont
  {M.}~\bibnamefont {Nantel}}, \bibinfo {author} {\bibfnamefont {L.~E.}\
  \bibnamefont {Nobrega}}, \bibinfo {author} {\bibfnamefont {R.~K.}\
  \bibnamefont {Plunkett}}, \bibinfo {author} {\bibfnamefont {S.}~\bibnamefont
  {Rajendran}}, \bibinfo {author} {\bibfnamefont {J.}~\bibnamefont {Rudolph}},
  \bibinfo {author} {\bibfnamefont {N.}~\bibnamefont {Sachdeva}}, \bibinfo
  {author} {\bibfnamefont {M.}~\bibnamefont {Safdari}}, \bibinfo {author}
  {\bibfnamefont {J.~K.}\ \bibnamefont {Santucci}}, \bibinfo {author}
  {\bibfnamefont {A.~G.}\ \bibnamefont {Schwartzman}}, \bibinfo {author}
  {\bibfnamefont {I.}~\bibnamefont {Shipsey}}, \bibinfo {author} {\bibfnamefont
  {H.}~\bibnamefont {Swan}}, \bibinfo {author} {\bibfnamefont {L.~R.}\
  \bibnamefont {Valerio}}, \bibinfo {author} {\bibfnamefont {A.}~\bibnamefont
  {Vasonis}}, \bibinfo {author} {\bibfnamefont {Y.}~\bibnamefont {Wang}}, \
  and\ \bibinfo {author} {\bibfnamefont {T.}~\bibnamefont {Wilkason}},\ }\href
  {\doibase 10.1088/2058-9565/abf719} {\bibfield  {journal} {\bibinfo
  {journal} {Quantum Science and Technology}\ }\textbf {\bibinfo {volume}
  {6}},\ \bibinfo {pages} {044003} (\bibinfo {year} {2021})}\BibitemShut
  {NoStop}%
\bibitem [{\citenamefont {Touboul}\ \emph {et~al.}(2017)\citenamefont
  {Touboul}, \citenamefont {M\'etris}, \citenamefont {Rodrigues}, \citenamefont
  {Andr\'e}, \citenamefont {Baghi}, \citenamefont {Berg\'e}, \citenamefont
  {Boulanger}, \citenamefont {Bremer}, \citenamefont {Carle}, \citenamefont
  {Chhun}, \citenamefont {Christophe}, \citenamefont {Cipolla}, \citenamefont
  {Damour}, \citenamefont {Danto}, \citenamefont {Dittus}, \citenamefont
  {Fayet}, \citenamefont {Foulon}, \citenamefont {Gageant}, \citenamefont
  {Guidotti}, \citenamefont {Hagedorn}, \citenamefont {Hardy}, \citenamefont
  {Huynh}, \citenamefont {Inchauspe}, \citenamefont {Kayser}, \citenamefont
  {Lala}, \citenamefont {L\"ammerzahl}, \citenamefont {Lebat}, \citenamefont
  {Leseur}, \citenamefont {Liorzou}, \citenamefont {List}, \citenamefont
  {L\"offler}, \citenamefont {Panet}, \citenamefont {Pouilloux}, \citenamefont
  {Prieur}, \citenamefont {Rebray}, \citenamefont {Reynaud}, \citenamefont
  {Rievers}, \citenamefont {Robert}, \citenamefont {Selig}, \citenamefont
  {Serron}, \citenamefont {Sumner}, \citenamefont {Tanguy},\ and\ \citenamefont
  {Visser}}]{Microscope17}%
  \BibitemOpen
  \bibfield  {author} {\bibinfo {author} {\bibfnamefont {P.}~\bibnamefont
  {Touboul}}, \bibinfo {author} {\bibfnamefont {G.}~\bibnamefont {M\'etris}},
  \bibinfo {author} {\bibfnamefont {M.}~\bibnamefont {Rodrigues}}, \bibinfo
  {author} {\bibfnamefont {Y.}~\bibnamefont {Andr\'e}}, \bibinfo {author}
  {\bibfnamefont {Q.}~\bibnamefont {Baghi}}, \bibinfo {author} {\bibfnamefont
  {J.}~\bibnamefont {Berg\'e}}, \bibinfo {author} {\bibfnamefont
  {D.}~\bibnamefont {Boulanger}}, \bibinfo {author} {\bibfnamefont
  {S.}~\bibnamefont {Bremer}}, \bibinfo {author} {\bibfnamefont
  {P.}~\bibnamefont {Carle}}, \bibinfo {author} {\bibfnamefont
  {R.}~\bibnamefont {Chhun}}, \bibinfo {author} {\bibfnamefont
  {B.}~\bibnamefont {Christophe}}, \bibinfo {author} {\bibfnamefont
  {V.}~\bibnamefont {Cipolla}}, \bibinfo {author} {\bibfnamefont
  {T.}~\bibnamefont {Damour}}, \bibinfo {author} {\bibfnamefont
  {P.}~\bibnamefont {Danto}}, \bibinfo {author} {\bibfnamefont
  {H.}~\bibnamefont {Dittus}}, \bibinfo {author} {\bibfnamefont
  {P.}~\bibnamefont {Fayet}}, \bibinfo {author} {\bibfnamefont
  {B.}~\bibnamefont {Foulon}}, \bibinfo {author} {\bibfnamefont
  {C.}~\bibnamefont {Gageant}}, \bibinfo {author} {\bibfnamefont {P.-Y.}\
  \bibnamefont {Guidotti}}, \bibinfo {author} {\bibfnamefont {D.}~\bibnamefont
  {Hagedorn}}, \bibinfo {author} {\bibfnamefont {E.}~\bibnamefont {Hardy}},
  \bibinfo {author} {\bibfnamefont {P.-A.}\ \bibnamefont {Huynh}}, \bibinfo
  {author} {\bibfnamefont {H.}~\bibnamefont {Inchauspe}}, \bibinfo {author}
  {\bibfnamefont {P.}~\bibnamefont {Kayser}}, \bibinfo {author} {\bibfnamefont
  {S.}~\bibnamefont {Lala}}, \bibinfo {author} {\bibfnamefont {C.}~\bibnamefont
  {L\"ammerzahl}}, \bibinfo {author} {\bibfnamefont {V.}~\bibnamefont {Lebat}},
  \bibinfo {author} {\bibfnamefont {P.}~\bibnamefont {Leseur}}, \bibinfo
  {author} {\bibfnamefont {F.}~\bibnamefont {Liorzou}}, \bibinfo {author}
  {\bibfnamefont {M.}~\bibnamefont {List}}, \bibinfo {author} {\bibfnamefont
  {F.}~\bibnamefont {L\"offler}}, \bibinfo {author} {\bibfnamefont
  {I.}~\bibnamefont {Panet}}, \bibinfo {author} {\bibfnamefont
  {B.}~\bibnamefont {Pouilloux}}, \bibinfo {author} {\bibfnamefont
  {P.}~\bibnamefont {Prieur}}, \bibinfo {author} {\bibfnamefont
  {A.}~\bibnamefont {Rebray}}, \bibinfo {author} {\bibfnamefont
  {S.}~\bibnamefont {Reynaud}}, \bibinfo {author} {\bibfnamefont
  {B.}~\bibnamefont {Rievers}}, \bibinfo {author} {\bibfnamefont
  {A.}~\bibnamefont {Robert}}, \bibinfo {author} {\bibfnamefont
  {H.}~\bibnamefont {Selig}}, \bibinfo {author} {\bibfnamefont
  {L.}~\bibnamefont {Serron}}, \bibinfo {author} {\bibfnamefont
  {T.}~\bibnamefont {Sumner}}, \bibinfo {author} {\bibfnamefont
  {N.}~\bibnamefont {Tanguy}}, \ and\ \bibinfo {author} {\bibfnamefont
  {P.}~\bibnamefont {Visser}},\ }\href {\doibase
  10.1103/PhysRevLett.119.231101} {\bibfield  {journal} {\bibinfo  {journal}
  {Phys. Rev. Lett.}\ }\textbf {\bibinfo {volume} {119}},\ \bibinfo {pages}
  {231101} (\bibinfo {year} {2017})}\BibitemShut {NoStop}%
\bibitem [{\citenamefont {WILLIAMS}\ \emph {et~al.}(2009)\citenamefont
  {WILLIAMS}, \citenamefont {TURYSHEV},\ and\ \citenamefont
  {BOGGS}}]{Williams09}%
  \BibitemOpen
  \bibfield  {author} {\bibinfo {author} {\bibfnamefont {J.~G.}\ \bibnamefont
  {WILLIAMS}}, \bibinfo {author} {\bibfnamefont {S.~G.}\ \bibnamefont
  {TURYSHEV}}, \ and\ \bibinfo {author} {\bibfnamefont {D.~H.}\ \bibnamefont
  {BOGGS}},\ }\href {\doibase 10.1142/s021827180901500x} {\bibfield  {journal}
  {\bibinfo  {journal} {International Journal of Modern Physics D}\ }\textbf
  {\bibinfo {volume} {18}},\ \bibinfo {pages} {1129–1175} (\bibinfo {year}
  {2009})}\BibitemShut {NoStop}%
\bibitem [{\citenamefont {Storey}\ and\ \citenamefont
  {Cohen-Tannoudji}(1994)}]{Storey}%
  \BibitemOpen
  \bibfield  {author} {\bibinfo {author} {\bibfnamefont {P.}~\bibnamefont
  {Storey}}\ and\ \bibinfo {author} {\bibfnamefont {C.}~\bibnamefont
  {Cohen-Tannoudji}},\ }\href {\doibase 10.1051/jp2:1994103} {\bibfield
  {journal} {\bibinfo  {journal} {{Journal de Physique II}}\ }\textbf {\bibinfo
  {volume} {4}},\ \bibinfo {pages} {1999} (\bibinfo {year} {1994})}\BibitemShut
  {NoStop}%
\bibitem [{\citenamefont {Wolf}\ and\ \citenamefont {Tourrenc}(1999)}]{Wolf}%
  \BibitemOpen
  \bibfield  {author} {\bibinfo {author} {\bibfnamefont {P.}~\bibnamefont
  {Wolf}}\ and\ \bibinfo {author} {\bibfnamefont {P.}~\bibnamefont
  {Tourrenc}},\ }\href {\doibase https://doi.org/10.1016/S0375-9601(98)00881-0}
  {\bibfield  {journal} {\bibinfo  {journal} {Physics Letters A}\ }\textbf
  {\bibinfo {volume} {251}},\ \bibinfo {pages} {241} (\bibinfo {year}
  {1999})}\BibitemShut {NoStop}%
\bibitem [{\citenamefont {Wolf}\ and\ \citenamefont {Borde}(2004)}]{Wolf04}%
  \BibitemOpen
  \bibfield  {author} {\bibinfo {author} {\bibfnamefont {P.}~\bibnamefont
  {Wolf}}\ and\ \bibinfo {author} {\bibfnamefont {C.~J.}\ \bibnamefont
  {Borde}},\ }\href@noop {} {\enquote {\bibinfo {title} {Recoil effects in
  microwave ramsey spectroscopy},}\ } (\bibinfo {year} {2004}),\ \Eprint
  {http://arxiv.org/abs/quant-ph/0403194} {arXiv:quant-ph/0403194 [quant-ph]}
  \BibitemShut {NoStop}%
\bibitem [{\citenamefont {Giese}\ \emph {et~al.}(2013)\citenamefont {Giese},
  \citenamefont {Roura}, \citenamefont {Tackmann}, \citenamefont {Rasel},\ and\
  \citenamefont {Schleich}}]{Giese13}%
  \BibitemOpen
  \bibfield  {author} {\bibinfo {author} {\bibfnamefont {E.}~\bibnamefont
  {Giese}}, \bibinfo {author} {\bibfnamefont {A.}~\bibnamefont {Roura}},
  \bibinfo {author} {\bibfnamefont {G.}~\bibnamefont {Tackmann}}, \bibinfo
  {author} {\bibfnamefont {E.~M.}\ \bibnamefont {Rasel}}, \ and\ \bibinfo
  {author} {\bibfnamefont {W.~P.}\ \bibnamefont {Schleich}},\ }\href {\doibase
  10.1103/PhysRevA.88.053608} {\bibfield  {journal} {\bibinfo  {journal} {Phys.
  Rev. A}\ }\textbf {\bibinfo {volume} {88}},\ \bibinfo {pages} {053608}
  (\bibinfo {year} {2013})}\BibitemShut {NoStop}%
\bibitem [{\citenamefont {Ahlers}\ \emph {et~al.}(2022)\citenamefont {Ahlers},
  \citenamefont {Badurina}, \citenamefont {Bassi}, \citenamefont {Battelier},
  \citenamefont {Beaufils}, \citenamefont {Bongs}, \citenamefont {Bouyer},
  \citenamefont {Braxmaier}, \citenamefont {Buchmueller}, \citenamefont
  {Carlesso}, \citenamefont {Charron}, \citenamefont {Chiofalo}, \citenamefont
  {Corgier}, \citenamefont {Donadi}, \citenamefont {Droz}, \citenamefont
  {Ecoffet}, \citenamefont {Ellis}, \citenamefont {Estève}, \citenamefont
  {Gaaloul}, \citenamefont {Gerardi}, \citenamefont {Giese}, \citenamefont
  {Grosse}, \citenamefont {Hees}, \citenamefont {Hensel}, \citenamefont {Herr},
  \citenamefont {Jetzer}, \citenamefont {Kleinsteinberg}, \citenamefont
  {Klempt}, \citenamefont {Lecomte}, \citenamefont {Lopes}, \citenamefont
  {Loriani}, \citenamefont {Métris}, \citenamefont {Martin}, \citenamefont
  {Martín}, \citenamefont {Müller}, \citenamefont {Nofrarias}, \citenamefont
  {Santos}, \citenamefont {Rasel}, \citenamefont {Robert}, \citenamefont
  {Saks}, \citenamefont {Salter}, \citenamefont {Schlippert}, \citenamefont
  {Schubert}, \citenamefont {Schuldt}, \citenamefont {Sopuerta}, \citenamefont
  {Struckmann}, \citenamefont {Tino}, \citenamefont {Valenzuela}, \citenamefont
  {von Klitzing}, \citenamefont {Wörner}, \citenamefont {Wolf}, \citenamefont
  {Yu},\ and\ \citenamefont {Zelan}}]{STE-QUEST}%
  \BibitemOpen
  \bibfield  {author} {\bibinfo {author} {\bibfnamefont {H.}~\bibnamefont
  {Ahlers}}, \bibinfo {author} {\bibfnamefont {L.}~\bibnamefont {Badurina}},
  \bibinfo {author} {\bibfnamefont {A.}~\bibnamefont {Bassi}}, \bibinfo
  {author} {\bibfnamefont {B.}~\bibnamefont {Battelier}}, \bibinfo {author}
  {\bibfnamefont {Q.}~\bibnamefont {Beaufils}}, \bibinfo {author}
  {\bibfnamefont {K.}~\bibnamefont {Bongs}}, \bibinfo {author} {\bibfnamefont
  {P.}~\bibnamefont {Bouyer}}, \bibinfo {author} {\bibfnamefont
  {C.}~\bibnamefont {Braxmaier}}, \bibinfo {author} {\bibfnamefont
  {O.}~\bibnamefont {Buchmueller}}, \bibinfo {author} {\bibfnamefont
  {M.}~\bibnamefont {Carlesso}}, \bibinfo {author} {\bibfnamefont
  {E.}~\bibnamefont {Charron}}, \bibinfo {author} {\bibfnamefont {M.~L.}\
  \bibnamefont {Chiofalo}}, \bibinfo {author} {\bibfnamefont {R.}~\bibnamefont
  {Corgier}}, \bibinfo {author} {\bibfnamefont {S.}~\bibnamefont {Donadi}},
  \bibinfo {author} {\bibfnamefont {F.}~\bibnamefont {Droz}}, \bibinfo {author}
  {\bibfnamefont {R.}~\bibnamefont {Ecoffet}}, \bibinfo {author} {\bibfnamefont
  {J.}~\bibnamefont {Ellis}}, \bibinfo {author} {\bibfnamefont
  {F.}~\bibnamefont {Estève}}, \bibinfo {author} {\bibfnamefont
  {N.}~\bibnamefont {Gaaloul}}, \bibinfo {author} {\bibfnamefont
  {D.}~\bibnamefont {Gerardi}}, \bibinfo {author} {\bibfnamefont
  {E.}~\bibnamefont {Giese}}, \bibinfo {author} {\bibfnamefont
  {J.}~\bibnamefont {Grosse}}, \bibinfo {author} {\bibfnamefont
  {A.}~\bibnamefont {Hees}}, \bibinfo {author} {\bibfnamefont {T.}~\bibnamefont
  {Hensel}}, \bibinfo {author} {\bibfnamefont {W.}~\bibnamefont {Herr}},
  \bibinfo {author} {\bibfnamefont {P.}~\bibnamefont {Jetzer}}, \bibinfo
  {author} {\bibfnamefont {G.}~\bibnamefont {Kleinsteinberg}}, \bibinfo
  {author} {\bibfnamefont {C.}~\bibnamefont {Klempt}}, \bibinfo {author}
  {\bibfnamefont {S.}~\bibnamefont {Lecomte}}, \bibinfo {author} {\bibfnamefont
  {L.}~\bibnamefont {Lopes}}, \bibinfo {author} {\bibfnamefont
  {S.}~\bibnamefont {Loriani}}, \bibinfo {author} {\bibfnamefont
  {G.}~\bibnamefont {Métris}}, \bibinfo {author} {\bibfnamefont
  {T.}~\bibnamefont {Martin}}, \bibinfo {author} {\bibfnamefont
  {V.}~\bibnamefont {Martín}}, \bibinfo {author} {\bibfnamefont
  {G.}~\bibnamefont {Müller}}, \bibinfo {author} {\bibfnamefont
  {M.}~\bibnamefont {Nofrarias}}, \bibinfo {author} {\bibfnamefont {F.~P.~D.}\
  \bibnamefont {Santos}}, \bibinfo {author} {\bibfnamefont {E.~M.}\
  \bibnamefont {Rasel}}, \bibinfo {author} {\bibfnamefont {A.}~\bibnamefont
  {Robert}}, \bibinfo {author} {\bibfnamefont {N.}~\bibnamefont {Saks}},
  \bibinfo {author} {\bibfnamefont {M.}~\bibnamefont {Salter}}, \bibinfo
  {author} {\bibfnamefont {D.}~\bibnamefont {Schlippert}}, \bibinfo {author}
  {\bibfnamefont {C.}~\bibnamefont {Schubert}}, \bibinfo {author}
  {\bibfnamefont {T.}~\bibnamefont {Schuldt}}, \bibinfo {author} {\bibfnamefont
  {C.~F.}\ \bibnamefont {Sopuerta}}, \bibinfo {author} {\bibfnamefont
  {C.}~\bibnamefont {Struckmann}}, \bibinfo {author} {\bibfnamefont {G.~M.}\
  \bibnamefont {Tino}}, \bibinfo {author} {\bibfnamefont {T.}~\bibnamefont
  {Valenzuela}}, \bibinfo {author} {\bibfnamefont {W.}~\bibnamefont {von
  Klitzing}}, \bibinfo {author} {\bibfnamefont {L.}~\bibnamefont {Wörner}},
  \bibinfo {author} {\bibfnamefont {P.}~\bibnamefont {Wolf}}, \bibinfo {author}
  {\bibfnamefont {N.}~\bibnamefont {Yu}}, \ and\ \bibinfo {author}
  {\bibfnamefont {M.}~\bibnamefont {Zelan}},\ }\href {\doibase
  10.48550/ARXIV.2211.15412} {\enquote {\bibinfo {title} {Ste-quest: Space time
  explorer and quantum equivalence principle space test},}\ } (\bibinfo {year}
  {2022})\BibitemShut {NoStop}%
\bibitem [{\citenamefont {Gauguet}\ \emph {et~al.}(2008)\citenamefont
  {Gauguet}, \citenamefont {Mehlst\"aubler}, \citenamefont {L\'ev\`eque},
  \citenamefont {Le~Gou\"et}, \citenamefont {Chaibi}, \citenamefont {Canuel},
  \citenamefont {Clairon}, \citenamefont {Dos~Santos},\ and\ \citenamefont
  {Landragin}}]{Gauguet08}%
  \BibitemOpen
  \bibfield  {author} {\bibinfo {author} {\bibfnamefont {A.}~\bibnamefont
  {Gauguet}}, \bibinfo {author} {\bibfnamefont {T.~E.}\ \bibnamefont
  {Mehlst\"aubler}}, \bibinfo {author} {\bibfnamefont {T.}~\bibnamefont
  {L\'ev\`eque}}, \bibinfo {author} {\bibfnamefont {J.}~\bibnamefont
  {Le~Gou\"et}}, \bibinfo {author} {\bibfnamefont {W.}~\bibnamefont {Chaibi}},
  \bibinfo {author} {\bibfnamefont {B.}~\bibnamefont {Canuel}}, \bibinfo
  {author} {\bibfnamefont {A.}~\bibnamefont {Clairon}}, \bibinfo {author}
  {\bibfnamefont {F.~P.}\ \bibnamefont {Dos~Santos}}, \ and\ \bibinfo {author}
  {\bibfnamefont {A.}~\bibnamefont {Landragin}},\ }\href {\doibase
  10.1103/PhysRevA.78.043615} {\bibfield  {journal} {\bibinfo  {journal} {Phys.
  Rev. A}\ }\textbf {\bibinfo {volume} {78}},\ \bibinfo {pages} {043615}
  (\bibinfo {year} {2008})}\BibitemShut {NoStop}%
\bibitem [{\citenamefont {Gillot}\ \emph {et~al.}(2016)\citenamefont {Gillot},
  \citenamefont {Cheng}, \citenamefont {Merlet},\ and\ \citenamefont {Pereira
  Dos~Santos}}]{Gillot16}%
  \BibitemOpen
  \bibfield  {author} {\bibinfo {author} {\bibfnamefont {P.}~\bibnamefont
  {Gillot}}, \bibinfo {author} {\bibfnamefont {B.}~\bibnamefont {Cheng}},
  \bibinfo {author} {\bibfnamefont {S.}~\bibnamefont {Merlet}}, \ and\ \bibinfo
  {author} {\bibfnamefont {F.}~\bibnamefont {Pereira Dos~Santos}},\ }\href
  {\doibase 10.1103/PhysRevA.93.013609} {\bibfield  {journal} {\bibinfo
  {journal} {Phys. Rev. A}\ }\textbf {\bibinfo {volume} {93}},\ \bibinfo
  {pages} {013609} (\bibinfo {year} {2016})}\BibitemShut {NoStop}%
\bibitem [{\citenamefont {Dimopoulos}\ \emph {et~al.}(2007)\citenamefont
  {Dimopoulos}, \citenamefont {Graham}, \citenamefont {Hogan},\ and\
  \citenamefont {Kasevich}}]{Dimopoulos07}%
  \BibitemOpen
  \bibfield  {author} {\bibinfo {author} {\bibfnamefont {S.}~\bibnamefont
  {Dimopoulos}}, \bibinfo {author} {\bibfnamefont {P.~W.}\ \bibnamefont
  {Graham}}, \bibinfo {author} {\bibfnamefont {J.~M.}\ \bibnamefont {Hogan}}, \
  and\ \bibinfo {author} {\bibfnamefont {M.~A.}\ \bibnamefont {Kasevich}},\
  }\href {\doibase 10.1103/PhysRevLett.98.111102} {\bibfield  {journal}
  {\bibinfo  {journal} {Phys. Rev. Lett.}\ }\textbf {\bibinfo {volume} {98}},\
  \bibinfo {pages} {111102} (\bibinfo {year} {2007})}\BibitemShut {NoStop}%
\bibitem [{\citenamefont {Takano}\ \emph {et~al.}(2017)\citenamefont {Takano},
  \citenamefont {Mizushima},\ and\ \citenamefont {Katori}}]{Takano17}%
  \BibitemOpen
  \bibfield  {author} {\bibinfo {author} {\bibfnamefont {T.}~\bibnamefont
  {Takano}}, \bibinfo {author} {\bibfnamefont {R.}~\bibnamefont {Mizushima}}, \
  and\ \bibinfo {author} {\bibfnamefont {H.}~\bibnamefont {Katori}},\ }\href
  {\doibase 10.7567/APEX.10.072801} {\bibfield  {journal} {\bibinfo  {journal}
  {Applied Physics Express}\ }\textbf {\bibinfo {volume} {10}},\ \bibinfo
  {pages} {072801} (\bibinfo {year} {2017})}\BibitemShut {NoStop}%
\bibitem [{\citenamefont {Peccei}\ and\ \citenamefont
  {Quinn}(1977)}]{Peccei_Quinn}%
  \BibitemOpen
  \bibfield  {author} {\bibinfo {author} {\bibfnamefont {R.~D.}\ \bibnamefont
  {Peccei}}\ and\ \bibinfo {author} {\bibfnamefont {H.~R.}\ \bibnamefont
  {Quinn}},\ }\href {\doibase 10.1103/PhysRevLett.38.1440} {\bibfield
  {journal} {\bibinfo  {journal} {Phys. Rev. Lett.}\ }\textbf {\bibinfo
  {volume} {38}},\ \bibinfo {pages} {1440} (\bibinfo {year}
  {1977})}\BibitemShut {NoStop}%
\bibitem [{\citenamefont {Flambaum}\ and\ \citenamefont
  {Samsonov}(2023)}]{Flambaum23}%
  \BibitemOpen
  \bibfield  {author} {\bibinfo {author} {\bibfnamefont {V.~V.}\ \bibnamefont
  {Flambaum}}\ and\ \bibinfo {author} {\bibfnamefont {I.~B.}\ \bibnamefont
  {Samsonov}},\ }\href {\doibase 10.1103/PhysRevD.108.075022} {\bibfield
  {journal} {\bibinfo  {journal} {Phys. Rev. D}\ }\textbf {\bibinfo {volume}
  {108}},\ \bibinfo {pages} {075022} (\bibinfo {year} {2023})}\BibitemShut
  {NoStop}%
\bibitem [{\citenamefont {Flambaum}\ and\ \citenamefont
  {Tedesco}(2006)}]{Flambaum06}%
  \BibitemOpen
  \bibfield  {author} {\bibinfo {author} {\bibfnamefont {V.~V.}\ \bibnamefont
  {Flambaum}}\ and\ \bibinfo {author} {\bibfnamefont {A.~F.}\ \bibnamefont
  {Tedesco}},\ }\href {\doibase 10.1103/PhysRevC.73.055501} {\bibfield
  {journal} {\bibinfo  {journal} {Phys. Rev. C}\ }\textbf {\bibinfo {volume}
  {73}},\ \bibinfo {pages} {055501} (\bibinfo {year} {2006})}\BibitemShut
  {NoStop}%
\bibitem [{\citenamefont {Beadle}\ \emph {et~al.}(2023)\citenamefont {Beadle},
  \citenamefont {Ellis}, \citenamefont {Quevillon},\ and\ \citenamefont
  {Vuong}}]{Beadle23}%
  \BibitemOpen
  \bibfield  {author} {\bibinfo {author} {\bibfnamefont {C.}~\bibnamefont
  {Beadle}}, \bibinfo {author} {\bibfnamefont {S.~A.~R.}\ \bibnamefont
  {Ellis}}, \bibinfo {author} {\bibfnamefont {J.}~\bibnamefont {Quevillon}}, \
  and\ \bibinfo {author} {\bibfnamefont {P.~N.~H.}\ \bibnamefont {Vuong}},\
  }\href@noop {} {\enquote {\bibinfo {title} {Quadratic coupling of the axion
  to photons},}\ } (\bibinfo {year} {2023}),\ \Eprint
  {http://arxiv.org/abs/2307.10362} {arXiv:2307.10362 [hep-ph]} \BibitemShut
  {NoStop}%
\bibitem [{\citenamefont {Kim}\ \emph {et~al.}(2024)\citenamefont {Kim},
  \citenamefont {Lenoci}, \citenamefont {Perez},\ and\ \citenamefont
  {Ratzinger}}]{Kim24}%
  \BibitemOpen
  \bibfield  {author} {\bibinfo {author} {\bibfnamefont {H.}~\bibnamefont
  {Kim}}, \bibinfo {author} {\bibfnamefont {A.}~\bibnamefont {Lenoci}},
  \bibinfo {author} {\bibfnamefont {G.}~\bibnamefont {Perez}}, \ and\ \bibinfo
  {author} {\bibfnamefont {W.}~\bibnamefont {Ratzinger}},\ }\href {\doibase
  10.1103/PhysRevD.109.015030} {\bibfield  {journal} {\bibinfo  {journal}
  {Phys. Rev. D}\ }\textbf {\bibinfo {volume} {109}},\ \bibinfo {pages}
  {015030} (\bibinfo {year} {2024})}\BibitemShut {NoStop}%
\bibitem [{\citenamefont {Zhao}\ \emph {et~al.}(2024)\citenamefont {Zhao},
  \citenamefont {Liu},\ and\ \citenamefont {Mei}}]{Zhao24}%
  \BibitemOpen
  \bibfield  {author} {\bibinfo {author} {\bibfnamefont {W.}~\bibnamefont
  {Zhao}}, \bibinfo {author} {\bibfnamefont {H.}~\bibnamefont {Liu}}, \ and\
  \bibinfo {author} {\bibfnamefont {X.}~\bibnamefont {Mei}},\ }\href@noop {}
  {\enquote {\bibinfo {title} {Ultralight scalar and axion dark matter
  detection with atom interferometers},}\ } (\bibinfo {year} {2024}),\ \Eprint
  {http://arxiv.org/abs/2401.17055} {arXiv:2401.17055 [hep-ph]} \BibitemShut
  {NoStop}%
\bibitem [{\citenamefont {Schelfhout}\ and\ \citenamefont
  {McFerran}(2021)}]{Schelfhout21}%
  \BibitemOpen
  \bibfield  {author} {\bibinfo {author} {\bibfnamefont {J.~S.}\ \bibnamefont
  {Schelfhout}}\ and\ \bibinfo {author} {\bibfnamefont {J.~J.}\ \bibnamefont
  {McFerran}},\ }\href {\doibase 10.1103/PhysRevA.104.022806} {\bibfield
  {journal} {\bibinfo  {journal} {Phys. Rev. A}\ }\textbf {\bibinfo {volume}
  {104}},\ \bibinfo {pages} {022806} (\bibinfo {year} {2021})}\BibitemShut
  {NoStop}%
\bibitem [{\citenamefont {Angeli}\ and\ \citenamefont
  {Marinova}(2013)}]{Angeli13}%
  \BibitemOpen
  \bibfield  {author} {\bibinfo {author} {\bibfnamefont {I.}~\bibnamefont
  {Angeli}}\ and\ \bibinfo {author} {\bibfnamefont {K.}~\bibnamefont
  {Marinova}},\ }\href {\doibase https://doi.org/10.1016/j.adt.2011.12.006}
  {\bibfield  {journal} {\bibinfo  {journal} {Atomic Data and Nuclear Data
  Tables}\ }\textbf {\bibinfo {volume} {99}},\ \bibinfo {pages} {69} (\bibinfo
  {year} {2013})}\BibitemShut {NoStop}%
\bibitem [{\citenamefont {Blaum}\ \emph {et~al.}(2013)\citenamefont {Blaum},
  \citenamefont {Dilling},\ and\ \citenamefont {Nörtershäuser}}]{Blaum13}%
  \BibitemOpen
  \bibfield  {author} {\bibinfo {author} {\bibfnamefont {K.}~\bibnamefont
  {Blaum}}, \bibinfo {author} {\bibfnamefont {J.}~\bibnamefont {Dilling}}, \
  and\ \bibinfo {author} {\bibfnamefont {W.}~\bibnamefont {Nörtershäuser}},\
  }\href {\doibase 10.1088/0031-8949/2013/T152/014017} {\bibfield  {journal}
  {\bibinfo  {journal} {Physica Scripta Volume T}\ }\textbf {\bibinfo {volume}
  {T152}} (\bibinfo {year} {2013}),\
  10.1088/0031-8949/2013/T152/014017}\BibitemShut {NoStop}%
\bibitem [{\citenamefont {{Arvanitaki}}\ \emph {et~al.}(2015)\citenamefont
  {{Arvanitaki}}, \citenamefont {{Huang}},\ and\ \citenamefont {{Van
  Tilburg}}}]{arvanitaki:2015qy}%
  \BibitemOpen
  \bibfield  {author} {\bibinfo {author} {\bibfnamefont {A.}~\bibnamefont
  {{Arvanitaki}}}, \bibinfo {author} {\bibfnamefont {J.}~\bibnamefont
  {{Huang}}}, \ and\ \bibinfo {author} {\bibfnamefont {K.}~\bibnamefont {{Van
  Tilburg}}},\ }\href {\doibase 10.1103/PhysRevD.91.015015} {\bibfield
  {journal} {\bibinfo  {journal} {\prd}\ }\textbf {\bibinfo {volume} {91}},\
  \bibinfo {eid} {015015} (\bibinfo {year} {2015})},\ \Eprint
  {http://arxiv.org/abs/1405.2925} {arXiv:1405.2925 [hep-ph]} \BibitemShut
  {NoStop}%
\bibitem [{\citenamefont {Stadnik}\ and\ \citenamefont
  {Flambaum}(2015)}]{Flambaum15}%
  \BibitemOpen
  \bibfield  {author} {\bibinfo {author} {\bibfnamefont {Y.~V.}\ \bibnamefont
  {Stadnik}}\ and\ \bibinfo {author} {\bibfnamefont {V.~V.}\ \bibnamefont
  {Flambaum}},\ }\href {\doibase 10.1103/PhysRevLett.114.161301} {\bibfield
  {journal} {\bibinfo  {journal} {Phys. Rev. Lett.}\ }\textbf {\bibinfo
  {volume} {114}},\ \bibinfo {pages} {161301} (\bibinfo {year}
  {2015})}\BibitemShut {NoStop}%
\bibitem [{\citenamefont {{Stadnik}}\ and\ \citenamefont
  {{Flambaum}}(2015)}]{stadnik:2015yu}%
  \BibitemOpen
  \bibfield  {author} {\bibinfo {author} {\bibfnamefont {Y.~V.}\ \bibnamefont
  {{Stadnik}}}\ and\ \bibinfo {author} {\bibfnamefont {V.~V.}\ \bibnamefont
  {{Flambaum}}},\ }\href {\doibase 10.1103/PhysRevLett.115.201301} {\bibfield
  {journal} {\bibinfo  {journal} {Physical Review Letters}\ }\textbf {\bibinfo
  {volume} {115}},\ \bibinfo {eid} {201301} (\bibinfo {year} {2015})},\ \Eprint
  {http://arxiv.org/abs/1503.08540} {arXiv:1503.08540} \BibitemShut {NoStop}%
\bibitem [{\citenamefont {Stadnik}\ and\ \citenamefont
  {Flambaum}(2016)}]{Flambaum16}%
  \BibitemOpen
  \bibfield  {author} {\bibinfo {author} {\bibfnamefont {Y.~V.}\ \bibnamefont
  {Stadnik}}\ and\ \bibinfo {author} {\bibfnamefont {V.~V.}\ \bibnamefont
  {Flambaum}},\ }\href {\doibase 10.1103/PhysRevA.93.063630} {\bibfield
  {journal} {\bibinfo  {journal} {Phys. Rev. A}\ }\textbf {\bibinfo {volume}
  {93}},\ \bibinfo {pages} {063630} (\bibinfo {year} {2016})}\BibitemShut
  {NoStop}%
\bibitem [{\citenamefont {Damour}\ \emph {et~al.}(2002)\citenamefont {Damour},
  \citenamefont {Piazza},\ and\ \citenamefont {Veneziano}}]{Damour02}%
  \BibitemOpen
  \bibfield  {author} {\bibinfo {author} {\bibfnamefont {T.}~\bibnamefont
  {Damour}}, \bibinfo {author} {\bibfnamefont {F.}~\bibnamefont {Piazza}}, \
  and\ \bibinfo {author} {\bibfnamefont {G.}~\bibnamefont {Veneziano}},\ }\href
  {\doibase 10.1103/PhysRevLett.89.081601} {\bibfield  {journal} {\bibinfo
  {journal} {Phys. Rev. Lett.}\ }\textbf {\bibinfo {volume} {89}},\ \bibinfo
  {pages} {081601} (\bibinfo {year} {2002})}\BibitemShut {NoStop}%
\bibitem [{\citenamefont {Hees}\ \emph {et~al.}(2018)\citenamefont {Hees},
  \citenamefont {Minazzoli}, \citenamefont {Savalle}, \citenamefont {Stadnik},\
  and\ \citenamefont {Wolf}}]{hees18}%
  \BibitemOpen
  \bibfield  {author} {\bibinfo {author} {\bibfnamefont {A.}~\bibnamefont
  {Hees}}, \bibinfo {author} {\bibfnamefont {O.}~\bibnamefont {Minazzoli}},
  \bibinfo {author} {\bibfnamefont {E.}~\bibnamefont {Savalle}}, \bibinfo
  {author} {\bibfnamefont {Y.~V.}\ \bibnamefont {Stadnik}}, \ and\ \bibinfo
  {author} {\bibfnamefont {P.}~\bibnamefont {Wolf}},\ }\href {\doibase
  10.1103/PhysRevD.98.064051} {\bibfield  {journal} {\bibinfo  {journal} {Phys.
  Rev. D}\ }\textbf {\bibinfo {volume} {98}},\ \bibinfo {pages} {064051}
  (\bibinfo {year} {2018})}\BibitemShut {NoStop}%
\bibitem [{\citenamefont {{Nitti}}\ and\ \citenamefont
  {{Piazza}}(2012)}]{nitti:2012ut}%
  \BibitemOpen
  \bibfield  {author} {\bibinfo {author} {\bibfnamefont {F.}~\bibnamefont
  {{Nitti}}}\ and\ \bibinfo {author} {\bibfnamefont {F.}~\bibnamefont
  {{Piazza}}},\ }\href {\doibase 10.1103/PhysRevD.86.122002} {\bibfield
  {journal} {\bibinfo  {journal} {\prd}\ }\textbf {\bibinfo {volume} {86}},\
  \bibinfo {eid} {122002} (\bibinfo {year} {2012})},\ \Eprint
  {http://arxiv.org/abs/1202.2105} {arXiv:1202.2105 [hep-th]} \BibitemShut
  {NoStop}%
\bibitem [{\citenamefont {Arvanitaki}\ \emph {et~al.}(2015)\citenamefont
  {Arvanitaki}, \citenamefont {Huang},\ and\ \citenamefont
  {Van~Tilburg}}]{Arvanitaki15}%
  \BibitemOpen
  \bibfield  {author} {\bibinfo {author} {\bibfnamefont {A.}~\bibnamefont
  {Arvanitaki}}, \bibinfo {author} {\bibfnamefont {J.}~\bibnamefont {Huang}}, \
  and\ \bibinfo {author} {\bibfnamefont {K.}~\bibnamefont {Van~Tilburg}},\
  }\href {\doibase 10.1103/PhysRevD.91.015015} {\bibfield  {journal} {\bibinfo
  {journal} {Phys. Rev. D}\ }\textbf {\bibinfo {volume} {91}},\ \bibinfo
  {pages} {015015} (\bibinfo {year} {2015})}\BibitemShut {NoStop}%
\bibitem [{\citenamefont {Asenbaum}\ \emph {et~al.}(2020)\citenamefont
  {Asenbaum}, \citenamefont {Overstreet}, \citenamefont {Kim}, \citenamefont
  {Curti},\ and\ \citenamefont {Kasevich}}]{Stanford20}%
  \BibitemOpen
  \bibfield  {author} {\bibinfo {author} {\bibfnamefont {P.}~\bibnamefont
  {Asenbaum}}, \bibinfo {author} {\bibfnamefont {C.}~\bibnamefont
  {Overstreet}}, \bibinfo {author} {\bibfnamefont {M.}~\bibnamefont {Kim}},
  \bibinfo {author} {\bibfnamefont {J.}~\bibnamefont {Curti}}, \ and\ \bibinfo
  {author} {\bibfnamefont {M.~A.}\ \bibnamefont {Kasevich}},\ }\href {\doibase
  10.1103/PhysRevLett.125.191101} {\bibfield  {journal} {\bibinfo  {journal}
  {Phys. Rev. Lett.}\ }\textbf {\bibinfo {volume} {125}},\ \bibinfo {pages}
  {191101} (\bibinfo {year} {2020})}\BibitemShut {NoStop}%
\bibitem [{\citenamefont {Rodrigues}\ \emph {et~al.}(2022)\citenamefont
  {Rodrigues}, \citenamefont {Touboul}, \citenamefont {Métris}, \citenamefont
  {Bedouet}, \citenamefont {Bergé}, \citenamefont {Carle}, \citenamefont
  {Chhun}, \citenamefont {Christophe}, \citenamefont {Foulon}, \citenamefont
  {Guidotti}, \citenamefont {Lala},\ and\ \citenamefont
  {Robert}}]{Rodrigues22}%
  \BibitemOpen
  \bibfield  {author} {\bibinfo {author} {\bibfnamefont {M.}~\bibnamefont
  {Rodrigues}}, \bibinfo {author} {\bibfnamefont {P.}~\bibnamefont {Touboul}},
  \bibinfo {author} {\bibfnamefont {G.}~\bibnamefont {Métris}}, \bibinfo
  {author} {\bibfnamefont {J.}~\bibnamefont {Bedouet}}, \bibinfo {author}
  {\bibfnamefont {J.}~\bibnamefont {Bergé}}, \bibinfo {author} {\bibfnamefont
  {P.}~\bibnamefont {Carle}}, \bibinfo {author} {\bibfnamefont
  {R.}~\bibnamefont {Chhun}}, \bibinfo {author} {\bibfnamefont
  {B.}~\bibnamefont {Christophe}}, \bibinfo {author} {\bibfnamefont
  {B.}~\bibnamefont {Foulon}}, \bibinfo {author} {\bibfnamefont {P.-Y.}\
  \bibnamefont {Guidotti}}, \bibinfo {author} {\bibfnamefont {S.}~\bibnamefont
  {Lala}}, \ and\ \bibinfo {author} {\bibfnamefont {A.}~\bibnamefont
  {Robert}},\ }\href {\doibase 10.1088/1361-6382/ac4b9a} {\bibfield  {journal}
  {\bibinfo  {journal} {Classical and Quantum Gravity}\ }\textbf {\bibinfo
  {volume} {39}},\ \bibinfo {pages} {204004} (\bibinfo {year}
  {2022})}\BibitemShut {NoStop}%
\bibitem [{\citenamefont {Pihan-Le~Bars}\ \emph {et~al.}(2019)\citenamefont
  {Pihan-Le~Bars}, \citenamefont {Guerlin}, \citenamefont {Hees}, \citenamefont
  {Peaucelle}, \citenamefont {Tasson}, \citenamefont {Bailey}, \citenamefont
  {Mo}, \citenamefont {Delva}, \citenamefont {Meynadier}, \citenamefont
  {Touboul}, \citenamefont {M\'etris}, \citenamefont {Rodrigues}, \citenamefont
  {Berg\'e},\ and\ \citenamefont {Wolf}}]{Pihan}%
  \BibitemOpen
  \bibfield  {author} {\bibinfo {author} {\bibfnamefont {H.}~\bibnamefont
  {Pihan-Le~Bars}}, \bibinfo {author} {\bibfnamefont {C.}~\bibnamefont
  {Guerlin}}, \bibinfo {author} {\bibfnamefont {A.}~\bibnamefont {Hees}},
  \bibinfo {author} {\bibfnamefont {R.}~\bibnamefont {Peaucelle}}, \bibinfo
  {author} {\bibfnamefont {J.~D.}\ \bibnamefont {Tasson}}, \bibinfo {author}
  {\bibfnamefont {Q.~G.}\ \bibnamefont {Bailey}}, \bibinfo {author}
  {\bibfnamefont {G.}~\bibnamefont {Mo}}, \bibinfo {author} {\bibfnamefont
  {P.}~\bibnamefont {Delva}}, \bibinfo {author} {\bibfnamefont
  {F.}~\bibnamefont {Meynadier}}, \bibinfo {author} {\bibfnamefont
  {P.}~\bibnamefont {Touboul}}, \bibinfo {author} {\bibfnamefont
  {G.}~\bibnamefont {M\'etris}}, \bibinfo {author} {\bibfnamefont
  {M.}~\bibnamefont {Rodrigues}}, \bibinfo {author} {\bibfnamefont
  {J.}~\bibnamefont {Berg\'e}}, \ and\ \bibinfo {author} {\bibfnamefont
  {P.}~\bibnamefont {Wolf}},\ }\href {\doibase 10.1103/PhysRevLett.123.231102}
  {\bibfield  {journal} {\bibinfo  {journal} {Phys. Rev. Lett.}\ }\textbf
  {\bibinfo {volume} {123}},\ \bibinfo {pages} {231102} (\bibinfo {year}
  {2019})}\BibitemShut {NoStop}%
\bibitem [{\citenamefont {Bergé}(2023)}]{Berge23}%
  \BibitemOpen
  \bibfield  {author} {\bibinfo {author} {\bibfnamefont {J.}~\bibnamefont
  {Bergé}},\ }\href {\doibase 10.1088/1361-6633/acd203} {\bibfield  {journal}
  {\bibinfo  {journal} {Reports on Progress in Physics}\ }\textbf {\bibinfo
  {volume} {86}},\ \bibinfo {pages} {066901} (\bibinfo {year}
  {2023})}\BibitemShut {NoStop}%
\bibitem [{\citenamefont {Barke}(2015)}]{Barke15}%
  \BibitemOpen
  \bibfield  {author} {\bibinfo {author} {\bibfnamefont {S.}~\bibnamefont
  {Barke}},\ }\emph {\bibinfo {title} {Inter-spacecraft frequency distribution
  for future gravitational wave observatories}},\ \href@noop {} {Ph.D.
  thesis},\ \bibinfo  {school} {Leibniz U.} (\bibinfo {year}
  {2015})\BibitemShut {NoStop}%
\bibitem [{\citenamefont {Hartnett}\ \emph {et~al.}(2012)\citenamefont
  {Hartnett}, \citenamefont {Nand},\ and\ \citenamefont {Lu}}]{Hartnett2012}%
  \BibitemOpen
  \bibfield  {author} {\bibinfo {author} {\bibfnamefont {J.~G.}\ \bibnamefont
  {Hartnett}}, \bibinfo {author} {\bibfnamefont {N.~R.}\ \bibnamefont {Nand}},
  \ and\ \bibinfo {author} {\bibfnamefont {C.}~\bibnamefont {Lu}},\ }\href
  {\doibase 10.1063/1.4709479} {\bibfield  {journal} {\bibinfo  {journal}
  {Applied Physics Letters}\ }\textbf {\bibinfo {volume} {100}},\ \bibinfo
  {pages} {183501} (\bibinfo {year} {2012})},\ \Eprint
  {http://arxiv.org/abs/https://pubs.aip.org/aip/apl/article-pdf/doi/10.1063/1.4709479/14246326/183501\_1\_online.pdf}
  {https://pubs.aip.org/aip/apl/article-pdf/doi/10.1063/1.4709479/14246326/183501\_1\_online.pdf}
  \BibitemShut {NoStop}%
\bibitem [{\citenamefont {Xie}\ \emph {et~al.}(2017)\citenamefont {Xie},
  \citenamefont {Bouchand}, \citenamefont {Nicolodi}, \citenamefont {Giunta},
  \citenamefont {Hänsel}, \citenamefont {Lezius}, \citenamefont {Joshi},
  \citenamefont {Datta}, \citenamefont {Alexandre}, \citenamefont {Lours},
  \citenamefont {Tremblin}, \citenamefont {Santarelli}, \citenamefont
  {Holzwarth},\ and\ \citenamefont {Le~Coq}}]{Xie2017a}%
  \BibitemOpen
  \bibfield  {author} {\bibinfo {author} {\bibfnamefont {X.}~\bibnamefont
  {Xie}}, \bibinfo {author} {\bibfnamefont {R.}~\bibnamefont {Bouchand}},
  \bibinfo {author} {\bibfnamefont {D.}~\bibnamefont {Nicolodi}}, \bibinfo
  {author} {\bibfnamefont {M.}~\bibnamefont {Giunta}}, \bibinfo {author}
  {\bibfnamefont {W.}~\bibnamefont {Hänsel}}, \bibinfo {author} {\bibfnamefont
  {M.}~\bibnamefont {Lezius}}, \bibinfo {author} {\bibfnamefont
  {A.}~\bibnamefont {Joshi}}, \bibinfo {author} {\bibfnamefont
  {S.}~\bibnamefont {Datta}}, \bibinfo {author} {\bibfnamefont
  {C.}~\bibnamefont {Alexandre}}, \bibinfo {author} {\bibfnamefont
  {M.}~\bibnamefont {Lours}}, \bibinfo {author} {\bibfnamefont {P.-A.}\
  \bibnamefont {Tremblin}}, \bibinfo {author} {\bibfnamefont {G.}~\bibnamefont
  {Santarelli}}, \bibinfo {author} {\bibfnamefont {R.}~\bibnamefont
  {Holzwarth}}, \ and\ \bibinfo {author} {\bibfnamefont {Y.}~\bibnamefont
  {Le~Coq}},\ }\href {https://doi.org/10.1038/nphoton.2016.215} {\bibfield
  {journal} {\bibinfo  {journal} {Nature Photonics}\ }\textbf {\bibinfo
  {volume} {11}},\ \bibinfo {pages} {44} (\bibinfo {year} {2017})}\BibitemShut
  {NoStop}%
\bibitem [{\citenamefont {Ludlow}\ \emph {et~al.}(2015)\citenamefont {Ludlow},
  \citenamefont {Boyd}, \citenamefont {Ye}, \citenamefont {Peik},\ and\
  \citenamefont {Schmidt}}]{Ludlow15}%
  \BibitemOpen
  \bibfield  {author} {\bibinfo {author} {\bibfnamefont {A.~D.}\ \bibnamefont
  {Ludlow}}, \bibinfo {author} {\bibfnamefont {M.~M.}\ \bibnamefont {Boyd}},
  \bibinfo {author} {\bibfnamefont {J.}~\bibnamefont {Ye}}, \bibinfo {author}
  {\bibfnamefont {E.}~\bibnamefont {Peik}}, \ and\ \bibinfo {author}
  {\bibfnamefont {P.~O.}\ \bibnamefont {Schmidt}},\ }\href {\doibase
  10.1103/RevModPhys.87.637} {\bibfield  {journal} {\bibinfo  {journal} {Rev.
  Mod. Phys.}\ }\textbf {\bibinfo {volume} {87}},\ \bibinfo {pages} {637}
  (\bibinfo {year} {2015})}\BibitemShut {NoStop}%
\bibitem [{\citenamefont {Le~Targat}\ \emph {et~al.}(2006)\citenamefont
  {Le~Targat}, \citenamefont {Baillard}, \citenamefont {Fouch\'e},
  \citenamefont {Brusch}, \citenamefont {Tcherbakoff}, \citenamefont {Rovera},\
  and\ \citenamefont {Lemonde}}]{LeTargat06}%
  \BibitemOpen
  \bibfield  {author} {\bibinfo {author} {\bibfnamefont {R.}~\bibnamefont
  {Le~Targat}}, \bibinfo {author} {\bibfnamefont {X.}~\bibnamefont {Baillard}},
  \bibinfo {author} {\bibfnamefont {M.}~\bibnamefont {Fouch\'e}}, \bibinfo
  {author} {\bibfnamefont {A.}~\bibnamefont {Brusch}}, \bibinfo {author}
  {\bibfnamefont {O.}~\bibnamefont {Tcherbakoff}}, \bibinfo {author}
  {\bibfnamefont {G.~D.}\ \bibnamefont {Rovera}}, \ and\ \bibinfo {author}
  {\bibfnamefont {P.}~\bibnamefont {Lemonde}},\ }\href {\doibase
  10.1103/PhysRevLett.97.130801} {\bibfield  {journal} {\bibinfo  {journal}
  {Phys. Rev. Lett.}\ }\textbf {\bibinfo {volume} {97}},\ \bibinfo {pages}
  {130801} (\bibinfo {year} {2006})}\BibitemShut {NoStop}%
\bibitem [{\citenamefont {Sterr}\ \emph {et~al.}(2004)\citenamefont {Sterr},
  \citenamefont {Degenhardt}, \citenamefont {Stoehr}, \citenamefont {Lisdat},
  \citenamefont {Schnatz}, \citenamefont {Helmcke}, \citenamefont {Riehle},
  \citenamefont {Wilpers}, \citenamefont {Oates},\ and\ \citenamefont
  {Hollberg}}]{Sterr04}%
  \BibitemOpen
  \bibfield  {author} {\bibinfo {author} {\bibfnamefont {U.}~\bibnamefont
  {Sterr}}, \bibinfo {author} {\bibfnamefont {C.}~\bibnamefont {Degenhardt}},
  \bibinfo {author} {\bibfnamefont {H.}~\bibnamefont {Stoehr}}, \bibinfo
  {author} {\bibfnamefont {C.}~\bibnamefont {Lisdat}}, \bibinfo {author}
  {\bibfnamefont {H.}~\bibnamefont {Schnatz}}, \bibinfo {author} {\bibfnamefont
  {J.}~\bibnamefont {Helmcke}}, \bibinfo {author} {\bibfnamefont
  {F.}~\bibnamefont {Riehle}}, \bibinfo {author} {\bibfnamefont
  {G.}~\bibnamefont {Wilpers}}, \bibinfo {author} {\bibfnamefont
  {C.}~\bibnamefont {Oates}}, \ and\ \bibinfo {author} {\bibfnamefont
  {L.}~\bibnamefont {Hollberg}},\ }\href {\doibase
  https://doi.org/10.1016/j.crhy.2004.08.005} {\bibfield  {journal} {\bibinfo
  {journal} {Comptes Rendus Physique}\ }\textbf {\bibinfo {volume} {5}},\
  \bibinfo {pages} {845} (\bibinfo {year} {2004})},\ \bibinfo {note}
  {fundamental metrology}\BibitemShut {NoStop}%
\bibitem [{\citenamefont {Atkinson}\ \emph {et~al.}(2019)\citenamefont
  {Atkinson}, \citenamefont {Schelfhout},\ and\ \citenamefont
  {McFerran}}]{Atkinson19}%
  \BibitemOpen
  \bibfield  {author} {\bibinfo {author} {\bibfnamefont {P.~E.}\ \bibnamefont
  {Atkinson}}, \bibinfo {author} {\bibfnamefont {J.~S.}\ \bibnamefont
  {Schelfhout}}, \ and\ \bibinfo {author} {\bibfnamefont {J.~J.}\ \bibnamefont
  {McFerran}},\ }\href {\doibase 10.1103/PhysRevA.100.042505} {\bibfield
  {journal} {\bibinfo  {journal} {Phys. Rev. A}\ }\textbf {\bibinfo {volume}
  {100}},\ \bibinfo {pages} {042505} (\bibinfo {year} {2019})}\BibitemShut
  {NoStop}%
\bibitem [{\citenamefont {Jones}\ \emph {et~al.}(2023)\citenamefont {Jones},
  \citenamefont {van Kann},\ and\ \citenamefont {McFerran}}]{Jones23}%
  \BibitemOpen
  \bibfield  {author} {\bibinfo {author} {\bibfnamefont {D.~M.}\ \bibnamefont
  {Jones}}, \bibinfo {author} {\bibfnamefont {F.}~\bibnamefont {van Kann}}, \
  and\ \bibinfo {author} {\bibfnamefont {J.~J.}\ \bibnamefont {McFerran}},\
  }\href {\doibase 10.1364/AO.488653} {\bibfield  {journal} {\bibinfo
  {journal} {Appl. Opt.}\ }\textbf {\bibinfo {volume} {62}},\ \bibinfo {pages}
  {3932} (\bibinfo {year} {2023})}\BibitemShut {NoStop}%
\bibitem [{\citenamefont {Witkowski}\ \emph {et~al.}(2019)\citenamefont
  {Witkowski}, \citenamefont {Kowzan}, \citenamefont {Munoz-Rodriguez},
  \citenamefont {Ciury{\l}o}, \citenamefont {\.{Z}uchowski}, \citenamefont
  {Mas{\l}owski},\ and\ \citenamefont {Zawada}}]{Witkowski19}%
  \BibitemOpen
  \bibfield  {author} {\bibinfo {author} {\bibfnamefont {M.}~\bibnamefont
  {Witkowski}}, \bibinfo {author} {\bibfnamefont {G.}~\bibnamefont {Kowzan}},
  \bibinfo {author} {\bibfnamefont {R.}~\bibnamefont {Munoz-Rodriguez}},
  \bibinfo {author} {\bibfnamefont {R.}~\bibnamefont {Ciury{\l}o}}, \bibinfo
  {author} {\bibfnamefont {P.~S.}\ \bibnamefont {\.{Z}uchowski}}, \bibinfo
  {author} {\bibfnamefont {P.}~\bibnamefont {Mas{\l}owski}}, \ and\ \bibinfo
  {author} {\bibfnamefont {M.}~\bibnamefont {Zawada}},\ }\href {\doibase
  10.1364/OE.27.011069} {\bibfield  {journal} {\bibinfo  {journal} {Opt.
  Express}\ }\textbf {\bibinfo {volume} {27}},\ \bibinfo {pages} {11069}
  (\bibinfo {year} {2019})}\BibitemShut {NoStop}%
\bibitem [{\citenamefont {Rudolph}\ \emph {et~al.}(2020)\citenamefont
  {Rudolph}, \citenamefont {Wilkason}, \citenamefont {Nantel}, \citenamefont
  {Swan}, \citenamefont {Holland}, \citenamefont {Jiang}, \citenamefont
  {Garber}, \citenamefont {Carman},\ and\ \citenamefont {Hogan}}]{Rudolph20}%
  \BibitemOpen
  \bibfield  {author} {\bibinfo {author} {\bibfnamefont {J.}~\bibnamefont
  {Rudolph}}, \bibinfo {author} {\bibfnamefont {T.}~\bibnamefont {Wilkason}},
  \bibinfo {author} {\bibfnamefont {M.}~\bibnamefont {Nantel}}, \bibinfo
  {author} {\bibfnamefont {H.}~\bibnamefont {Swan}}, \bibinfo {author}
  {\bibfnamefont {C.~M.}\ \bibnamefont {Holland}}, \bibinfo {author}
  {\bibfnamefont {Y.}~\bibnamefont {Jiang}}, \bibinfo {author} {\bibfnamefont
  {B.~E.}\ \bibnamefont {Garber}}, \bibinfo {author} {\bibfnamefont {S.~P.}\
  \bibnamefont {Carman}}, \ and\ \bibinfo {author} {\bibfnamefont {J.~M.}\
  \bibnamefont {Hogan}},\ }\href {\doibase 10.1103/PhysRevLett.124.083604}
  {\bibfield  {journal} {\bibinfo  {journal} {Phys. Rev. Lett.}\ }\textbf
  {\bibinfo {volume} {124}},\ \bibinfo {pages} {083604} (\bibinfo {year}
  {2020})}\BibitemShut {NoStop}%
\bibitem [{\citenamefont {Miuchi}\ \emph {et~al.}(2020)\citenamefont {Miuchi},
  \citenamefont {Baracchini}, \citenamefont {Lane}, \citenamefont {Spooner},\
  and\ \citenamefont {Vahsen}}]{Miuchi20}%
  \BibitemOpen
  \bibfield  {author} {\bibinfo {author} {\bibfnamefont {K.}~\bibnamefont
  {Miuchi}}, \bibinfo {author} {\bibfnamefont {E.}~\bibnamefont {Baracchini}},
  \bibinfo {author} {\bibfnamefont {G.}~\bibnamefont {Lane}}, \bibinfo {author}
  {\bibfnamefont {N.~J.~C.}\ \bibnamefont {Spooner}}, \ and\ \bibinfo {author}
  {\bibfnamefont {S.~E.}\ \bibnamefont {Vahsen}},\ }\href {\doibase
  10.1088/1742-6596/1468/1/012044} {\bibfield  {journal} {\bibinfo  {journal}
  {Journal of Physics: Conference Series}\ }\textbf {\bibinfo {volume}
  {1468}},\ \bibinfo {pages} {012044} (\bibinfo {year} {2020})}\BibitemShut
  {NoStop}%
\bibitem [{\citenamefont {van Leeuwen}(2007)}]{vanLeeuwen07}%
  \BibitemOpen
  \bibfield  {author} {\bibinfo {author} {\bibfnamefont {F.}~\bibnamefont {van
  Leeuwen}},\ }\href {\doibase 10.1051/0004-6361:20078357} {\bibfield
  {journal} {\bibinfo  {journal} {Astronomy \& Astrophysics}\ }\textbf
  {\bibinfo {volume} {474}},\ \bibinfo {pages} {653} (\bibinfo {year}
  {2007})}\BibitemShut {NoStop}%
\bibitem [{\citenamefont {Berg\'e}\ \emph {et~al.}(2018)\citenamefont
  {Berg\'e}, \citenamefont {Brax}, \citenamefont {M\'etris}, \citenamefont
  {Pernot-Borr\`as}, \citenamefont {Touboul},\ and\ \citenamefont
  {Uzan}}]{Microscope18}%
  \BibitemOpen
  \bibfield  {author} {\bibinfo {author} {\bibfnamefont {J.}~\bibnamefont
  {Berg\'e}}, \bibinfo {author} {\bibfnamefont {P.}~\bibnamefont {Brax}},
  \bibinfo {author} {\bibfnamefont {G.}~\bibnamefont {M\'etris}}, \bibinfo
  {author} {\bibfnamefont {M.}~\bibnamefont {Pernot-Borr\`as}}, \bibinfo
  {author} {\bibfnamefont {P.}~\bibnamefont {Touboul}}, \ and\ \bibinfo
  {author} {\bibfnamefont {J.-P.}\ \bibnamefont {Uzan}},\ }\href {\doibase
  10.1103/PhysRevLett.120.141101} {\bibfield  {journal} {\bibinfo  {journal}
  {Phys. Rev. Lett.}\ }\textbf {\bibinfo {volume} {120}},\ \bibinfo {pages}
  {141101} (\bibinfo {year} {2018})}\BibitemShut {NoStop}%
\bibitem [{\citenamefont {Gu\'ena}\ \emph {et~al.}(2012)\citenamefont
  {Gu\'ena}, \citenamefont {Abgrall}, \citenamefont {Rovera}, \citenamefont
  {Rosenbusch}, \citenamefont {Tobar}, \citenamefont {Laurent}, \citenamefont
  {Clairon},\ and\ \citenamefont {Bize}}]{Guena12}%
  \BibitemOpen
  \bibfield  {author} {\bibinfo {author} {\bibfnamefont {J.}~\bibnamefont
  {Gu\'ena}}, \bibinfo {author} {\bibfnamefont {M.}~\bibnamefont {Abgrall}},
  \bibinfo {author} {\bibfnamefont {D.}~\bibnamefont {Rovera}}, \bibinfo
  {author} {\bibfnamefont {P.}~\bibnamefont {Rosenbusch}}, \bibinfo {author}
  {\bibfnamefont {M.~E.}\ \bibnamefont {Tobar}}, \bibinfo {author}
  {\bibfnamefont {P.}~\bibnamefont {Laurent}}, \bibinfo {author} {\bibfnamefont
  {A.}~\bibnamefont {Clairon}}, \ and\ \bibinfo {author} {\bibfnamefont
  {S.}~\bibnamefont {Bize}},\ }\href {\doibase 10.1103/PhysRevLett.109.080801}
  {\bibfield  {journal} {\bibinfo  {journal} {Phys. Rev. Lett.}\ }\textbf
  {\bibinfo {volume} {109}},\ \bibinfo {pages} {080801} (\bibinfo {year}
  {2012})}\BibitemShut {NoStop}%
\bibitem [{\citenamefont {Angstmann}\ \emph {et~al.}(2004)\citenamefont
  {Angstmann}, \citenamefont {Dzuba},\ and\ \citenamefont
  {Flambaum}}]{Angstmann04}%
  \BibitemOpen
  \bibfield  {author} {\bibinfo {author} {\bibfnamefont {E.~J.}\ \bibnamefont
  {Angstmann}}, \bibinfo {author} {\bibfnamefont {V.~A.}\ \bibnamefont
  {Dzuba}}, \ and\ \bibinfo {author} {\bibfnamefont {V.~V.}\ \bibnamefont
  {Flambaum}},\ }\href {\doibase 10.1103/PhysRevA.70.014102} {\bibfield
  {journal} {\bibinfo  {journal} {Phys. Rev. A}\ }\textbf {\bibinfo {volume}
  {70}},\ \bibinfo {pages} {014102} (\bibinfo {year} {2004})}\BibitemShut
  {NoStop}%
\bibitem [{\citenamefont {Bloom}\ \emph {et~al.}(2014)\citenamefont {Bloom},
  \citenamefont {Nicholson}, \citenamefont {Williams}, \citenamefont
  {Campbell}, \citenamefont {Bishof}, \citenamefont {Zhang}, \citenamefont
  {Zhang}, \citenamefont {Bromley},\ and\ \citenamefont {Ye}}]{Bloom14}%
  \BibitemOpen
  \bibfield  {author} {\bibinfo {author} {\bibfnamefont {B.~J.}\ \bibnamefont
  {Bloom}}, \bibinfo {author} {\bibfnamefont {T.~L.}\ \bibnamefont
  {Nicholson}}, \bibinfo {author} {\bibfnamefont {J.~R.}\ \bibnamefont
  {Williams}}, \bibinfo {author} {\bibfnamefont {S.~L.}\ \bibnamefont
  {Campbell}}, \bibinfo {author} {\bibfnamefont {M.}~\bibnamefont {Bishof}},
  \bibinfo {author} {\bibfnamefont {X.}~\bibnamefont {Zhang}}, \bibinfo
  {author} {\bibfnamefont {W.}~\bibnamefont {Zhang}}, \bibinfo {author}
  {\bibfnamefont {S.~L.}\ \bibnamefont {Bromley}}, \ and\ \bibinfo {author}
  {\bibfnamefont {J.}~\bibnamefont {Ye}},\ }\href {\doibase
  10.1038/nature12941} {\bibfield  {journal} {\bibinfo  {journal} {Nature}\
  }\textbf {\bibinfo {volume} {506}},\ \bibinfo {pages} {71} (\bibinfo {year}
  {2014})}\BibitemShut {NoStop}%
\bibitem [{\citenamefont {Hinkley}\ \emph {et~al.}(2013)\citenamefont
  {Hinkley}, \citenamefont {Sherman}, \citenamefont {Phillips}, \citenamefont
  {Schioppo}, \citenamefont {Lemke}, \citenamefont {Beloy}, \citenamefont
  {Pizzocaro}, \citenamefont {Oates},\ and\ \citenamefont
  {Ludlow}}]{Hinkley13}%
  \BibitemOpen
  \bibfield  {author} {\bibinfo {author} {\bibfnamefont {N.}~\bibnamefont
  {Hinkley}}, \bibinfo {author} {\bibfnamefont {J.~A.}\ \bibnamefont
  {Sherman}}, \bibinfo {author} {\bibfnamefont {N.~B.}\ \bibnamefont
  {Phillips}}, \bibinfo {author} {\bibfnamefont {M.}~\bibnamefont {Schioppo}},
  \bibinfo {author} {\bibfnamefont {N.~D.}\ \bibnamefont {Lemke}}, \bibinfo
  {author} {\bibfnamefont {K.}~\bibnamefont {Beloy}}, \bibinfo {author}
  {\bibfnamefont {M.}~\bibnamefont {Pizzocaro}}, \bibinfo {author}
  {\bibfnamefont {C.~W.}\ \bibnamefont {Oates}}, \ and\ \bibinfo {author}
  {\bibfnamefont {A.~D.}\ \bibnamefont {Ludlow}},\ }\href {\doibase
  10.1126/science.1240420} {\bibfield  {journal} {\bibinfo  {journal}
  {Science}\ }\textbf {\bibinfo {volume} {341}},\ \bibinfo {pages} {1215}
  (\bibinfo {year} {2013})},\ \Eprint
  {http://arxiv.org/abs/https://www.science.org/doi/pdf/10.1126/science.1240420}
  {https://www.science.org/doi/pdf/10.1126/science.1240420} \BibitemShut
  {NoStop}%
\bibitem [{\citenamefont {Derevianko}(2018)}]{Derevianko}%
  \BibitemOpen
  \bibfield  {author} {\bibinfo {author} {\bibfnamefont {A.}~\bibnamefont
  {Derevianko}},\ }\href {\doibase 10.1103/PhysRevA.97.042506} {\bibfield
  {journal} {\bibinfo  {journal} {Phys. Rev. A}\ }\textbf {\bibinfo {volume}
  {97}},\ \bibinfo {pages} {042506} (\bibinfo {year} {2018})}\BibitemShut
  {NoStop}%
\bibitem [{\citenamefont {Evans}\ \emph {et~al.}(2019)\citenamefont {Evans},
  \citenamefont {O'Hare},\ and\ \citenamefont {McCabe}}]{Evans19}%
  \BibitemOpen
  \bibfield  {author} {\bibinfo {author} {\bibfnamefont {N.~W.}\ \bibnamefont
  {Evans}}, \bibinfo {author} {\bibfnamefont {C.~A.~J.}\ \bibnamefont
  {O'Hare}}, \ and\ \bibinfo {author} {\bibfnamefont {C.}~\bibnamefont
  {McCabe}},\ }\href {\doibase 10.1103/PhysRevD.99.023012} {\bibfield
  {journal} {\bibinfo  {journal} {Phys. Rev. D}\ }\textbf {\bibinfo {volume}
  {99}},\ \bibinfo {pages} {023012} (\bibinfo {year} {2019})}\BibitemShut
  {NoStop}%
\bibitem [{\citenamefont {Centers}\ \emph {et~al.}(2021)\citenamefont
  {Centers}, \citenamefont {Blanchard}, \citenamefont {Conrad}, \citenamefont
  {Figueroa}, \citenamefont {Garcon}, \citenamefont {Gramolin}, \citenamefont
  {Kimball}, \citenamefont {Lawson}, \citenamefont {Pelssers}, \citenamefont
  {Smiga}, \citenamefont {Sushkov}, \citenamefont {Wickenbrock}, \citenamefont
  {Budker},\ and\ \citenamefont {Derevianko}}]{Centers21}%
  \BibitemOpen
  \bibfield  {author} {\bibinfo {author} {\bibfnamefont {G.~P.}\ \bibnamefont
  {Centers}}, \bibinfo {author} {\bibfnamefont {J.~W.}\ \bibnamefont
  {Blanchard}}, \bibinfo {author} {\bibfnamefont {J.}~\bibnamefont {Conrad}},
  \bibinfo {author} {\bibfnamefont {N.~L.}\ \bibnamefont {Figueroa}}, \bibinfo
  {author} {\bibfnamefont {A.}~\bibnamefont {Garcon}}, \bibinfo {author}
  {\bibfnamefont {A.~V.}\ \bibnamefont {Gramolin}}, \bibinfo {author}
  {\bibfnamefont {D.~F.~J.}\ \bibnamefont {Kimball}}, \bibinfo {author}
  {\bibfnamefont {M.}~\bibnamefont {Lawson}}, \bibinfo {author} {\bibfnamefont
  {B.}~\bibnamefont {Pelssers}}, \bibinfo {author} {\bibfnamefont {J.~A.}\
  \bibnamefont {Smiga}}, \bibinfo {author} {\bibfnamefont {A.~O.}\ \bibnamefont
  {Sushkov}}, \bibinfo {author} {\bibfnamefont {A.}~\bibnamefont
  {Wickenbrock}}, \bibinfo {author} {\bibfnamefont {D.}~\bibnamefont {Budker}},
  \ and\ \bibinfo {author} {\bibfnamefont {A.}~\bibnamefont {Derevianko}},\
  }\href {\doibase 10.1038/s41467-021-27632-7} {\bibfield  {journal} {\bibinfo
  {journal} {Nature Communications}\ }\textbf {\bibinfo {volume} {12}}
  (\bibinfo {year} {2021}),\ 10.1038/s41467-021-27632-7}\BibitemShut {NoStop}%
\bibitem [{\citenamefont {{Foster}}\ \emph {et~al.}(2018)\citenamefont
  {{Foster}}, \citenamefont {{Rodd}},\ and\ \citenamefont
  {{Safdi}}}]{foster:2018aa}%
  \BibitemOpen
  \bibfield  {author} {\bibinfo {author} {\bibfnamefont {J.~W.}\ \bibnamefont
  {{Foster}}}, \bibinfo {author} {\bibfnamefont {N.~L.}\ \bibnamefont
  {{Rodd}}}, \ and\ \bibinfo {author} {\bibfnamefont {B.~R.}\ \bibnamefont
  {{Safdi}}},\ }\href {\doibase 10.1103/PhysRevD.97.123006} {\bibfield
  {journal} {\bibinfo  {journal} {\prd}\ }\textbf {\bibinfo {volume} {97}},\
  \bibinfo {eid} {123006} (\bibinfo {year} {2018})},\ \Eprint
  {http://arxiv.org/abs/1711.10489} {arXiv:1711.10489} \BibitemShut {NoStop}%
\bibitem [{\citenamefont {Budker}\ \emph {et~al.}(2014)\citenamefont {Budker},
  \citenamefont {Graham}, \citenamefont {Ledbetter}, \citenamefont
  {Rajendran},\ and\ \citenamefont {Sushkov}}]{Budker14}%
  \BibitemOpen
  \bibfield  {author} {\bibinfo {author} {\bibfnamefont {D.}~\bibnamefont
  {Budker}}, \bibinfo {author} {\bibfnamefont {P.~W.}\ \bibnamefont {Graham}},
  \bibinfo {author} {\bibfnamefont {M.}~\bibnamefont {Ledbetter}}, \bibinfo
  {author} {\bibfnamefont {S.}~\bibnamefont {Rajendran}}, \ and\ \bibinfo
  {author} {\bibfnamefont {A.~O.}\ \bibnamefont {Sushkov}},\ }\href {\doibase
  10.1103/PhysRevX.4.021030} {\bibfield  {journal} {\bibinfo  {journal} {Phys.
  Rev. X}\ }\textbf {\bibinfo {volume} {4}},\ \bibinfo {pages} {021030}
  (\bibinfo {year} {2014})}\BibitemShut {NoStop}%
\bibitem [{\citenamefont {McMillan}(2011)}]{McMillan11}%
  \BibitemOpen
  \bibfield  {author} {\bibinfo {author} {\bibfnamefont {P.~J.}\ \bibnamefont
  {McMillan}},\ }\href {\doibase 10.1111/j.1365-2966.2011.18564.x} {\bibfield
  {journal} {\bibinfo  {journal} {Monthly Notices of the Royal Astronomical
  Society}\ }\textbf {\bibinfo {volume} {414}},\ \bibinfo {pages} {2446}
  (\bibinfo {year} {2011})}\BibitemShut {NoStop}%
\bibitem [{\citenamefont {Schulthess}\ \emph {et~al.}(2022)\citenamefont
  {Schulthess} \emph {et~al.}}]{Beam_EDM}%
  \BibitemOpen
  \bibfield  {author} {\bibinfo {author} {\bibfnamefont {I.}~\bibnamefont
  {Schulthess}} \emph {et~al.},\ }\href {\doibase
  10.1103/PhysRevLett.129.191801} {\bibfield  {journal} {\bibinfo  {journal}
  {Phys. Rev. Lett.}\ }\textbf {\bibinfo {volume} {129}},\ \bibinfo {pages}
  {191801} (\bibinfo {year} {2022})},\ \Eprint
  {http://arxiv.org/abs/2204.01454} {arXiv:2204.01454 [hep-ex]} \BibitemShut
  {NoStop}%
\bibitem [{\citenamefont {Roussy}\ \emph {et~al.}(2021)\citenamefont {Roussy}
  \emph {et~al.}}]{HfF}%
  \BibitemOpen
  \bibfield  {author} {\bibinfo {author} {\bibfnamefont {T.~S.}\ \bibnamefont
  {Roussy}} \emph {et~al.},\ }\href {\doibase 10.1103/PhysRevLett.126.171301}
  {\bibfield  {journal} {\bibinfo  {journal} {Phys. Rev. Lett.}\ }\textbf
  {\bibinfo {volume} {126}},\ \bibinfo {pages} {171301} (\bibinfo {year}
  {2021})},\ \Eprint {http://arxiv.org/abs/2006.15787} {arXiv:2006.15787
  [hep-ph]} \BibitemShut {NoStop}%
\bibitem [{\citenamefont {Zhang}\ \emph {et~al.}(2022)\citenamefont {Zhang},
  \citenamefont {Banerjee}, \citenamefont {Leyser}, \citenamefont {Perez},
  \citenamefont {Schiller}, \citenamefont {Budker},\ and\ \citenamefont
  {Antypas}}]{Rb_quartz}%
  \BibitemOpen
  \bibfield  {author} {\bibinfo {author} {\bibfnamefont {X.}~\bibnamefont
  {Zhang}}, \bibinfo {author} {\bibfnamefont {A.}~\bibnamefont {Banerjee}},
  \bibinfo {author} {\bibfnamefont {M.}~\bibnamefont {Leyser}}, \bibinfo
  {author} {\bibfnamefont {G.}~\bibnamefont {Perez}}, \bibinfo {author}
  {\bibfnamefont {S.}~\bibnamefont {Schiller}}, \bibinfo {author}
  {\bibfnamefont {D.}~\bibnamefont {Budker}}, \ and\ \bibinfo {author}
  {\bibfnamefont {D.}~\bibnamefont {Antypas}},\ }\href@noop {} {\  (\bibinfo
  {year} {2022})},\ \Eprint {http://arxiv.org/abs/2212.04413} {arXiv:2212.04413
  [physics.atom-ph]} \BibitemShut {NoStop}%
\bibitem [{\citenamefont {Abel}\ \emph {et~al.}(2017)\citenamefont {Abel},
  \citenamefont {Ayres}, \citenamefont {Ban}, \citenamefont {Bison},
  \citenamefont {Bodek}, \citenamefont {Bondar}, \citenamefont {Daum},
  \citenamefont {Fairbairn}, \citenamefont {Flambaum}, \citenamefont
  {Geltenbort}, \citenamefont {Green}, \citenamefont {Griffith}, \citenamefont
  {van~der Grinten}, \citenamefont {Grujic}, \citenamefont {Harris},
  \citenamefont {Hild}, \citenamefont {Iaydjiev}, \citenamefont {Ivanov},
  \citenamefont {Kasprzak}, \citenamefont {Kermaidic}, \citenamefont {Kirch},
  \citenamefont {Koch}, \citenamefont {Komposch}, \citenamefont {Koss},
  \citenamefont {Kozela}, \citenamefont {Krempel}, \citenamefont {Lauss},
  \citenamefont {Lefort}, \citenamefont {Lemi\`ere}, \citenamefont {Marsh},
  \citenamefont {Mohanmurthy}, \citenamefont {Mtchedlishvili}, \citenamefont
  {Musgrave}, \citenamefont {Piegsa}, \citenamefont {Pignol}, \citenamefont
  {Rawlik}, \citenamefont {Rebreyend}, \citenamefont {Ries}, \citenamefont
  {Roccia}, \citenamefont {Rozpedzik}, \citenamefont {Schmidt-Wellenburg},
  \citenamefont {Severijns}, \citenamefont {Shiers}, \citenamefont {Stadnik},
  \citenamefont {Weis}, \citenamefont {Wursten}, \citenamefont {Zejma},\ and\
  \citenamefont {Zsigmond}}]{nEDM}%
  \BibitemOpen
  \bibfield  {author} {\bibinfo {author} {\bibfnamefont {C.}~\bibnamefont
  {Abel}}, \bibinfo {author} {\bibfnamefont {N.~J.}\ \bibnamefont {Ayres}},
  \bibinfo {author} {\bibfnamefont {G.}~\bibnamefont {Ban}}, \bibinfo {author}
  {\bibfnamefont {G.}~\bibnamefont {Bison}}, \bibinfo {author} {\bibfnamefont
  {K.}~\bibnamefont {Bodek}}, \bibinfo {author} {\bibfnamefont
  {V.}~\bibnamefont {Bondar}}, \bibinfo {author} {\bibfnamefont
  {M.}~\bibnamefont {Daum}}, \bibinfo {author} {\bibfnamefont {M.}~\bibnamefont
  {Fairbairn}}, \bibinfo {author} {\bibfnamefont {V.~V.}\ \bibnamefont
  {Flambaum}}, \bibinfo {author} {\bibfnamefont {P.}~\bibnamefont
  {Geltenbort}}, \bibinfo {author} {\bibfnamefont {K.}~\bibnamefont {Green}},
  \bibinfo {author} {\bibfnamefont {W.~C.}\ \bibnamefont {Griffith}}, \bibinfo
  {author} {\bibfnamefont {M.}~\bibnamefont {van~der Grinten}}, \bibinfo
  {author} {\bibfnamefont {Z.~D.}\ \bibnamefont {Grujic}}, \bibinfo {author}
  {\bibfnamefont {P.~G.}\ \bibnamefont {Harris}}, \bibinfo {author}
  {\bibfnamefont {N.}~\bibnamefont {Hild}}, \bibinfo {author} {\bibfnamefont
  {P.}~\bibnamefont {Iaydjiev}}, \bibinfo {author} {\bibfnamefont {S.~N.}\
  \bibnamefont {Ivanov}}, \bibinfo {author} {\bibfnamefont {M.}~\bibnamefont
  {Kasprzak}}, \bibinfo {author} {\bibfnamefont {Y.}~\bibnamefont {Kermaidic}},
  \bibinfo {author} {\bibfnamefont {K.}~\bibnamefont {Kirch}}, \bibinfo
  {author} {\bibfnamefont {H.-C.}\ \bibnamefont {Koch}}, \bibinfo {author}
  {\bibfnamefont {S.}~\bibnamefont {Komposch}}, \bibinfo {author}
  {\bibfnamefont {P.~A.}\ \bibnamefont {Koss}}, \bibinfo {author}
  {\bibfnamefont {A.}~\bibnamefont {Kozela}}, \bibinfo {author} {\bibfnamefont
  {J.}~\bibnamefont {Krempel}}, \bibinfo {author} {\bibfnamefont
  {B.}~\bibnamefont {Lauss}}, \bibinfo {author} {\bibfnamefont
  {T.}~\bibnamefont {Lefort}}, \bibinfo {author} {\bibfnamefont
  {Y.}~\bibnamefont {Lemi\`ere}}, \bibinfo {author} {\bibfnamefont {D.~J.~E.}\
  \bibnamefont {Marsh}}, \bibinfo {author} {\bibfnamefont {P.}~\bibnamefont
  {Mohanmurthy}}, \bibinfo {author} {\bibfnamefont {A.}~\bibnamefont
  {Mtchedlishvili}}, \bibinfo {author} {\bibfnamefont {M.}~\bibnamefont
  {Musgrave}}, \bibinfo {author} {\bibfnamefont {F.~M.}\ \bibnamefont
  {Piegsa}}, \bibinfo {author} {\bibfnamefont {G.}~\bibnamefont {Pignol}},
  \bibinfo {author} {\bibfnamefont {M.}~\bibnamefont {Rawlik}}, \bibinfo
  {author} {\bibfnamefont {D.}~\bibnamefont {Rebreyend}}, \bibinfo {author}
  {\bibfnamefont {D.}~\bibnamefont {Ries}}, \bibinfo {author} {\bibfnamefont
  {S.}~\bibnamefont {Roccia}}, \bibinfo {author} {\bibfnamefont
  {D.}~\bibnamefont {Rozpedzik}}, \bibinfo {author} {\bibfnamefont
  {P.}~\bibnamefont {Schmidt-Wellenburg}}, \bibinfo {author} {\bibfnamefont
  {N.}~\bibnamefont {Severijns}}, \bibinfo {author} {\bibfnamefont
  {D.}~\bibnamefont {Shiers}}, \bibinfo {author} {\bibfnamefont {Y.~V.}\
  \bibnamefont {Stadnik}}, \bibinfo {author} {\bibfnamefont {A.}~\bibnamefont
  {Weis}}, \bibinfo {author} {\bibfnamefont {E.}~\bibnamefont {Wursten}},
  \bibinfo {author} {\bibfnamefont {J.}~\bibnamefont {Zejma}}, \ and\ \bibinfo
  {author} {\bibfnamefont {G.}~\bibnamefont {Zsigmond}},\ }\href {\doibase
  10.1103/PhysRevX.7.041034} {\bibfield  {journal} {\bibinfo  {journal} {Phys.
  Rev. X}\ }\textbf {\bibinfo {volume} {7}},\ \bibinfo {pages} {041034}
  (\bibinfo {year} {2017})}\BibitemShut {NoStop}%
\bibitem [{\citenamefont {O'Hare}(2020)}]{AxionLimits}%
  \BibitemOpen
  \bibfield  {author} {\bibinfo {author} {\bibfnamefont {C.}~\bibnamefont
  {O'Hare}},\ }\href {\doibase 10.5281/zenodo.3932430} {\enquote {\bibinfo
  {title} {cajohare/axionlimits: Axionlimits},}\ }\bibinfo {howpublished}
  {\url{https://cajohare.github.io/AxionLimits/}} (\bibinfo {year}
  {2020})\BibitemShut {NoStop}%
\bibitem [{\citenamefont {Dzuba}\ \emph {et~al.}(1999)\citenamefont {Dzuba},
  \citenamefont {Flambaum},\ and\ \citenamefont {Webb}}]{Dzuba99}%
  \BibitemOpen
  \bibfield  {author} {\bibinfo {author} {\bibfnamefont {V.~A.}\ \bibnamefont
  {Dzuba}}, \bibinfo {author} {\bibfnamefont {V.~V.}\ \bibnamefont {Flambaum}},
  \ and\ \bibinfo {author} {\bibfnamefont {J.~K.}\ \bibnamefont {Webb}},\
  }\href {\doibase 10.1103/PhysRevLett.82.888} {\bibfield  {journal} {\bibinfo
  {journal} {Phys. Rev. Lett.}\ }\textbf {\bibinfo {volume} {82}},\ \bibinfo
  {pages} {888} (\bibinfo {year} {1999})}\BibitemShut {NoStop}%
\bibitem [{\citenamefont {Hees}\ \emph {et~al.}(2016)\citenamefont {Hees},
  \citenamefont {Gu\'ena}, \citenamefont {Abgrall}, \citenamefont {Bize},\ and\
  \citenamefont {Wolf}}]{Hees16}%
  \BibitemOpen
  \bibfield  {author} {\bibinfo {author} {\bibfnamefont {A.}~\bibnamefont
  {Hees}}, \bibinfo {author} {\bibfnamefont {J.}~\bibnamefont {Gu\'ena}},
  \bibinfo {author} {\bibfnamefont {M.}~\bibnamefont {Abgrall}}, \bibinfo
  {author} {\bibfnamefont {S.}~\bibnamefont {Bize}}, \ and\ \bibinfo {author}
  {\bibfnamefont {P.}~\bibnamefont {Wolf}},\ }\href {\doibase
  10.1103/PhysRevLett.117.061301} {\bibfield  {journal} {\bibinfo  {journal}
  {Phys. Rev. Lett.}\ }\textbf {\bibinfo {volume} {117}},\ \bibinfo {pages}
  {061301} (\bibinfo {year} {2016})}\BibitemShut {NoStop}%
\bibitem [{\citenamefont {Kennedy}\ \emph {et~al.}(2020)\citenamefont
  {Kennedy}, \citenamefont {Oelker}, \citenamefont {Robinson}, \citenamefont
  {Bothwell}, \citenamefont {Kedar}, \citenamefont {Milner}, \citenamefont
  {Marti}, \citenamefont {Derevianko},\ and\ \citenamefont {Ye}}]{Kennedy20}%
  \BibitemOpen
  \bibfield  {author} {\bibinfo {author} {\bibfnamefont {C.~J.}\ \bibnamefont
  {Kennedy}}, \bibinfo {author} {\bibfnamefont {E.}~\bibnamefont {Oelker}},
  \bibinfo {author} {\bibfnamefont {J.~M.}\ \bibnamefont {Robinson}}, \bibinfo
  {author} {\bibfnamefont {T.}~\bibnamefont {Bothwell}}, \bibinfo {author}
  {\bibfnamefont {D.}~\bibnamefont {Kedar}}, \bibinfo {author} {\bibfnamefont
  {W.~R.}\ \bibnamefont {Milner}}, \bibinfo {author} {\bibfnamefont {G.~E.}\
  \bibnamefont {Marti}}, \bibinfo {author} {\bibfnamefont {A.}~\bibnamefont
  {Derevianko}}, \ and\ \bibinfo {author} {\bibfnamefont {J.}~\bibnamefont
  {Ye}},\ }\href {\doibase 10.1103/PhysRevLett.125.201302} {\bibfield
  {journal} {\bibinfo  {journal} {Phys. Rev. Lett.}\ }\textbf {\bibinfo
  {volume} {125}},\ \bibinfo {pages} {201302} (\bibinfo {year}
  {2020})}\BibitemShut {NoStop}%
\bibitem [{\citenamefont {Hodgson}\ \emph {et~al.}(1997)\citenamefont
  {Hodgson}, \citenamefont {Gadioli},\ and\ \citenamefont
  {Gadioli~Erba}}]{Introductory_Nucl_Phys}%
  \BibitemOpen
  \bibfield  {author} {\bibinfo {author} {\bibfnamefont {P.}~\bibnamefont
  {Hodgson}}, \bibinfo {author} {\bibfnamefont {E.}~\bibnamefont {Gadioli}}, \
  and\ \bibinfo {author} {\bibfnamefont {E.}~\bibnamefont {Gadioli~Erba}},\
  }\href@noop {} {\emph {\bibinfo {title} {Introductory Nuclear Physics}}}\
  (\bibinfo  {publisher} {Oxford, Clarendon, United Kingdom},\ \bibinfo {year}
  {1997})\ p.\ \bibinfo {pages} {723}\BibitemShut {NoStop}%
\end{thebibliography}%
\end{document}